\documentclass[final,letterpaper,leqno, 11pt]{article}

\usepackage{secdot}
\usepackage[letterpaper,left=1in,right=1in, top = 1.2in, bottom=1.2in]{geometry}
\usepackage{titling}
\usepackage{amssymb, amsmath, amsfonts, mathrsfs, cite}
\usepackage[amsmath,thmmarks]{ntheorem}
\usepackage{graphicx} 
\usepackage{booktabs}
\usepackage{color, colortbl}
\usepackage[T1]{fontenc}
\usepackage{enumerate}

\usepackage{subcaption} 
\usepackage{natbib}
\setcitestyle{aysep={}}

\definecolor{Maroon}{cmyk}{.4,1,.3,.2}
\definecolor{LightMaroon}{cmyk}{.04,0.1,.03,.02}
\definecolor{Gray}{cmyk}{0,0,0,.5}
\definecolor{Green}{cmyk}{1,0,1,0}
\definecolor{Red}{cmyk}{0,1,.8,0}
\definecolor{Orange}{cmyk}{0,.55,1,0}
\definecolor{LightOrange}{cmyk}{0,.12,0.3,0.01}
\definecolor{Blue}{cmyk}{.6,.1,.1,.1}

\theoremstyle{plain}
\theoremheaderfont{\normalfont\bfseries}
\theorembodyfont{\normalfont\itshape}
\theoremseparator{.}
\theoremsymbol{}

\newcommand{\QEDbox}{\textcolor{black}{\rule{1.5ex}{1.5ex}}}
\theoremstyle{nonumberplain}
\theoremheaderfont{\normalfont\bfseries} 
\theorembodyfont{\normalfont}
\theoremseparator{.}
\theoremsymbol{\QEDbox}

\newcommand{\newthm}[2]{
	\theoremstyle{plain}
	\theoremheaderfont{\normalfont\bfseries}
	\theorembodyfont{\normalfont\itshape}
	\theoremseparator{.}
	\theoremsymbol{}
	\newtheorem{#1}[theorem]{#2}
}
\newthm{lemma}{Lemma}
\newthm{corollary}{Corollary}
\newthm{proposition}{Proposition}
\newthm{definition}{Definition}
\newthm{assumption}{Assumption}
\newcommand{\newremark}[2]{
	\theoremstyle{plain}
	\theoremheaderfont{\normalfont\bfseries}
	\theorembodyfont{\normalfont}
	\theoremseparator{.}
	\theoremsymbol{}
	\newtheorem{#1}[theorem]{#2}
}
\newremark{remark}{Remark}

\newcommand{\dt}{\partial_t}
\newcommand{\dx}{\partial_x}

\newcommand{\inner}[3]{\left( #2  ,  #3 \right)_{#1}} 

\DeclareMathOperator{\divergence}{div}
\DeclareMathOperator{\sech}{sech}

\def\XXint#1#2#3{{\setbox0=\hbox{$#1{#2#3}{\int}$}
		\vcenter{\hbox{$#2#3$}}\kern-.5\wd0}}

\def\YYint#1#2#3{{\setbox0=\hbox{$#1{#2#3}{\int}$}
		\lower1ex\hbox{$#2#3$}\kern-.46\wd0}}
\def\YYYint#1#2#3{{\setbox0=\hbox{$#1{#2#3}{\int}$}
		\lower0.35ex\hbox{$#2#3$}\kern-.48\wd0}}

\def\ZZint#1#2#3{{\setbox0=\hbox{$#1{#2#3}{\int}$}
		\raise1.15ex\hbox{$#2#3$}\kern-.57\wd0}}
\def\ZZZint#1#2#3{{\setbox0=\hbox{$#1{#2#3}{\int}$}
		\raise0.85ex\hbox{$#2#3$}\kern-.53\wd0}}

\newcommand{\rd}{\mathrm{d}}

\newcommand{\rmm}{\mathrm{m}}

\newcommand{\rA}{\mathrm{A}}
\newcommand{\rB}{\mathrm{B}}

\newcommand{\rD}{\mathrm{D}}

\newcommand{\rH}{\mathrm{H}}

\newcommand{\rK}{\mathrm{K}}

\newcommand{\rM}{\mathrm{M}}
\newcommand{\rN}{\mathrm{N}}
\newcommand{\rQ}{\mathrm{Q}}
\newcommand{\rR}{\mathrm{R}}
\newcommand{\rS}{\mathrm{S}}

\newcommand{\rU}{\mathrm{U}}
\newcommand{\rV}{\mathrm{V}}

\newcommand{\scC}{\textsc{c}}

\newcommand{\scT}{\textsc{t}}

\newcommand{\scP}{\textsc{p}}
\newcommand{\scM}{\textsc{m}}

\newcommand{\bbR}{\mathbb{R}}

\newcommand{\bbN}{\mathbb{N}}

\newcommand{\tT}{\mathtt{T}}
\newcommand{\tX}{\mathtt{X}}
\newcommand{\tN}{\mathtt{N}}
\newcommand{\tQ}{\mathtt{Q}}

\begin{document}

\title{Competition, Trait Variance Dynamics, and the Evolution of a Species' Range
	\thanks{This work was supported by the NSF grant DMS 1615126 to JRM. }
}

\author{Farshad Shirani\thanks{Department of Mathematics and Statistics, Georgetown University, Washington, DC 20057}
	\and Judith R. Miller\protect\footnotemark[2]
}

\maketitle

\begin{abstract}

Geographic ranges of communities of species evolve in response to environmental, ecological, and evolutionary forces.
Understanding the effects of these forces on species' range dynamics is a major goal of spatial ecology. 
Previous mathematical models have jointly captured the dynamic changes in species' population distributions and the selective evolution of fitness-related phenotypic traits in the presence of an environmental gradient. 
These models inevitably include some unrealistic assumptions, and biologically reasonable ranges of values for their parameters are not easy to specify. 
As a result, simulations of the seminal models of this type can lead to markedly different conclusions about the behavior of such populations, including the possibility of maladaptation setting stable range boundaries. 
Here, we  harmonize such results by developing and simulating a continuum model of range evolution in a community of species that interact competitively while diffusing over an environmental gradient. 
Our model extends existing models by incorporating both competition and freely changing intraspecific trait variance.
Simulations of this model predict a spatial profile of species' trait variance that is consistent
with experimental measurements available in the literature.
Moreover, they reaffirm interspecific competition as an effective factor in limiting species' ranges, even when trait variance is not artificially constrained. 
These theoretical results can inform the design of, as yet rare, empirical studies to clarify the evolutionary causes of range stabilization. 


\end{abstract}

\section{Introduction}

Identifying the biotic, genetic, and environmental factors that determine a species' range is a fundamental problem in spatial ecology. 
Sometimes the reason why a range limit is where it is is obvious: a terrestrial species cannot colonize the ocean, for example. Yet often species' range limits do not coincide with a gross environmental discontinuity \citep{Gaston:Oxford:2003}.

Numerous theoretical models have therefore been developed to explain species' range dynamics. These models incorporate a variety of factors and processes, such as dispersal limitations, Allee effects, landscape heterogeneity, environmental stress gradients, niche limitations, gene flow, genetic drift, population genetic structure, and biotic interactions such as competition and predation
\citep{Sexton:AnnualReviewEcology:2009, Bridle:TrendsEcolEvol:2007, Miller:Ecology:2020, Angert:AnnualReview:2020,  Holt:Oikos:2005, Godsoe:TheoreticalEcology:2017, Louthan:TrendsEcolEvol:2015, Case:Oikos:2005}. 
Here, we focus on two factors that are commonly said to limit species' ranges by halting range expansions or stabilizing a range boundary: competition and (mal)adaptation to a heterogeneous environment.

\subsection{Competition, adaptation and species' range dynamics}

Empirical and theoretical studies have confirmed that competing species often segregate in different parts of the habitat available to them and resist encroachment by competitors in areas where they are established \citep{Pigot:EcologyLetters:2013}.
By contrast, less evidence for failure to adapt to a continuously varying environment as a range-limiting factor exists \citep{Micheletti:MolecularEcology:2020, Angert:AnnualReview:2020, Colautti:MolecularEcology:2015, Benning:Naturalist:2019}. 
This is largely due to the difficulty of establishing values for key parameters in evolutionary  models, especially the optimum value of a trait under selection as a function of time and space. 
However, a few empirical studies in both altitudinal and latitudinal environmental gradients support the hypothesis that a population spreading in a heterogeneous environment can be halted by maladaptation, even in the absence of environmental discontinuities \citep{Dawson:MolecularEcology:2010, Sanford:Ecology:2006}.

Theoretical studies of the joint action of competition and adaptation are therefore valuable, as they can point to conditions under which adaptation, or failure to adapt, become important factors in limiting ranges. 
Such studies should therefore help researchers develop tests for selection as a range-limiting factor.
Ultimately, they should inform environmental management decisions, such as allocating conservation effort among marginal or central populations, planning for the impact of climate change on the structure and survival of natural communities, and controlling invasive species.

\subsection{Models of adaptation to an environmental gradient}

The model we present here builds on a sequence of models inspired by an influential idea---genetic swamping---suggested by 
\citet{Haldane:RoyalSociety:1956} and \citet{Mayr:AnimalSpeciesEvolution:1963}. 
They hypothesized that the intraspecific feedback between core-to-edge maladaptive gene flow and population density of a species can limit its range.
Building on the work by \citet{pease1989model},
this process was mathematically modeled by \citet{Kirkpatrick:Naturalist:1997} for a single species in a one-dimensional geographic space---as a system of two partial differential equations that represent joint evolution of the species' population density and the mean value of a quantitative phenotypic trait along a linear environmental gradient in the trait optimum.
The quantitative trait is assumed to be directly associated with the fitness of the species' individuals.
Wing loading and body size in flying insects \citep{Takahashi:MolecularEcology:2016}, or specific leaf area and height in plants \citep{Ackerly:EcologyLetters:2007}, are examples of such fitness-related traits.  
Stabilizing selection on the quantitative trait in the \citeauthor{Kirkpatrick:Naturalist:1997}~(KB) model enables the species to adapt to new areas and expand its range.
However, \citeauthor{Kirkpatrick:Naturalist:1997} showed numerically that with sufficiently steep environmental gradients, gene flow from the densely populated center of the species' range can prevent local adaptation at the periphery and result in an evolutionarily stable range limit---that is, their model demonstrated range pinning through genetic swamping.
A derivation of the KB model as well as analytical results on the existence of range expansion traveling waves and some localized stationary solutions for this model are provided by \citet{Miller:MathBiology:2014}, \citet{Miller:MathBiology:2019}, and \citet{Mirrahimi:PopulationBiology:2013}.

A number of models have been constructed and (usually numerically) analyzed that build on the KB model. Each of these modeling studies takes some factor or factors omitted from the KB model, checks whether it makes range pinning easier or harder to achieve, and in some cases also studies its effect on quantitative properties like invasion speed. 
Such variants include individual-based models, models with stochastic differential equations, and models with discrete time and/or space \citep{duputie2012genetic,polechova2009species,alleaume2006geographical,polechova2018sky,filin2008relation,Bridle:TrendsEcolEvol:2007,Polechova:PNAS:2015,kanarek2010allee}. 
Some models incorporate interspecific competition \citep{leimar2008evolution,goldberg2006ecological} and some incorporate mutation \citep{behrman2011species}.

Among these varied studies, one key extension of the KB model was developed by \citet{Case:Naturalist:2000}. They gave a community context to the KB model by considering multiple species and competitive interactions between them.  
Through numerical simulations of their model, \citeauthor{Case:Naturalist:2000} showed that the interaction of an environmental gradient and gene flow with the frequency-dependent selection generated by phenotypic competition between species can result in 
range limits at much shallower environmental gradients than without competition. 
They also showed that their model generates a number of other important evolutionary phenomena, such as species character displacement in sympatry and range shifts in response to climate change. 
In addition, \citeauthor{Case:Naturalist:2000} challenged the conclusions of \citet{Kirkpatrick:Naturalist:1997} by arguing that competition would be an important force setting range boundaries, even in conditions favorable to genetic swamping, for most realistic parameter combinations. 
Indeed, they argued that realistic trait optimum gradients would rarely be steep enough to limit ranges in the absence of competition \citep{Case:Naturalist:2000}.

Shortly after \citeauthor{Case:Naturalist:2000}'s study appeared, \citeauthor{Barton:BookChapter:2001} extended the original KB model, in which additive genetic variance was held constant, to a model in which this variance was allowed to evolve along with population density and trait mean \citep{Barton:BookChapter:2001}. 
He developed different variants of the model, in which he also incorporated mutation as a constant forcing term in the equations governing the variance.
He concluded from numerical simulations that range pinning via genetic swamping was impossible for these models. 
Instead, he found that genetic variance inevitably rose past a critical value, providing ``fuel'' that allowed even a temporarily pinned population to spread 
over the whole domain. 
In the same book chapter, \citeauthor{Barton:BookChapter:2001} raised two arguments against \citeauthor{Case:Naturalist:2000}'s conclusion that competition would usually play a greater role than genetic swamping in setting range limits. 
Specifically, he argued that adaptation typically involves several traits, which would increase the selective pressures in a swamping model, and that \citeauthor{Case:Naturalist:2000} were too restrictive in their judgment as to how steep a spatial trait optimum gradient might realistically be \citep{Barton:BookChapter:2001}.

\subsection{Present work}

Juxtaposing the work of \citet{Case:Naturalist:2000} with that of \citet{Barton:BookChapter:2001} presents a conflict that must be resolved. 
That range limits can arise in continuously varying environments is well established \citep{Sexton:AnnualReviewEcology:2009}. 
When Barton made the KB model (arguably) more realistic by taking account of mutational and other changes in trait variance, the revised models predicted that such range limits could not be set by genetic swamping \citep{Barton:BookChapter:2001}. 
This would imply that other key drivers of range pinning must exist. 
Competition suggests itself as such a driver, particularly in view of \citeauthor{Case:Naturalist:2000}'s findings \citep{Case:Naturalist:2000}. 
However, Case and Taper held variance constant in their model, so the challenge to the genetic swamping hypothesis posed by \citeauthor{Barton:BookChapter:2001}'s findings \citep{Barton:BookChapter:2001} cannot be settled merely by appealing to \citeauthor{Case:Naturalist:2000}.
Rather, we must test whether competition, combined with selection favoring an optimum trait value that varies in space, can still set range limits when trait variance is allowed to evolve together with population density and trait mean.

Here, we resolve the tension between the disparate conclusions of \citet{Kirkpatrick:Naturalist:1997}, \citet{Case:Naturalist:2000}, and \citet{Barton:BookChapter:2001} by incorporating both competition and nonconstant trait variance in a model of adaptation and spread in an environmental gradient. 
We take particular care to establish plausible ranges for parameter values, since these ranges play a role in the debate over whether genetic swamping is commonly a dominant influence on range limits. 
We explore the behavior of solutions to our model numerically, finding that the results of \citet{Case:Naturalist:2000} are robust when the assumption of constant trait variance is eliminated and even when mutation, as a perpetual source of genetic variance, is incorporated in the model through a forcing term as in the work of \citet{Barton:BookChapter:2001}. 
We focus primarily on the establishment of stable boundaries between the ranges of two competitors initially established in allopatry. 
However, we also briefly investigate other scenarios, including the ability of a newly arrived species to establish itself and spread in a habitat already occupied by a competitor. 

\section{Model description} \label{sec:Discription}

We present our model of species range dynamics in a community of $\rN$ competing species as a system of coupled PDEs.
The derivation of the model is provided in detail in Appendix A, and relies on the following main assumptions about the species' dispersal and reproduction rates, as well as trait distributions and selection.
Mathematical formulations of these assumptions are given in Appendix \ref{sec:Assumptions}.

\begin{enumerate}[i)]
	
\item{Each species disperses by diffusion in a rectangular or linear habitat.}
\item {Trait values within each species are normally distributed at each occupied point in space at all times.\footnote{
	It is argued that normal distribution of phenotypes, which is also assumed in the ancestors of our model, is a reasonable assumption when most of the genetic variation in a species is maintained by migration (gene flow) rather than by mutation \citep{Barton:BookChapter:2001, Barton:GeneticalResearch:1999}.
	This is often the case when a species adaptively expands its range over an environmental gradient, as we primarily study here. } }
\item{Traits are subject to directional and stabilizing selection toward an optimal value that may vary over space and time.}
\item The reproduction rate of individuals with a given phenotype depends (predominantly) on the population density of individuals with the same phenotype.
\footnote{In our model, the reproduction of individuals with phenotype $p$ has been modeled through a logistic growth term that depends on the population density of individuals only with phenotype $p$\,; see equations \eqref{eq:GrowthRate} and \eqref{eq:PhenPopuVariation} in Appendix A. 
Although this logistic population growth fits in with an asexual reproduction system more trivially, it can also approximately accommodate a sexual reproduction system as long as the rate of production of offspring with phenotype $p$ is predominantly proportional to the density of parents with phenotype $p$.
This can approximately occur under our assumption of normal (unimodal) phenotype distribution within each population, provided the populations are sufficiently panmictic. 
}

\item{The strength of competition between the species is determined by each species' pattern of utilizing a common resource.}
\item The probability of mutational changes from one phenotype to another phenotype depends on the difference between the phenotypes. 
Moreover, these phenotypic changes due to mutation follow a distribution with zero mean and constant variance.

\end{enumerate}

To specify the model, we start by defining the $\rmm$-dimensional habitat $\Omega \subset \bbR^{\rmm}$ to be an open rectangle. At position $x=(x_1,\dots, x_{\rmm}) \in \Omega$ and time $t \in [0, T]$, $T>0$, 
we further let $n_i(x,t)$ denote the population density of the $i$th species, and let 
$q_i(x,t)$ and $v_i(x,t)$ denote the mean value and variance of a quantitative phenotypic trait in the $i$th species, respectively. 
Note that, by definition, $n_i(x,t)$ and $v_i(x,t)$ are nonnegative quantities.
For brevity, we define a vector $u$ containing all these state variables:
\begin{equation*}
u = (n_1, q_1, v_1, \dots, n_{\rN}, q_{\rN}, v_{\rN}).
\end{equation*}
Now, letting $\rD_i(x)$ and $G_i(x,t)$ denote the diffusion coefficient and mean growth rate, respectively, of the $i$th species, we write the equation for the population density $n_i$ as
\begin{equation}
	\dt n_i(x,t) = \divergence(\rD_i(x) \dx n_i(x,t)) + G_i(x,u(x,t)) n_i(x,t). \label{eq:PopulationDensity}
\end{equation}
The functional form of $G_i$ is given below.
Likewise, equations for the trait mean $q_i$ and variance $v_i$ of the $i$th species are given as 
\begin{equation}
\dt q_i(x,t) = \divergence(\rD_i(x) \dx q_i(x,t)) + 2 \inner{\bbR^{\rmm}}{\dx \log n_i(x,t)}{\rD_i(x) \dx q_i(x,t)} + H_i(x,u(x,t)), \label{eq:TraitMean}
\end{equation}
and
\begin{align}
\dt v_i(x,t) &= \divergence(\rD_i(x) \dx v_i(x,t)) + 2 \inner{\bbR^{\rmm}}{\dx \log n_i(x,t)}{\rD_i(x) \dx v_i(x,t)} \nonumber\\
	&\quad + 2 \inner{\bbR^{\rmm}}{\dx q_i(x,t)}{\rD_i(x) \dx q_i(x,t)} + W_i(x,u(x,t)). \label{eq:TraitVariance}
\end{align}
Here, the nonlinear mappings $G_i$, $H_i$, and $W_i$ are defined as 
\begin{align} 
	G_i(x,u) &= \rR_i(x)  - \frac{\rR_i(x)}{\rK_i(x)} \sum_{j=1}^{\rN} M_{ij}(u) C_{ij}(u) n_j 
	- \frac{\rS}{2}  \left[(q_i - \rQ(x))^2 + v_i\right], \label{eq:G}\\
	H_i(x,u) &= (\rR_i(x) - G_i(x,u)) q_i - \frac{\rR_i(x)}{\rK_i(x)} \sum_{j=1}^{\rN} L_{ij}(u) M_{ij}(u) C_{ij}(u) n_j 
	+ E_i(x,u), \label{eq:H} \\
	W_i(x,u) &= (\rR_i(x) - G_i(x,u)) (v_i - q_i^2) - \frac{\rR_i(x)}{\rK_i(x)} \sum_{j=1}^{\rN} P_{ij}(u) M_{ij}(u) C_{ij}(u) n_j 
	+ Y_i(x,u), \label{eq:W}
\end{align}
where, letting $\Lambda_{ij}:=\sqrt{ \smash[b]{ \rV_i / \bar{\rV}_{ij}}}$ with $\bar{\rV}_{ij}:= \frac{1}{2}(\rV_i + \rV_j)$,
\begin{align} 
	C_{ij}(u) &:= \sqrt{\frac{2 \bar{\rV}_{ij}}{v_i + v_j + 2 \bar{\rV}_{ij}}} \Lambda_{ij}  \exp(\kappa^2 \bar{\rV}_{ij}), \nonumber\\
	M_{ij}(u) &:= \exp \left(-\frac{ (q_i - q_j + 2 \kappa \bar{\rV}_{ij})^2 }{2 (v_i + v_j + 2 \bar{\rV}_{ij})} \right), \nonumber\\
	L_{ij}(u) &:= \frac{v_i (q_j - 2 \kappa \bar{\rV}_{ij}) + (v_j + 2 \bar{\rV}_{ij})q_i}{v_i + v_j + 2 \bar{\rV}_{ij}}, \nonumber\\
	P_{ij}(u) &:= \frac{v_i(v_j + 2 \bar{\rV}_{ij})}{v_i + v_j + 2 \bar{\rV}_{ij}} + L_{ij}(u) (L_{ij}(u) - 2 q_i), \nonumber\\
	E_i(x,u) &:= \frac{\rS}{2 } \left[ 2 \rQ(x) v_i + 2 \rQ(x) q_i^2 - \rQ^2(x) q_i - 3 v_i q_i - q_i^3 \right], \nonumber\\
	Y_i(x,u) &:= \frac{\rS}{2} \left[ 2 \rQ(x) v_i q_i - 2 \rQ(x) q_i^3 - \rQ^2(x) (v_i - q_i^2) - 3 v_i^2  + q_i^4 \right] \mathbf{ +\rU}, \hspace{3em} i,j \in \{1,\dots, \rN\}.  \label{eq:NonlinearTerms}
\end{align}
In \eqref{eq:PopulationDensity}--\eqref{eq:TraitVariance}, the partial derivative with respect to $t$ is denoted by $\dt$, the gradient with respect to $x$ is denoted by $\dx$, the divergence with respect to $x$ is denoted by $\divergence$, and the standard inner product in $\bbR^{\rmm}$ is denoted by $\inner{\bbR^{\rmm}}{\cdot}{\cdot}$. 
Definitions of the model parameters and plausible ranges for their values are given in Table \ref{tb:Parameters}. 
Further discussion on parameter units and the choice of typical values for the computational results of this paper are provided in Section \ref{sec:Parameters}. 
Note that $\rD_i(x) \in \bbR^{\rmm \times \rmm}$, whereas the rest of the parameters are scalar valued.
Moreover, $\rS$, $\rU$,  $\rV_i$, and $\kappa$ are assumed to be constant all over the habitat, whereas $\rD_i$, $\rK_i$, $\rR_i$, and $\rQ$ can be variable in space. 
All these model parameters may also vary in time, although their dependence on $t$ is not explicitly shown in the equations.

\begin{table}[t!]
	\caption{ 
		Definition and plausible range of values of the parameters of the model \eqref{eq:PopulationDensity}--\eqref{eq:NonlinearTerms}.
		The typical values given here are the values used in the numerical studies of Sections \ref{sec:SingleSpecies} and \ref{sec:TwoSpecies}. 
		When $\rmm >1$, the range of values specified for $\rD_i(x)$ can be considered for each entry of $\rD_i(x)\subset \bbR^{\rmm \times \rmm}$. 
		Typically, $\rD_i$ is assumed to be diagonal. 
		The additional typical value $0.02$ provided for $\rU$ is suggested in Section \ref{sec:ParameterValues} based on estimates available in the literature.  However, $\rU = 0$ is used for the numerical simulations of Sections \ref{sec:SingleSpecies} and \ref{sec:TwoSpecies}.  
		The typical value ``Linear'' specified for $\rQ$ means that $\rQ$ is typically considered to be a linear function of $x$ over $\Omega$. 
	}
	\label{tb:Parameters}
	
	\begin{center} \small 
		\renewcommand{\arraystretch}{1.3}
		\begin{tabular}{lllll}
			\hline
			\textbf{Parameter} & \textbf{Definition} & \textbf{Range} & \textbf{Typical} & \textbf{Unit} \\ 
			\hline
			$\rmm$ & Spatial dimension of the geographic space & $\{1,2,3\}$ & $1$, $2$ & ---\\
			$\rN$ & Number of species & $\bbN$ & $1$, $2$ & ---\\
			$\rD_i(x)$ & Diffusion coefficient of the $i$th species dispersal  & $[0,25]$ & $1$ & ${\tX}^2 / \tT$\\ 
			$\rK_i(x)$ & Carrying capacity of the environment for $i$th species  & $(0, 10]$ & $1$ & $\tN / \tX^{\rmm}$ \\
			$\rR_i(x)$ & Maximum population growth rate of $i$th species  & $[0.1, 10]$ &$2$ & $1/{\tT}$\\
			$\rV_i$ & Variance of phenotype utilization within $i$th species & $[0.25, 25]$ & $4$ &  ${\tQ}^2$\\ 
			$\kappa$ & Asymmetric impact factor of phenotypic competitions  & $[0, 1]$ & $0$ & $1/\tQ$\\
			$\rS$ & Measure of the strength of stabilizing selection & $[0,2]$ & $0.2$ & ${\tQ}^{-2} / \tT$\\ 
			$\rU$ & Rate of increase in trait variance due to mutation & $[0,0.2]$ & $0 (0.02)$ & ${\tQ}^2 / \tT$\\
			$\rQ(x)$ & Optimal trait value for the environment & $[0, \infty)$ & Linear & $\tQ$\\
			$|\rd_x \rQ(x)|$ & Magnitude of the gradient of the optimal trait & $[0, 10]$ & $0.2$ & $\tQ / \tX$\\
			\hline
		\end{tabular}
	\end{center}
\end{table}

\begin{remark}[Boundary conditions] \label{rmk:BoundaryConditions} 
	In general, different boundary conditions can be imposed on the model \eqref{eq:PopulationDensity}--\eqref{eq:NonlinearTerms} based on the spatial dimension and specific environmental conditions of a problem under study.
	For the general purpose of the computational studies performed in the present work,
	we assume that there is typically no phenotypic flux through the boundary of a one-dimensional habitat
	$\Omega = (a,b)$. 
	This, as explained in Appendix \ref{sec:EquationsDerivation}, simply implies the following homogeneous Neumann conditions
	\begin{equation} \label{eq:ReflectingBC}
	\dx n_i = 0, \quad \dx q_i = 0, \quad \dx v_i = 0, \quad i = 1,\dots,\rN \qquad \text{on } \{a, b\}\times[0,T].
	\end{equation} 
	This boundary condition is equivalent to reflectively extending the equations over $\bbR$.
	For a two-dimensional rectangular habitat, the same reflecting boundary condition can be considered at all boundary lines.
	However, the environmental gradient in trait optimum, such as latitudinal gradient, is often assumed to occur only along one spatial dimension, say in the $x_1$ direction.
	In this case, the reflecting boundary condition described above can be typically considered at habitat boundaries along the $x_1$-axis.
	Across the boundary lines in the $x_2$ direction, periodic boundary conditions can be used provided the other model parameters also take the same values at opposing points on these boundary lines.
	In particular, periodic extension is not recommended along a spatial dimension that presents monotonic changes in the environmental trait optimum $\rQ$,
	because the periodic extension of $\rQ$ in this case will have jumps at the boundaries along this spatial dimension. 
	Since the species' trait means $q_i$ tend to converge to $\rQ$ in response to the force of natural selection, numerically computed solutions of $q_i$ will develop large gradients, $\dx q_i$, near these boundaries.  
	This can result in singularities in numerical computations of the solutions, particularly due to the term $\inner{\bbR^{\rmm}}{\dx q_i}{\rD_i \dx q_i}$ in \eqref{eq:TraitVariance} which will take growing values near such boundaries.
	\qed
\end{remark}

\section{Model parameters} \label{sec:Parameters}

The behavior of the model described in Section \ref{sec:Discription} will depend on the values of its parameters, which will differ from one species to another.
Although the model is fairly abstract, meaningful ranges of parameter values can still be identified. 
In this section, we discuss parameter values and their units as given in Table \ref{tb:Parameters}.
We also specify the typical values of the parameters that are used in Sections \ref{sec:SingleSpecies} and \ref{sec:TwoSpecies} for numerical studies of the model.

\subsection{Parameter units} \label{sec:Units}
Biological species are diverse in their level of abundance, growth rate, dispersal range, and the nature of their functional traits.
This implies that an appropriate choice of units for the physical quantities of the model such as time, space, populations density, and trait values must be species-dependent.
For this, we first select one of the species from the community of $\rN$ species in the model as a \emph{representative species}. 
The physical units are then chosen (in principle) based on specific measurements made on this species.
The representative species can be selected, for example, as the species that is best adapted to the environment, is most widely spread, or has the widest trait niche among the community.
  
Let $\tT$ denote the unit of time. 
We set $1 \; \tT$ to be equal to the mean generation time of the representative species. 
This is a natural choice of time unit to analyze population dynamics of species over evolutionary time scales; see, for example, the time unit used by \citet{Estes:Naturalist:2007}.
It also makes the model compatible with common experimental approaches in estimating parameters such as the the strength of phenotypic selection, which is often estimated by measuring changes in the mean phenotype of the population in one generation.
Moreover, choosing generation time as unit of time can make the predictions of the model comparable with the results obtained from discrete-time individual-based models which usually describe the evolution of the population at generation time steps; see, for example, the models used by \citet{Polechova:PNAS:2015} and \citet{Bridle:EcologyLetters:2010}.    

To set the unit of space, denoted by $\tX$, we first consider a one-dimensional habitat, that is, $x \in \bbR$.
It can be seen in equations \eqref{eq:PopulationDensity}-\eqref{eq:TraitVariance} that rescaling the space as $x \mapsto kx$ leaves the equations unchanged, provided the diffusion coefficients $\rD_i$ are rescaled accordingly as $\rD_i \mapsto \rD_i / k^2$.
Using this flexibility in the equations and having set the unit of time, we choose the unit of space so that the dispersal coefficient of the representative population becomes unity, that is, $1 \;\tX$ is the root mean square dispersal distance of the population in $1\; \tT$ divided by $\sqrt 2$.  
Since the population dispersal may vary at different locations of the habitat, the measurements for setting the unit can be done based on a local subpopulation at the core of the population or at regions where dispersal is not affected by environmental barriers. 
For multi-dimensional habitats, the same approach can be used to set the unit of space for each spatial dimension independently.
We denote the units associated with all dimensions by $\tX$, noting that $\tX$ may refer to different physical scales for different spatial axes.

Similarly, rescaling the population density of the species as $n_i \mapsto k n_i$ does not change the equations of the model provided the carrying capacity of the environment is rescaled accordingly as $\rK_i \mapsto k \rK_i$.  
Therefore, having set the unit of space, we choose the unit of measurement for population abundances, denoted by $\tN$, so that the carrying capacity of the environment for the representative population becomes unity.
That is, $1 \; \tN$ is equal to the carrying capacity of the environment for $1 \; \tX^{\rmm}$ unit of habitat volume.  
The required measurement of the carrying capacity can be done locally at the core of the representative population where it has the largest and most stable population density.  

Finally, we denote the unit of measurement for the quantitative trait by $\tQ$, and set $1 \; \tQ$ to be equal to one standard deviation of the trait values at the core of the representative population, where the population likely shows highest variance in the individual's trait values.
This is a common choice of unit for quantitative traits, which provides generality for quantitative models of evolutionary processes by making them independent of the diverse nature of the quantitative traits across different species
\citep{Kingsolver:Naturalist:2001, Lande:Evolution:1983, Estes:Naturalist:2007, Kirkpatrick:Naturalist:1997, Case:Naturalist:2000}.

\subsection{Parameter values} \label{sec:ParameterValues}

Based on the units chosen in Section \ref{sec:Units} for measuring physical quantities of the model, plausible ranges of parameter values can be suggested as follows.

\subsubsection*{Range of values for carrying capacities and dispersal coefficients}
The choices of units described in Section \ref{sec:Units} suggest a typical value of $1$ for $\rK_i$ and diagonal entries of $\rD_i$.
To take into account the heterogeneity of the environment and variations among species, we suggest a range of values for these parameters within one order of magnitude above and below these typical values.
Note that $\rD_i \geq 0$, whereas the equations of the model require $\rK_i$ to be nonzero, that is $\rK_i >0$.

\subsubsection*{Range of values for maximum growth rates}

The maximum population growth rate $\rR_i$ is attained by a local population of the $i$th species that is fully adapted to the environment, has access to abundant resources, is not involved in any effective interspecific competition with other species, and has low density so that the effect of intraspecific competition is minimal.
It is shown in the literature that if the generation time is chosen as the unit of time $\tT$, as is the case here, then the maximum intrinsic growth rate will be a demographic invariant within some homogeneous taxonomic groups \citep{Niel:ConservationBiology:2005}.
For a variety of taxa such as mammals, birds, sharks, and turtles, the maximum population growth per generation is shown to be approximately equal to $1$, \citep{Hatch:EcologyEvolution:2019, Dillingham:EcologicalApplications:2016, Niel:ConservationBiology:2005}.
Under optimal laboratory conditions, an estimate of the intrinsic rate of natural increase for a shorter-lived species such as a fruit fly is given by \citet[Table~2]{Emiljanowicz:Entomology:2014} as $0.179$ per capita per day.
With the estimate of mean generation time provided for this species as $\tT = 30.6$ days, the maximum population growth of the species can then be estimated as $0.179 \times 30.6 = 5.48 \, \tT^{-1}$.
For a variety of other species of insects, the estimates given by \citet[Table 8.2]{Pianka:EvolutionaryEcology:2000} and \citet{Birch:AnimalEcology:1948} for the maximum population growth rate vary within the range of $0.9\, \tT^{-1}$ to $9.6 \, \tT^{-1}$.   
Moreover, the estimates given by \citet[Table~8.2]{Pianka:EvolutionaryEcology:2000} for two species of Protozoa  lie within $0.1$-$0.6$ $\tT^{-1}$.
Therefore, considering these sample values, we suggest the range of values for $\rR_i$ to be between $0.1\, \tT^{-1}$ and $10\,\tT^{-1}$. 
We choose a typical value of $\rR_i = 2\, \tT^{-1}$ for the numerical studies of this paper.   

\subsubsection*{Range of values for the strength of stabilization selection}

Estimates of the strength of phenotypic selection are available in the literature for a variety of species \citep{Kingsolver:Naturalist:2001, Stinchcombe:Evolution:2008}.
These estimates are usually provided in the form of standardized linear (directional) and quadratic (stabilizing/disruptive) selection gradients, as defined by \citet{Lande:Evolution:1983}.  
In order to be able to use these estimates for suggesting a plausible range of values for $\rS$, we must first identify the relation between this parameter and the standardized selection gradients.
For this, we first establish an adaptive landscape associated with the model,
based on the approximate evolution of a single population in the absence of population dispersal, interspecific competition, and mutation.

Let the population density of individuals with phenotype $p$ be denoted by $n_p(t):=n(t)\phi(t,p)$, where $\phi$ gives the relative frequency of $p$ as defined in Appendix \ref{sec:GrowthRate}. 
Note that the dependence of variables on $x$ and the numeration index $i=1$ are dropped for the single ($\rN=1$) local population under consideration. 
Over a small time step $\tau$, the population density evolves approximately as
\begin{equation*}
	n_p(t+\tau) - n_p(t) = \tau g(t,p) n_p(t), 
\end{equation*}
where $g(t,p)$ gives the intrinsic growth rate of the population as defined by equation \eqref{eq:GrowthRate} in Appendix \ref{sec:GrowthRate}.
Therefore, after one generation time we have $n_p(t+1) \approx (1 + g(t,p)) n_p(t)$, which identifies the fitness function $f(t,p):=1 + g(t,p)$ for individuals with phenotype $p$.
Corresponding to this phenotypic fitness function, an adaptive landscape can be defined as
\begin{equation*}
	F(n,q,v):=\int_{\bbR}f(t,p)\phi(t,p)\rd p = 1 + \int_{\bbR}g(t,p)\phi(t,p)\rd p,
\end{equation*}
which particularly relates the mean value of the fitness across the population to the mean value of the phenotypic trait \citep{Hendry:EcoEvoDynamics:2016}.
The mean growth rate $G(n,q,v) := \int_{\bbR}g(t,p)\phi(t,p)\rd p$ can be obtained from \eqref{eq:G} by setting $\rN =1$ and $\kappa = 0$, which yields
\begin{equation} \label{eq:Landscape}
		F(n,q,v) = 1 + G(n,q,v) = 1 + \rR \left(1 - \frac{n}{\rK} \sqrt{\frac{\rV}{v+\rV}} \right)  
	- \frac{\rS}{2}  \left[(q - \rQ)^2 + v\right].
\end{equation}
See Appendix \ref{sec:EquationsDerivation} for the derivation of $G$ in the general case.

Standard measures of selection are then obtained by calculating the slope and curvature of the logarithmically scaled adaptive landscape along the mean trait axis \citep{Estes:Naturalist:2007, Lande:Evolution:1979}.
That is, $\partial_q \log F(n,q,v)$ provides a measure of linear selection, and $\partial_q^2 \log F(n,q,v)$ provides a measure of quadratic selection. 
These estimates are related to the standardized directional selection gradient $\beta$ and the standardized stabilizing/disruptive selection gradient $\gamma$ as 
$\partial_q \log F = \beta$ and $\partial_q^2 \log F = \gamma - \beta^2$;
see \citet{Phillips:Evolution:1989} and the derivation given by \citet[Suppl. Appx.]{Estes:Naturalist:2007}.
Now, using \eqref{eq:Landscape}, a first-order approximation of $\log F$ along the $q$-axis  gives $\log F \approx G$.
Therefore, we have   
\begin{align} \label{eq:SelectionFormulas}
	\beta &= \partial_q \log F(n,q,v) \approx \partial_q G(n,q,v) = -\rS (q-\rQ), \nonumber \\
	\gamma - \beta^2 &=\partial_q^2 \log F(n,q,v) \approx \partial_q^2 G(n,q,v) = -\rS.
\end{align}
Then, plausible values for $\rS$ can be obtained as $\rS\approx - \gamma + \beta^2 $, using the estimates of $\beta$ and $\gamma$ available in the literature.
Note that our assumption of stabilizing selection, specified as assumption (\ref{assmpt:Selection}) in Appendix \ref{sec:Assumptions}, implies that 
$\gamma <0$ and $\rS > 0$. 

\begin{figure}[t!]
	\centering
	\parbox{0.48\textwidth}{
		\includegraphics[width=0.48\textwidth]{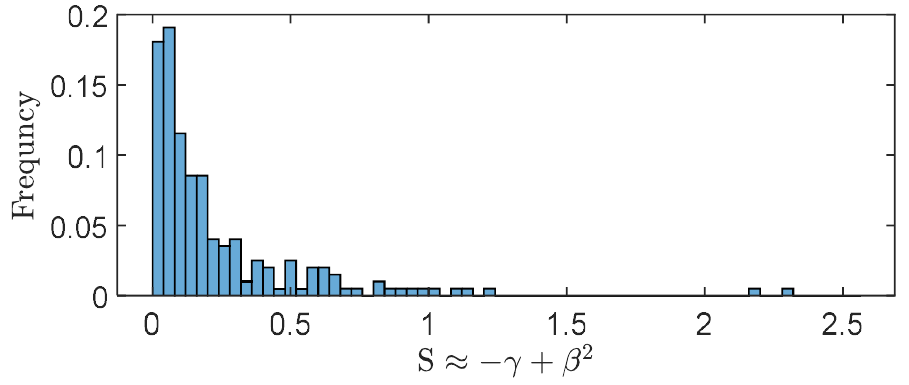}
	}
	\hfill
	\parbox{0.48\textwidth}{ 
		\begin{tabular}{llll}		
			\hline  \\[-0.95em]
			& $-\gamma$ & $\beta^2$ & $ \rS \approx -\gamma +\beta^2$  \\[1pt]
			\hline \\[-0.95em]
			Minimum & $7 \times 10^{-6}$ & $0$ & $1.07 \times 10^{-4}$ \\[1pt]
			Maximum & $2.1960$ & $2.04$ & $2.58$ \\[1pt]
			Mean & $0.18$ & $0.08$ & $0.26$ \\[1pt]
			Median & $0.08$ & $0.01$ & $0.13$ \\[1pt]
			$95$th Percentile & $0.64$ & $0.27$ & $0.92$ \\[1pt]
			\hline
		\end{tabular}
	}
	\caption{Distribution of the estimated values of $\rS$ based on the dataset provided by \citet{Kingsolver:Dryad:2008}} \label{fig:selectionestimate}.
\end{figure}

Estimates of linear or quadratic selection provided by $63$ studies for a total number of $62$ species are analyzed by \citet{Kingsolver:Naturalist:2001} and the resulting dataset is made available by \citet{Kingsolver:Dryad:2008}.
Here, we choose those species in this database for which both estimates of $\gamma$ and $\beta$ are available and have $\gamma<0$. 
This results in a dataset of $199$ estimates, based on which we obtain a set of realistic values for $\rS$.
The distribution of these estimated values are shown in Figure \ref{fig:selectionestimate}, along with basic statistical measures of the components of the estimates. 
The results show a median value of $0.13$ and a $95$th percentile of $0.92$ for $\rS$.
However, a fair amount of these estimates are likely to be underestimates due to the error made in some of the studies when using quadratic regression coefficients for estimating $\gamma$.
This error was identified by \citet{Stinchcombe:Evolution:2008}, and results in an underestimation of $|\gamma|$ by a factor of $2$.
Specifically, analyzing $673$ estimates of $\gamma$ from $32$ additional studies, \citet{Stinchcombe:Evolution:2008} show an underestimation of $25-40\%$ in the typical value of $\gamma$.
Therefore, taking into account the possibility of underestimation, we suggest a typical value of $0.2$ and a maximum value of $2$ for the parameter $\rS$ of the model.  

\begin{remark}[Constraint on the values of $\rS$ and $\rR$]
	Equations of the model impose a constraint on the maximum value that $\rS$ can take, or the minimum value that $\rR$ can take, with respect to each other.
	To find this constraint, consider a single population of minimal density $n \ll \rK$, so that the effect of intraspecific competition is negligible.
	Suppose that the population is well-adapted to the environment, that is, $q-\rQ \approx 0$. 
	For this population, we expect a mean fitness of $F > 1$, or equivalently, a mean growth rate of $G>0$ in \eqref{eq:Landscape}, so that this minimally constrained population can grow.
	This gives the constraint 
	\begin{equation} \label{eq:Constraint_SR}
	\rR -\frac{\rS}{2} v  > 0,
	\end{equation}
	where $v$ is the variance of trait values within the population. 
	Since $v$ can increase during the growth of the population, as a rough estimate we expect that \eqref{eq:Constraint_SR} is satisfied at least for $v=1 \, \tQ^2$, that is the variance of trait values in the representative population used to determine the unit of the trait, as described in Section~\ref{sec:Units}.  
	This gives the approximate constraint $\rS < 2 \rR$.
	\qed	
\end{remark}

\subsection*{Range of values for the mutational rate of increase in trait variance}

The amount of increase in genetic variance of phenotypic traits per one generation of mutation---after being standardized with (divided by) the estimates of environmental
variance of the trait---is known as \emph{mutational heritability} in the literature. 
Estimates of this dimensionless quantity are available for a variety of species \citep{Houle:Genetics:1996} and can be used here to obtain biologically reasonable values for $\rU$.
Note that the environmental trait variance is $(1 - \rH^2)$ times the total phenotypic variance, where $\rH^2$ denotes the broad sense heritability \citep{Visscher:Nature:2008}. 
Moreover, since the standard deviation of the phenotypes at the core of a representative population is chosen here as the unit of trait, as described in Section \ref{sec:Units},
the typical value of the total trait variance in our model is expected to be approximately $1 \, \tQ^2$.
Therefore, the estimates of mutational heritabilities provided by \citet[Table 1]{Houle:Genetics:1996}, after being multiplied by the factor $(1 - \rH^2)$, give plausible values for $\rU$.

The standardized values given by \citet[Table 1]{Houle:Genetics:1996} are approximately in the range $[0.1, 30]\times 10^{-3}$. 
Typical values of heritability for fitness-related traits, as given by \citet[Figure 1]{Visscher:Nature:2008} for a number of different species, range from $0.05$ to $0.3$.
These results can suggest $[0.1 \times (1-0.3), 30 \times (1-0.05)] \times 10^{-3} \approx [7\times 10^{-5}, 0.029] \, \tQ^2 /\tT$ as an approximate range of values for $\rU$.  
Specifically, the standardized estimates of mutational rate of increase provided by \citet[Table 1]{Houle:Genetics:1996} for \emph{Daphnia pulex} are in the range $[0.8, 3.3] \times 10^{-3}$, and for \emph{Drosophila melanogaster} are in  the range $[0.1,13.5]\times 10^{-3}$. 
Values of heritability given by \citet[Figure 1]{Visscher:Nature:2008} for fitness-related traits in \emph{Daphnia} and \emph{Drosophila} are approximately $0.3$ and $0.18$, respectively.
Noting that not all the traits considered by \citet[Table 1]{Houle:Genetics:1996} are fitness traits, these results give the estimates  
$[0.8\times (1-0.3), 3.3\times (1-0.3)] \times 10^{-3} \approx [0.0006, 0.0023] \, \tQ^2 /\tT$ and 
$[0.1\times (1-0.18), 13.5\times (1-0.18)] \times 10^{-3} \approx [8\times 10^{-5}, 0.01] \, \tQ^2 /\tT$ for the range of values that $\rU$ can take for \emph{Daphnia} and \emph{Drosophila}, respectively.

In addition to providing estimates of the mutational rate of increase in trait variance relative to the environmental variance, \citet{Houle:Genetics:1996} also provide estimates of the rate relative to the standing genetic variance.
Noting that genetic variance is equal to $\rH^2$ times the total phenotypic variance \citep{Visscher:Nature:2008}, we can use similar calculations as given above to obtain additional estimates of values for $\rU$. 
The estimates provided by \citet[Figure 4]{Houle:Genetics:1996} are in the range $[10^{-2.75}, 10^{-1.25}] \approx [0.002, 0.056]$, relative to the genetic variance.
Therefore, considering the values of heritability from $0.05$ to $0.3$, as given above, we obtain a range of possible values for $\rU$ as $[0.002 \times 0.05, 0.056 \times 0.3] \approx [0.0001, 0.017] \, \tQ^2 /\tT$.  
It should be noted that, since in our model we implicitly assume that variation in phenotypes is mainly caused by genetic effects, that is $\rH^2 \approx 1$, we expect to use slightly larger values than those estimated here for te parameter $\rU$ in our model. 

An extremely rough estimate for the range of values of $\rU$ can be obtained from the spatially homogeneous equilibrium value of trait variance in \eqref{eq:PopulationDensity}--\eqref{eq:TraitVariance}.
For a perfectly adapted solitary species at spatially homogeneous equilibrium, and in the absence of environmental gradients and intraspecific competition, the equilibrium trait variance $v^{\ast}$ is maintained by the mutation-selection balance $\rU = \rS {v^{\ast}}^2$; see equation \eqref{eq:TraitVariance_Single} and the discussions in Section \ref{sec:SingleSpecies}.  
With a typical value of $v^{\ast} = 1 \, \tQ^2$, this gives estimates of values for $\rU$ equal to the strength of stabilizing selection $\rS$, which as discussed above may range from $0$ to $2$.
However, gene flow over an environmental cline and intraspecific competition are indeed  prominent sources of producing genetic variation, which were ignored in obtaining this rough estimate of the range of values for $\rU$.
Therefore, a value of $\rU = 2 \, \tQ^2 /\tT$ will be too extreme to be realistic, although in shallow environmental gradients we may expect the value of $\rU$ to be relatively close to $\rS$.

Finally, considering altogether the estimates and sample values provided above, we suggest a plausible range of values for $\rU$ as $[0, 0.2] \, \tQ^2 /\tT$, with a typical value of $0.02$.
However, we set $\rU = 0$ in all of the computational studies presented in this paper, except for the results discussed in Remark \ref{rmk:Mutation} in Section \ref{sec:SingleSpecies}. 
This is because the inflation caused by mutation in species' trait variance does not affect the conclusions of our numerical studies; see Remark~\ref{rmk:Mutation} for details.

\subsubsection*{Range of values for the variance of phenotype utilization distributions}

We suggest plausible values for the variance parameter $\rV_i$ of phenotype utilization distributions, as defined by \eqref{eq:UtilizationDistribution} in Appendix \ref{sec:CompetionKernels},
by first noting that resource utilization curves, as described in Appendix \ref{sec:CompetionKernels}, are often used to quantify species' niches \citep{Colwell:Ecolofy:1971, Roughgarden:PopulationGenetics:1979, Pianka:PNAS:1974}.
In particular, the variance of utilization curves can quantify the within-phenotype component of a species' niche breadth.
For categorical resources such as food type or microhabitats, comparative quantification of species' niche can be performed for a fairly large community of species; see, for example, the quantification by \citet{Pianka:AnnualReviewEcolology:1973}.
However, estimation of the parameters of species' environmental niche may not be practicable for a continuum of quantitative resources.
The subsequent resource-phenotype identification described in Appendix \ref{sec:CompetionKernels} is also hard to establish.

By contrast, the trait-based niche quantification approach proposed by \citet{Ackerly:EcologyLetters:2007} and advocated by \citet{Violle:PlantEcology:2009} seems to provide a more straightforward method for estimating the variance of phenotype utilization. 
In this approach, a species' niche is defined directly based on trait values instead of environmental parameters.
The breadth of a species' trait niche is then simply measured as the total intraspecific trait variation across the species over the entire range of its habitat. 
To suggest a typical value for $\rV_i$, we assume that the representative population is composed of generalist individuals, so that $\rV_i$ is comparable to the species' trait niche breadth. 
Based on the specific choice of trait unit described in Section \ref{sec:Units}, the standard deviation of the trait values at the core of the population is approximately $1 \, \tQ$.
However, the intraspecific trait variation can increase due to environmental gradients as the population spreads and fills more of its niche. 
To take this partially into account, we  roughly consider the width of phenotype utilization curves to be twice as large as the standard deviation of the trait at the core, giving a typical value of $4$ for $\rV_i$.
We suggest a range of values between $0.25$ and $25$ to include variations due to adaptive evolution and variations among the species in the community.  

\subsubsection*{Range of values for the asymmetric competition factor}
 
For $\kappa$, we choose a typical value of $0$, which implies symmetric intraspecific competition between phenotypes.
Since $\kappa$ is the rate of an exponential growth in the total phenotype utilization, given by \eqref{eq:UtilizationFunction} in Appendix \ref{sec:CompetionKernels}, we expect $\kappa$ to be relatively small.
Therefore, we suggest a maximum value of $1$, which according to \eqref{eq:CompetitionKernel} implies an asymmetric competition of factor $2 \bar{\rV}_{ij}$ between phenotypes.  

\subsubsection*{Range of values for the environmental trait optimum}

To suggest ranges of values and patterns of variation for the environmental trait optimum $\rQ$, first note that the equations of the model \eqref{eq:PopulationDensity}--\eqref{eq:W} are invariant to the additive changes 
\begin{equation*}
	\rQ \mapsto \rQ + c, \quad q_i \mapsto q_i + c,\quad  i = 1, \dots \rN,
\end{equation*}
where $c$ is a constant. 
This can be seen by noting that $C_{ij}$ and $M_{ij}$ in \eqref{eq:NonlinearTerms} are invariant to this additive change, whereas 
\begin{align*}
	L_{ij} &\mapsto L_{ij}+c,\\
	P_{ij} &\mapsto P_{ij} - c (2 q_i + c),\\
	E_i &\mapsto E_i - \frac{\rS}{2} [(q_i - \rQ)^2 +v_i] c,\\
	Y_i &\mapsto Y_i + \frac{\rS}{2} [(q_i - \rQ)^2 +v_i] c (2 q_i + c).
\end{align*}
Substituting these changes in \eqref{eq:G}--\eqref{eq:W} gives the invariance of $G_i$, $H_i$, and $W_i$, and subsequently the invariance of \eqref{eq:PopulationDensity}--\eqref{eq:TraitVariance}. 
Using this invariance property of the equations, we can assume without loss of generality that $\rQ$ is nonnegative everywhere on $\Omega$. 
If $\rQ$ takes negative values in practice, it can always be shifted up by a positive constant $c$ so that it becomes nonnegative everywhere. 
This shift simply shifts the $q_i$ component of the solutions by the same constant $c$, and has no impact on the evolutionary behavior of the model. 
Moreover, the constant $c$ can always be chosen so that it shifts the minimum value of $\rQ$ to zero. 

Finally, the spatial pattern of variation in $\rQ$ is assumed to be monotonic along one spatial dimension.
This can, for example, represent latitudinal or elevational clines in the optimal trait.
Here, we further assume that these monotonic changes are linear.
The slope of this linear trait optimum gradient, measured here in units of phenotypic standard deviations per ($1/\sqrt{2}$ times) root mean square dispersal distance in one generation time, is a key to the differing conclusions by \citet{Case:Naturalist:2000} and \citet{Kirkpatrick:Naturalist:1997}.
An estimate for the slope of optimum gradient might in principle be obtained by measuring the trait values at the core of a well-established representative population which can be considered to be closely adapted to the environmental optimum.
Although such measurements are available for some species over certain geographic regions, for example for a species of damselflies in Japan \citep{Takahashi:MolecularEcology:2016}, they cannot by themselves be used to provide estimates of the gradient of $\rQ$ for the model presented here.
This is because the unit of space $\tX$ used in the available experimental studies, for example latitude degrees as used by \citet{Takahashi:MolecularEcology:2016}, is not consistent with the choice of $\tX$ suggested in Section \ref{sec:Units}---which requires both measurements of dispersal distance and generation time.
Therefore, due to the lack of consistent experimental results, here we intuitively suggest a range of values from $0$ to $10$ for the magnitude of the gradient of $\rQ$, with the expectation that the values near $10\, \tQ/\tX$ will practically represent an environmental barrier.
We choose a typical value of $0.2 \, \tQ/\tX$, which means one unit of standard deviation change in $\rQ$ per five units of space.   

It would be worthwhile then to compare this choice of a range for trait optimum gradient with those suggested by \citet{Kirkpatrick:Naturalist:1997} and \citet{Case:Naturalist:2000}.
It should be noted, however, that the compatibility of those suggested ranges of values with the values that are expressed based on our choices of units is not fully ensured, but is likely to hold after a certain rescaling as described below.   

\citet{Kirkpatrick:Naturalist:1997}, based on what they acknowledged to be slim available evidence, suggested that a plausible range for standardized environmental gradient could have a [maximum] value of at least $0.25$. 
They denote this standardized gradient by $b^{\ast}$ and define it as the optimum gradient measured in units of phenotypic standard deviations per dispersal [distance].
Therefore, the range of values suggested by \citet{Kirkpatrick:Naturalist:1997} would be comparable to our suggested values, after being divided by $\sqrt{2}$, if we assume that they measure one unit of dispersal distance over one generation time. 
Although the unit of time is not clearly specified by \citet{Kirkpatrick:Naturalist:1997}, this assumption is likely to be valid based on the way they define $b^{\ast}$ and the dispersal coefficient. 

\citet{Case:Naturalist:2000} redefine $b^{\ast}$, which we denote by $b^{\ast}_{\scC\scT}$ to avoid confusion, as the optimum gradient expressed in units of phenotypic standard deviation per physical distance. 
However, they do not clearly specify the unit of distance.
They suggested that a plausible range for $b^{\ast}_{\scC\scT}$ could be $[0.0001,0.01]$ which, according to the sample values they provide, is likely to be based on the choice of kilometers as the unit of physical distance.
Therefore, to convert the range $[0.0001,0.01] \, \tQ/\mathrm{km}$ to a range compatible with our choices of units, we need to rescale it
by plausible values of the root mean square dispersal distance measured in km over one generation time, as well as the factor $1/\sqrt{2}$.
There is evidence in the literature 
that mammalian species may show maximum natal dispersal distances ranging from a few tens of meters to roughly 300 km \citep{whitmee2013predicting}. 
Moreover, for a large number of species in four different taxonomic groups, estimates for both dispersal distance (in km) per year and generation time (in year) are provided by \citet{Ohashi:Nature:2019}.
This allows us to find estimates of dispersal distance in km per generation for the species of this dataset.
The statistical distribution of such estimates are given in Table \ref{tb:DispersalDistances}.
Based on these sample ranges of values, we will therefore take values of dispersal distance (per generation) in the broad range from $0.01$ km to $300$ km to be plausible, at least for terrestrial species. 

Now, multiplying the suggested range $[0.0001,0.01] \, \tQ/\mathrm{km}$ for $b^{\ast}_{\scC\scT}$ by $1/\sqrt{2}$ and our minimum and maximum plausible values of dispersal distance per generation, we obtain the range 
$[0.0001 \times 0.01 / \sqrt{2}, 0.01 \times 300 / \sqrt{2}] \approx [7 \times 10^{-7}, 2.1] \, \tQ/\tX$
that might have been considered plausible by \citet{Case:Naturalist:2000} based on our choices of units. 
Note that this range includes the value $0.25 /\sqrt{2} \approx 0.18 \, \tQ/\tX $ put forth by \citet{Kirkpatrick:Naturalist:1997}, but clearly extends several orders of magnitude below that value. 

\begin{table}[t!]
	\caption{Statistical distribution of estimated values of dispersal distance (in km) per generation for 
		512 species of Amphibians, 385 species of Reptiles, 4829 species of Birds, and 1160 species of Mammals.
		The estimated values are calculated using the dataset provided by \citet{Ohashi:Nature:2019}.}
	\label{tb:DispersalDistances}
	\begin{center} \small 
		\renewcommand{\arraystretch}{1.3}
		\begin{tabular}{lcccccc}		
			\hline 
			& Minimum & Maximum & Mean & Median & $5$th Percentile & $95$th Percentile\\
			\hline
			Amphibians & $0.0085$ & $17.1$ & $0.41$ & $0.25$ & $0.083$ & $1.00$\\
			Reptiles & $0.010$ & $16.0$ & $0.65$ & $0.15$ & $0.028$ & $1.83$ \\
			Birds & $3.18$ & $682$ & $38.3$ & $17.7$ & $6.63$ & $148$ \\
			Mammals & $0.10$ & $154$ & $2.84$ & $0.38$ & $0.15$ & $12.0$\\
			\hline
		\end{tabular}
	\end{center}
\end{table}

\section{Range dynamics of a single species} \label{sec:SingleSpecies}

To demonstrate general predictions of the model on the range dynamics and intraspecific trait variations of a species, we qualitatively study the solutions of the model for a solitary species over a one-dimensional habitat.
This can, for example, represent the spread of an invasive aquatic species in a river, or the distribution of a species with an ecological niche restricted along a coastline.
Therefore,  $\rmm = 1$ and $\rN = 1$, and we use the typical values given in Table \ref{tb:Parameters} for the rest of the model parameters, with certain alterations independently specified in each study.
Other than the trait optimum $\rQ$, which is considered to be linearly increasing over $\Omega$, the rest of the parameters are assumed to be constant.
As a result, the equations of the model \eqref{eq:PopulationDensity}--\eqref{eq:NonlinearTerms} are reduced to
\begin{align}     
	\dt n &= \rD \dx^2 n + \left( \rR - \frac{\rR}{\rK} \sqrt{\frac{\rV}{v+\rV}} n  
		- \frac{\rS}{2}  \left[(q - \rQ)^2 + v\right] \right) n, \label{eq:PopulationDensity_Single}\\
	\dt q &= \rD \dx^2 q + 2 \rD \dx (\log n) \dx q-\rS (q-\rQ) v, \label{eq:TraitMean_Single}\\
	\dt v &= \rD \dx^2 v + 2 \rD \dx (\log n) \dx v + 2 \rD |\dx q|^2 
		+ \frac{\rR}{\rK} \sqrt{\frac{\rV}{v+\rV}} \frac{n v^2}{2(v+\rV)} - \rS v^2 + \rU, \label{eq:TraitVariance_Single}
\end{align}
where, since $\rN = 1$, the numeration index $i=1$ of the variables and parameters is dropped for notational simplicity.
Moreover, the dependence of $n$, $q$, and $v$ on $x$ and $t$, as well as the dependence of $\rQ$ on $x$, are
not shown for the simplicity of exposition. 

We set the habitat as $\Omega = (-50\; \tX,50\; \tX) \subset \bbR$ for all computational studies of this section, and we consider reflecting boundary conditions as described in Remark \ref{rmk:BoundaryConditions}.
The details of the numerical scheme used to compute the solutions are given in Appendix B.
The implementation of the numerical simulations in MATLAB R2021a is available online as Supplementary Material 1.
Note that, the qualitative behavior of the model shown in this section for a one-dimensional habitat are also equivalently observable in a two-dimensional habitat. 
 
\subsection{Range expansion under constant environmental gradient} \label{sec:Typical_Single}

\begin{figure}[t!]
	\centering 
	\begin{subfigure}[t]{0.4\textwidth} 
		\includegraphics[width= 1\textwidth]{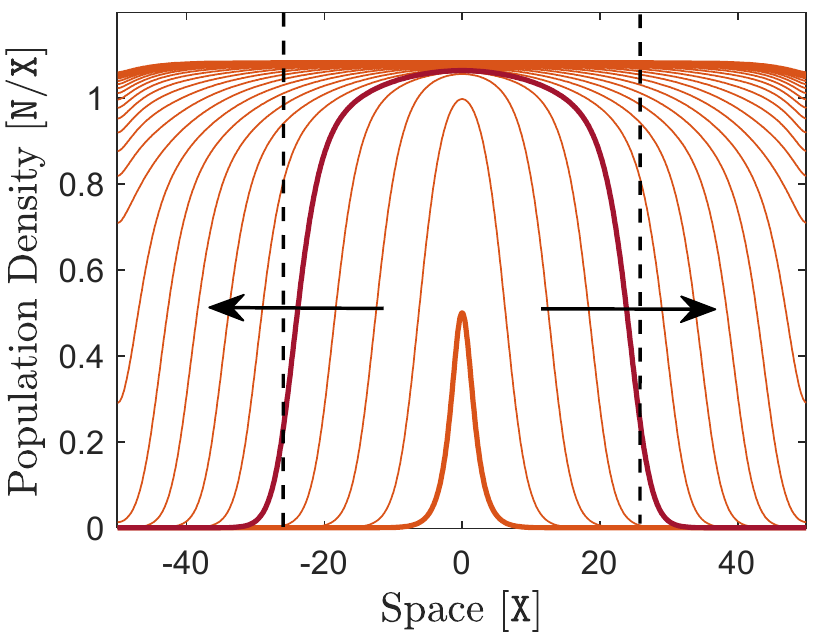}
		\caption{} \label{fig:n_nominal_1D_single}
	\end{subfigure}
	\hfil
	\begin{subfigure}[t]{0.4\textwidth} 
		\includegraphics[width= 1\textwidth]{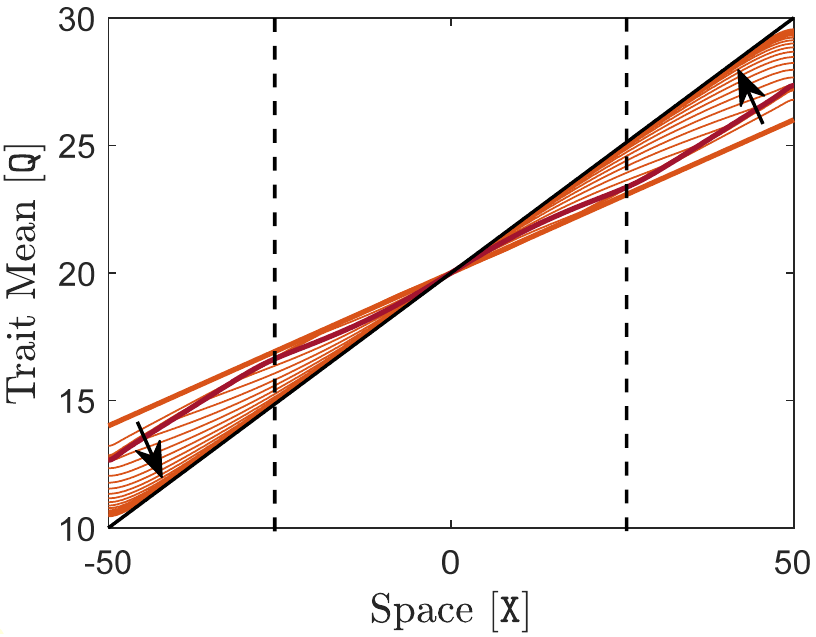}
		\caption{} \label{fig:q_nominal_1D_single}
	\end{subfigure}
	\vfil
	\begin{subfigure}[t]{0.4\textwidth} 
		\includegraphics[width= 1\textwidth]{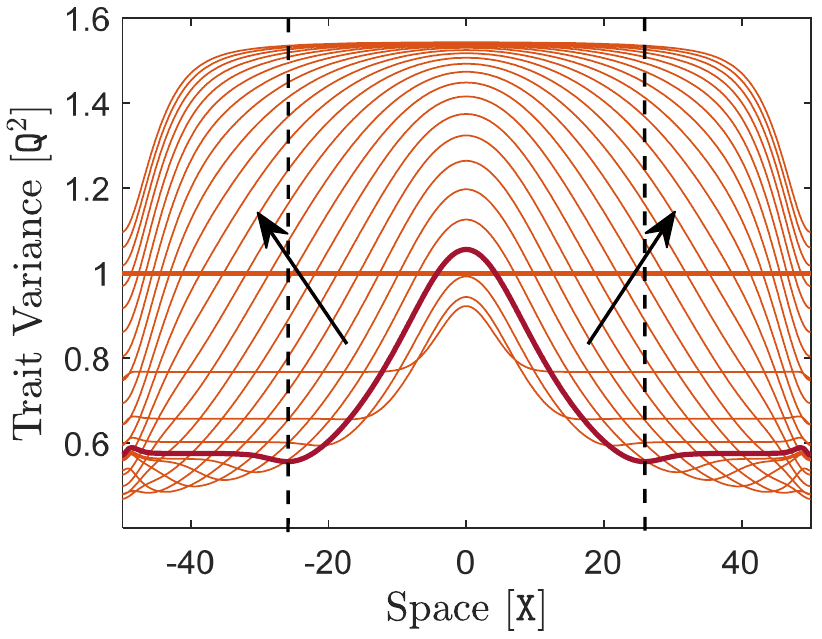}
		\caption{} \label{fig:v_nominal_1D_single}
	\end{subfigure}
	\hfil
	\begin{subfigure}[t]{0.4\textwidth} 
		\includegraphics[width= 1\textwidth]{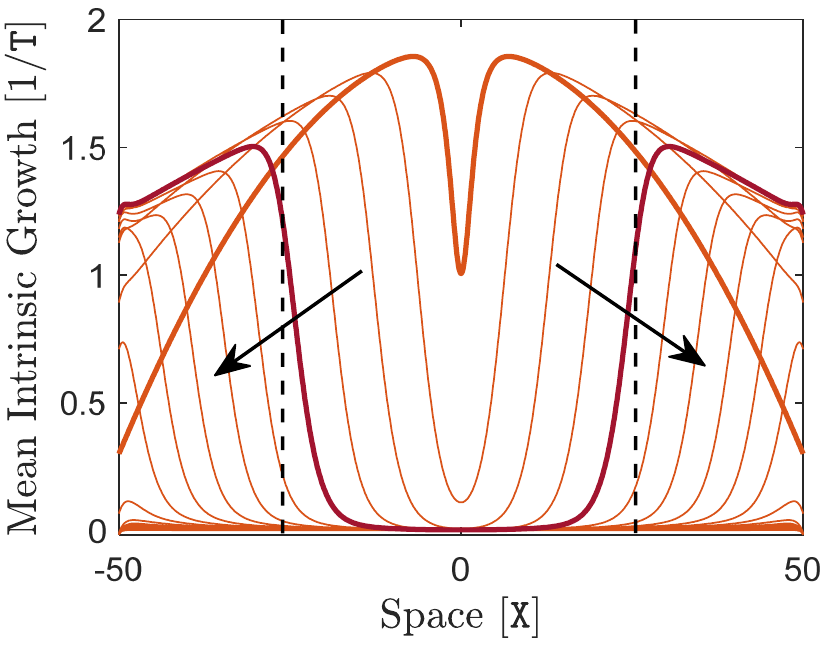}
		\caption{} \label{fig:G_nominal_1D_single}
	\end{subfigure}	
	\caption{Typical range dynamics of a single species in a one-dimensional habitat with a linear trait optimum. 
		Here, $\rmm = 1$, $\rN = 1$, and the rest of the model parameters take their typical values given in Table \ref{tb:Parameters}.
		Curves are shown at every $2 \; \tT$, and the thick orange curves indicate the initial curves at $t=0 \;\tT$. 
		Arrows show the direction of evolution in time.
		In each graph, a sample curve at $t = 8 \; \tT$ is highlighted in red.
		Dashed lines indicate the effective edges of the population at $t = 8 \; \tT$, associated with the inflection points on the curve of population density, at which the trait variance takes its minimum value.
		The solid black line in (b) shows the environmental trait optimum $\rQ$.}
	\label{fig:Nominal_1D_single}
\end{figure}

We assume the environmental trait optimum has a constant gradient $\rd_x \rQ = 0.2 \; \tQ/\tX$, which is the slope of the black line in Figure \ref{fig:q_nominal_1D_single}.
The species is introduced at the center of the habitat with a density given as $n(x,0) = 0.5 \sech( |x| / \sqrt{2} )$. 
It is assumed that this initial population is perfectly adapted to the environment at the center, that is, $q(0,0) = \rQ(0)$, and has a linearly varying trait mean of slope $\dx q(x,0) = 0.6\, \rd_x \rQ$ for all $x \in \Omega$.
We further assume that the initial population has a constant trait variance of $v(x,0) = 1 \; \tQ^2$. 

Figure \ref{fig:Nominal_1D_single} shows the solutions of the model \eqref{eq:PopulationDensity_Single}--\eqref{eq:TraitVariance_Single} over the computation time horizon of $T = 50\; \tT$. 
It can be seen that the species' population density initially grows to an upper limit relatively fast, and then the population invades the entire habitat in the form of a traveling wave. 
As the population spreads over the habitat, it successfully adapts to new areas in response to the force of natural selection, and its mean trait gradually converges to the optimal trait.
The initially constant profile of the population's trait variance  evolves quickly to a bell-shaped profile, showing a larger trait variance at the core of the population with a gradual decline towards the edges.  
The maximum trait variance at the population center evolves fairly slowly to an upper bound as the population expands its range across the habitat.

To investigate further the invasive range dynamics of this solitary species, a sample curve is highlighted in the solutions shown in Figure \ref{fig:Nominal_1D_single}. 
At the effective edges of the population, that is considered to correspond to the inflection points on the curve of the population density, the diffusion term $\rD  \dx^2 n$ in \eqref{eq:PopulationDensity_Single} is zero.
However, the mean intrinsic growth rate $G$, given by the term inside the parentheses in \eqref{eq:PopulationDensity_Single} and shown in Figure \ref{fig:G_nominal_1D_single}, is positive at these edge positions.
This implies that the population density $n$ is not stationary at the edges, which confirms the fact that the sample solution curve is a traveling wave. 

The curve of trait mean highlighted in Figure \ref{fig:q_nominal_1D_single} shows substantial adaptation at the core of the population, but considerable failure in adaptation near the edges.
This adaptation profile is mainly due to the effect of gene flow 
and is observed, after few generations, even if the initial population is perfectly adapted everywhere, that is, even if $q(x,0) = \rQ(x)$ for all $x\in \Omega$.
This profile can be explained as follows. 
The density of the population is nearly uniform at its core, and hence there is a fairly symmetric gene flow to a core location from its adjacent areas located on the upper and lower parts of the cline. 
Due to this symmetry, gene flow does not significantly affect the mean value of the trait at core locations and adaptation is successfully maintained by natural selection.
Near the edge, however, the population density varies sharply and the gene flow is highly asymmetric. 
At the left edge shown in Figure \ref{fig:q_nominal_1D_single}, for example, gene flow is predominantly from the upper points on the cline.   
Therefore, near the left edge, the mean value of the trait is increased above the optimal value due to the effect of this asymmetric gene flow.
As a result, the curve of trait mean is gradually flattened and departs from the optimal curve as it approaches the left edge. 
A similar process, but in opposite direction, flattens the curve below the optimal value at the right edge.

The bell shape of the trait variance is a result of gene flow and the specific adaptation pattern described above.
At core locations, the population is well-adapted to the optimal cline and, due to the relatively large gradient of the cline, gene flow from adjacent areas generate large phenotypic variations among central individuals. 
Near the edges, however, the population fails to adapt to the optimal cline and the curve of trait mean flattens.
Therefore, due to the low gradient in the trait mean near the edges, gene flow from adjacent areas does not substantially contribute to phenotypic variations in marginal individuals.
As a result, trait variance constantly declines from core to edges, in parallel with the decline in the gradient of the trait mean.  
       
The profile of variations in the trait mean and the trait variance described above are consistent with experimental observations available in the literature.
For example,  \citet{Takahashi:MolecularEcology:2016} have shown geographic variations in abdomen length and wing loading of two closely related species of damselflies along a latitudinal cline in Japan.
The patterns of variation in both of these phenotypic traits indicate that the species are well adapted to the environmental gradient at the core of their range, whereas they are significantly maladapted at their range margins. 
Additionally, variations in heterozygosity and the Garza-Williamson index, as the two indicators of genetic diversity measured by \citet{Takahashi:MolecularEcology:2016}, show drastic decline in species' genetic and phenotypic variation at their range margins where maladaptation is observed.

\subsection{Effect of steep environmental gradients} \label{sec:Gradients_Single}

A major difference between the predictions of the present model and predictions of the models that assume constant trait variance \citep{ Kirkpatrick:Naturalist:1997, Case:Naturalist:2000} appears in response to steep environmental gradients.
To show this, we first repeat the computation of Section \ref{sec:Typical_Single} for a species under a steep environmental gradient of $\rd_x \rQ = 2 \; \tQ/\tX$, that is $10$ times larger than the typical gradient considered in Section \ref{sec:Typical_Single}.
Here, we initialize the computation with a better adapted population with $\dx q(x,0) = 0.9\, \rd_x \rQ$, to avoid the numerical singularities that would otherwise occur at initial iterations of the computation---mainly due to  large trait mismatch, $|q-\rQ|$, occurring near the boundary of the habitat because the environmental gradient is very steep. 
The rest of the computation parameters take the same values as used in Section \ref{sec:Typical_Single}.

\begin{figure}[t]
	\centering 
	\begin{subfigure}[t]{0.4\textwidth} 
		\includegraphics[width= 1\textwidth]{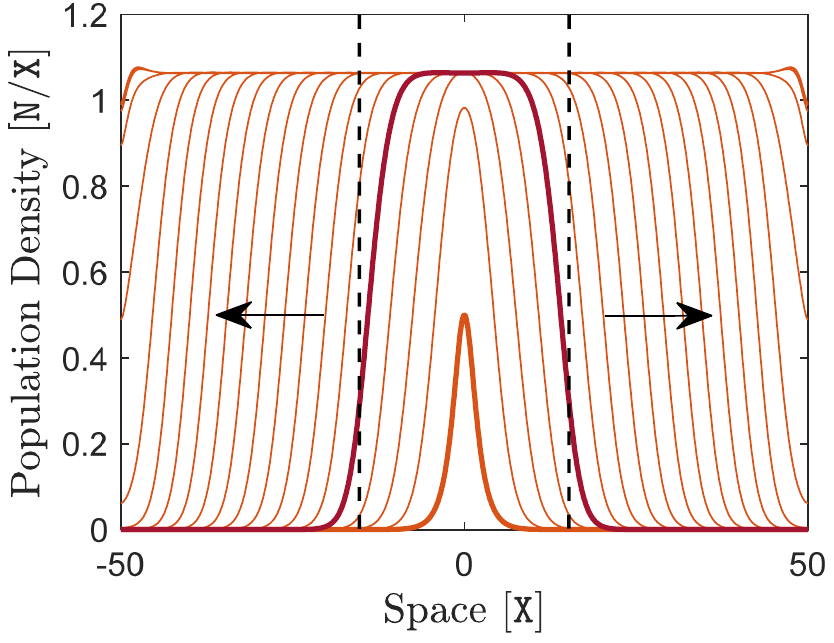}
		\caption{} \label{fig:n_Qopt_2_1D_single}
	\end{subfigure}
	\hfil
	\begin{subfigure}[t]{0.4\textwidth} 
		\includegraphics[width= 1\textwidth]{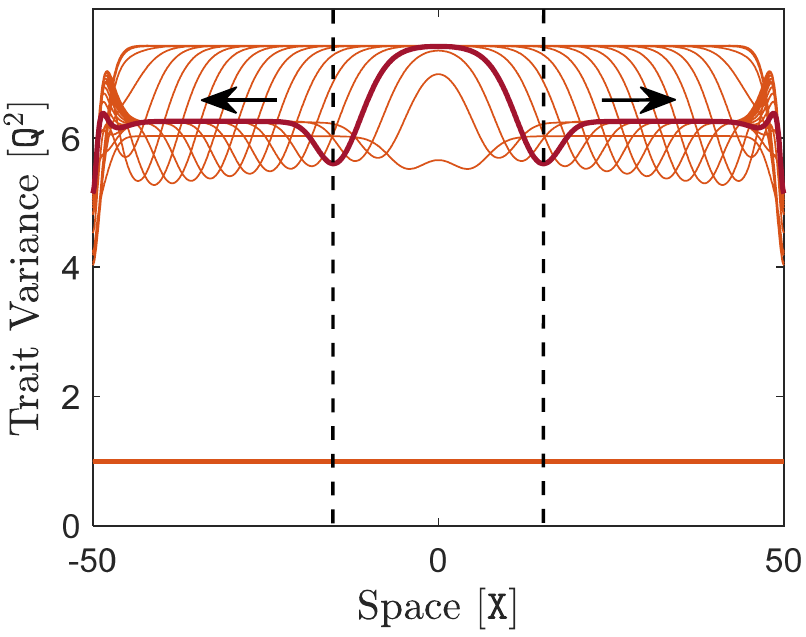}
		\caption{} \label{fig:v_Qopt_2_1D_single}
	\end{subfigure}	
	\caption{Range dynamics of a species in a one-dimensional habitat under a steep environmental gradient of $\rd_x \rQ = 2 \, \tQ/\tX$.
		The rest of the model parameters take their typical values given in Table \ref{tb:Parameters}.
		The	same description as given in Figure \ref{fig:Nominal_1D_single} holds here for the curves, arrows, and dashed lines.
		The evolution curves of $q$ and $G$ are not shown as they qualitatively resemble the curves shown in Figures \ref{fig:q_nominal_1D_single} and \ref{fig:G_nominal_1D_single}, respectively. } \label{fig:GradientHigh_1D_single}
\end{figure}

The population density and trait variance of the species is shown in Figure \ref{fig:GradientHigh_1D_single}.
It can be seen that the species is still able to expand its range in the form of a traveling wave, but with a lower speed compared to the wave speed at the typical gradient value.
More importantly, the initially small value of the trait variance, $v(x,0) = 1\, \tQ^2$, evolves quickly to a significantly larger value.
In fact, since the population here is adapting itself to an optimal phenotype with much steeper variation along the habitat,
the mean phenotypes diffused to a given location within the species' range have a much wider range of values.
As a result, gene flow in this case generates large phenotypic variations in the population.
Similar to the variance profile described in Section \ref{sec:Typical_Single}, the trait variance declines at the range margins due to the effect of asymmetric gene flow.
Note that, for $\rd_x \rQ = 2 \, \tQ/\tX$ considered here, the models that assume constant trait variance show an evolutionarily stable limited range, for which the mean intrinsic growth rate vanishes to zero at the inflection points on the range edges, and remains negative beyond. 
Examples of such limited range dynamics are provided by \citet{Kirkpatrick:Naturalist:1997} and hence are not presented here.

Next, to see in more detail the differences resulting from the evolution of phenotypic variance in the range dynamics of the species, we additionally generate a constant-variance version of the model and repeat the computations described above for both models, and with different values of the environmental gradient. 
The constant-variance model is obtained by fixing $v(x,t) = \rV_{\scP}$ in \eqref{eq:PopulationDensity_Single}--\eqref{eq:TraitVariance_Single}, that is, by omitting \eqref{eq:TraitVariance_Single} and replacing $v$ by $\rV_{\scP}$ in \eqref{eq:PopulationDensity_Single} and \eqref{eq:TraitMean_Single}.
The resulting model is equivalent to the model presented by  \citet{Case:Naturalist:2000} and \citet{Kirkpatrick:Naturalist:1997}.  
We set $v(x,0) = 1 \, \tQ^2$ for our variable-variance model and, correspondingly, $\rV_{\scP} = 1 \, \tQ^2$ for the constant-variance model.

\begin{figure}[t!]
	\centering 
	\begin{subfigure}[t]{0.4\textwidth} 
		\includegraphics[width= 1\textwidth]{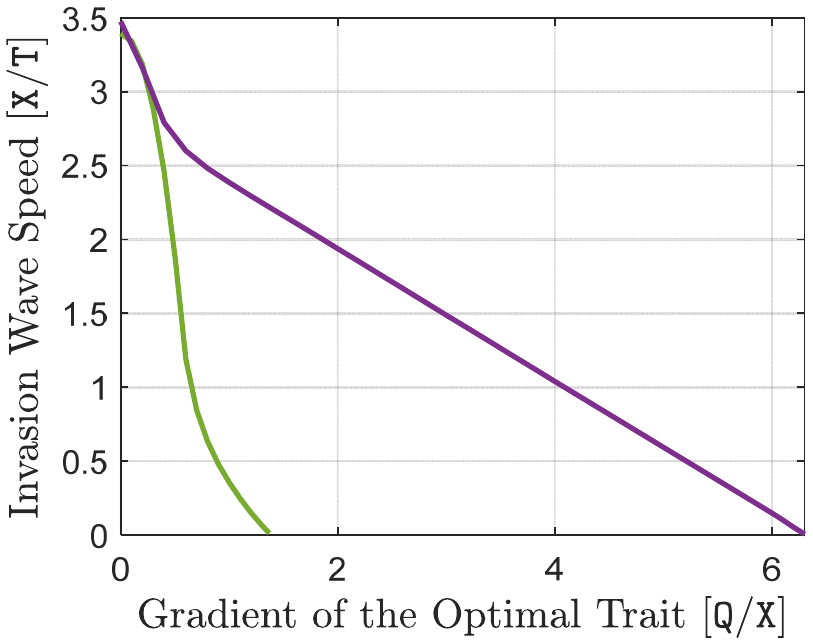}
		\caption{} \label{fig:Speed_vs_Gradient_1D_single}
	\end{subfigure}
	\hfil
	\begin{subfigure}[t]{0.4\textwidth} 
		\includegraphics[width= 1\textwidth]{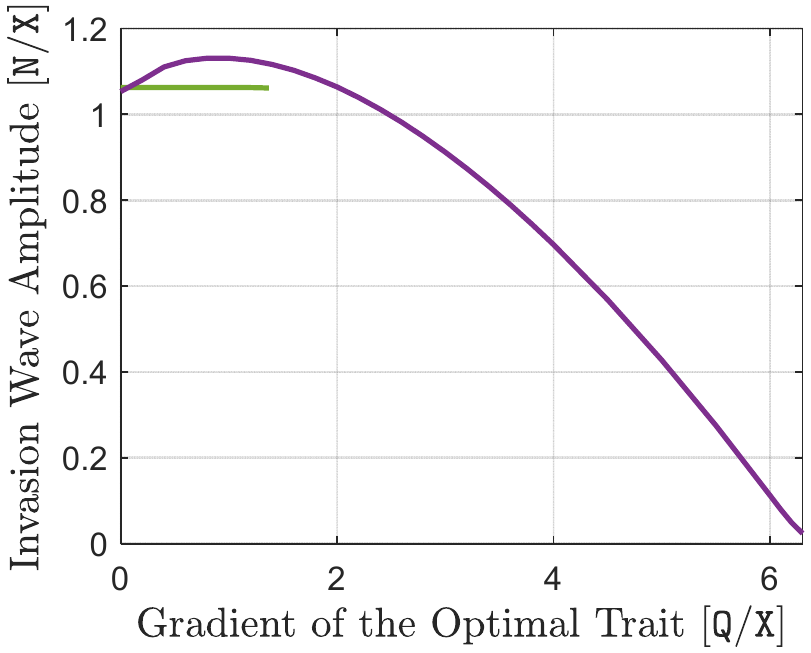}
		\caption{} \label{fig:Amplitude_vs_Gradient_1D_single}
	\end{subfigure}
	\caption{Impact of the gradient of the environmental trait optimum, $\rd_x \rQ$, on a species' invasion waves over a one-dimensional habitat. 
		Here, $\rmm = 1$, $\rN = 1$, and the rest of the model parameters take their typical values given in Table \ref{tb:Parameters}.
		The curves shown in green are obtained based on the assumption of constant trait variance, as in the models of \citet{Case:Naturalist:2000} and \citet{ Kirkpatrick:Naturalist:1997}, with $v(x,t)$ being fixed at $\rV_{\scP} = 1$.
		The curves shown in purple are obtained when the trait variance is free to evolve in time based on the present model \eqref{eq:PopulationDensity_Single}--\eqref{eq:TraitVariance_Single}.
		(a) Variations in the traveling speed of the invasion waves. 
		(b) Variations in the amplitude of the invasion waves.} \label{fig:GradientVariation_1D_single}
\end{figure}
 
For each value of the environmental gradient, solutions of the equations are calculated for a sufficiently large period of time.
The approximate speed and amplitude\footnote{Here, we define the amplitude of a population's traveling wave solution as the (steady) peak value of the population's density during its range expansion regime. 
In the results presented in Figure \ref{fig:Amplitude_vs_Gradient_1D_single}, the traveling wave amplitudes are approximately calculated as the density of the population at its center when it has reached a nearly steady value. 
}
of the population's traveling waves are then shown in Figure \ref{fig:GradientVariation_1D_single}.  
Note that, at every gradient value, the position of a point on the edges of the population density waves initially changes nonlinearly in time.
However, when the effect of initial transient dynamics vanishes, the variations in the edge positions eventually become linear with respect to time.
The slope of these linear variations is shown in Figure \ref{fig:Speed_vs_Gradient_1D_single} as an approximate value of the wave speed at each gradient value.
Moreover, the amplitude of the traveling wave at the final time of computations for each gradient value is shown in Figure~\ref{fig:Amplitude_vs_Gradient_1D_single} as an approximate value of the wave amplitude.    

Figure \ref{fig:GradientVariation_1D_single} shows that the constant-variance model predicts a relatively sharp decline in the speed of invasion waves as the environmental gradient increases, but no change in waves' amplitude.
\citet{Kirkpatrick:Naturalist:1997} and \citet{Case:Naturalist:2000} show that a stable limited range is formed if the gradient is increased beyond the value at which wave speed becomes zero.
Eventually, the population becomes extinct if the gradient is further increased to very large values.
However, removing the assumption of constant phenotypic variance, as in the present model, results in essentially different predictions.
It can be seen in Figure \ref{fig:GradientVariation_1D_single} that,
although the speed of invasion waves declines (almost linearly) when the environmental gradient increases, the species is still able to expand its range under much steeper gradients than what is predicted under the constant variance assumption.
Nonetheless, the amplitude of the species' expansion waves declines substantially at steep gradients and the population eventually vanishes at extreme gradients.
This implies that physical barriers, such as mountains or oceans, which impose extreme environmental gradients on the species' habitat can still prevent the species' range expansion.     

The approximate wave amplitudes shown in Figure \ref{fig:Amplitude_vs_Gradient_1D_single} can be alternatively calculated using the equations of a fully adapted homogeneous population.   
As the range dynamics of the species shown in Figures \ref{fig:Nominal_1D_single} and \ref{fig:GradientHigh_1D_single} suggests, population density and trait variance eventually become homogeneous over $\Omega$ as $t \to \infty$, and trait mean converges uniformly to $\rQ$.
We denote this equilibrium state by $(n^{\ast}, q^{\ast}, v^{\ast})$, where $n^{\ast}>0$ and $v^{\ast} \geq 0$.
At this equilibrium, we have $\dx n^{\ast} = 0$, $q^\ast = \rQ$, and $\dx v^{\ast} = 0$.
Moreover, $\dx^2 q^{\ast} = 0$ since $\rQ$ is assumed to be linear here.
Therefore, from \eqref{eq:PopulationDensity_Single}--\eqref{eq:TraitVariance_Single} we obtain
\begin{equation}     \label{eq:Equilibrium_Density}
	n^{\ast} = \frac{\rK}{\rR} \sqrt{\frac{v^{\ast}+\rV}{\rV}}   
	\left(\rR - \frac{\rS}{2}  v^{\ast}\right), 
\end{equation}
and
\begin{equation} \label{eq:Equilibrium_TraitVariance}
	5\rS {v^{\ast}}^3 + (4 \rS \rV - 2 \rR) {v^{\ast}}^2 - 4(\rU + 2 \rD |\rd_x \rQ|^2) v^{\ast} 
	- 4 (\rU + 2\rD |\rd_x \rQ|^2) \rV = 0. 
\end{equation}
Letting $v^{\ast} = \rV_{\scP} =1$ in \eqref{eq:Equilibrium_Density} for the constant-variance model, we simply obtain $n^{\ast} = 1.062$ as the constant wave amplitude shown in Figure \ref{fig:Amplitude_vs_Gradient_1D_single}.
For the variable variance model, the cubic algebraic equation \eqref{eq:Equilibrium_TraitVariance} has one and only one positive root for all nonzero values of $\rU + 2 \rD |\rd_x \rQ|^2$. 
The graph of this nonzero root with respect to changes in the gradient $\rd_x \rQ$ appears to be approximately a straight line with positive slope.
Moreover, substituting this root into \eqref{eq:Equilibrium_Density} for different values of $\rd_x \rQ$ provides a graph of $n^{\ast}$ with respect to $\rd_x \rQ$. 
With the parameter values considered in this section, this graph gives an approximate curve of wave amplitudes as shown in Figure \ref{fig:Amplitude_vs_Gradient_1D_single}. 
Note that $n^{\ast}=0$ in \eqref{eq:Equilibrium_Density} when $v^{\ast} = 2 \rR/\rS$, which is a solution of \eqref{eq:Equilibrium_TraitVariance} with $|\rd_x \rQ| =  \sqrt{(2\rR^2/\rS \rD) - \rU/2\rD} = 6.325$.
This gives the value of the environmental gradient at which the wave amplitude becomes zero and the population fails to survive.

\begin{remark}[Effect of large dispersal coefficients]
	The dispersal coefficient $\rD$ and the square of environmental gradient $|\rd_x \rQ|^2$ appear together in the equilibrium equation \eqref{eq:Equilibrium_TraitVariance}.
	This implies that, for a fixed value of $\rd_x \rQ$, the invasion wave amplitude has the same pattern of variation with respect to $\sqrt{\rD}$ as it has with respect to $|\rd_x \rQ|$ shown in Figure \ref{fig:Amplitude_vs_Gradient_1D_single}. 
	Basically, this is because, as described in Section \ref{sec:Units}, a change of $\sqrt{\rD} \mapsto k \sqrt{\rD}$ in the dispersal coefficient can be equivalently absorbed by a rescaling of $x \mapsto x/k$ in the space, and consequently by a change of  $\rd_x \rQ \mapsto k \rd_x \rQ $ in the environmental gradient.
	Note that, by the same calculations as performed above, the species fails to survive if its dispersal coefficient is greater than 
	$(2\rR^2/\rS|\rd_x \rQ|^2) - \rU/2|\rd_x \rQ|^2$.   
	\qed   
\end{remark}
 
\begin{remark}[Effects of maximum growth rate and strength of stabilizing selection]
	For smaller values of $\rR$ or larger values of $\rS$, the extreme environmental gradient $|\rd_x \rQ|_{\max} = \sqrt{(2\rR^2/\rS \rD) - \rU/2\rD}$, above which the population cannot survive, becomes smaller.
	This implies that a slowly growing species under strong stabilizing selection is at a higher risk of extinction in the environments for which the species' optimal trait has sharp spatial variations.  
	Reduced dispersal in such environments can help the species survive. 
	\qed
\end{remark}

\begin{remark} [Effect of genetic mutations] \label{rmk:Mutation} 
	The computational results shown in Figure \ref{fig:GradientVariation_1D_single} were obtained in the absence of genetic mutations, that is, by considering the typical value $\rU = 0 \, \tQ^2/\tT$ given in Table \ref{tb:Parameters}.  
	When $\rU \neq 0$, equation \eqref{eq:TraitVariance_Single} implies that the trait variance is inflated at the constant absolute rate $\rU$ due to mutation.
	With an exceedingly large value of $\rU = 2 \, \tQ^2 / \tT$, which may appear to be unrealistic, we computed the solutions of \eqref{eq:PopulationDensity_Single}--\eqref{eq:TraitVariance_Single} using the same simulation layout as described above for the results shown in Figure \ref{fig:GradientVariation_1D_single}. 
	We verified that the curves of wave speeds and wave amplitudes did not differ significantly from those of Figure \ref{fig:GradientVariation_1D_single}, meaning that the conclusions of this section are not affected by the effect of mutation.
	To see this further, note that the critical value of the trait optimum gradient
	$|\rd_x \rQ|_{\max} = \sqrt{(2\rR^2 / \rS \rD) - \rU/2\rD}$, beyond which the species fails to survive, will be decreased significantly by the effect of mutation only if $\rU$ is sufficiently large compared with $4\rR^2 / \rS$.
	However, for the typical values given in Table \ref{tb:Parameters}, we have $4\rR^2 / \rS = 80 \, \tQ^2/{\tT}$, which can be of about three orders of magnitude larger than realistic values of $\rU$.
	For slowly growing species under strong phenotypic selection, however, the effect of mutation may considerably decrease the critical value of the trait optimum gradient. 	  
	\qed
\end{remark}

\subsection{Effect of abrupt environmental fluctuations}
An abrupt change in the curve of optimal phenotype may occur due to a rapid climate change and can largely affect a species' geographic range and abundance, particularly when it occurs frequently.
To observe the predictions of the model in response to these changes in climate, here we initialize the computations at $t = 0\, \tT$ using the solution curves computed in Section \ref{sec:Typical_Single} at $t = 4\, \tT$.
However, we uniformly shift up the line of trait optimum $\rQ$ by a factor of $5 \, \tQ$ at $t = 0\, \tT$.  
As a result, a high level of maladaptation is immediately induced in the population and the population's density and expansion speed decline quickly.  
These impacts are more severe near the right margin of the population's range, where, as discussed in Section \ref{sec:Typical_Single}, the population initially has a trait mean below the trait optimum and hence faces a greater maladaptation as the optimum is shifted up.
 
\begin{figure}[t!]
	\centering 
	\begin{subfigure}[t]{0.4\textwidth} 
		\includegraphics[width= 1\textwidth]{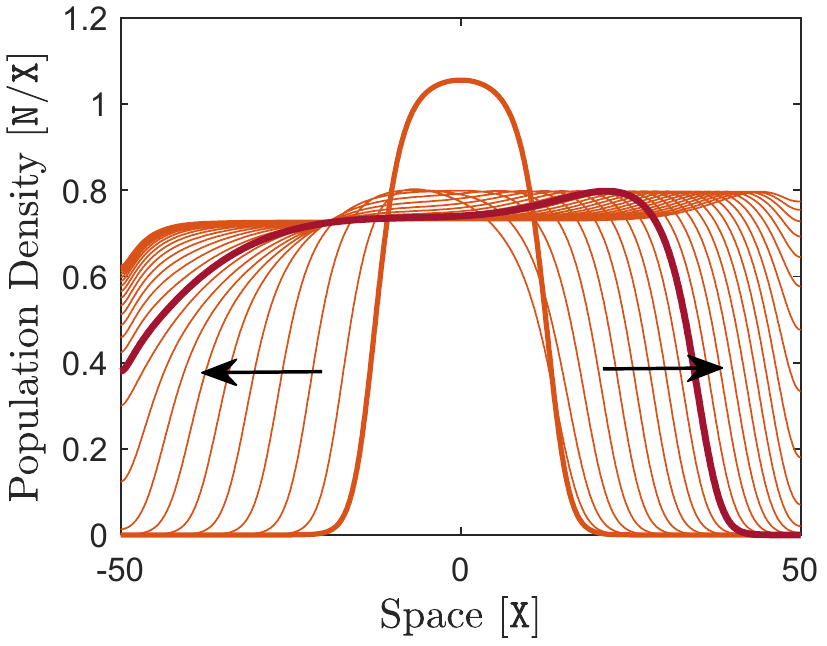}
		\caption{} \label{fig:n_fluctuation_1D_single}
	\end{subfigure}
	\hfil
	\begin{subfigure}[t]{0.4\textwidth} 
		\includegraphics[width= 1\textwidth]{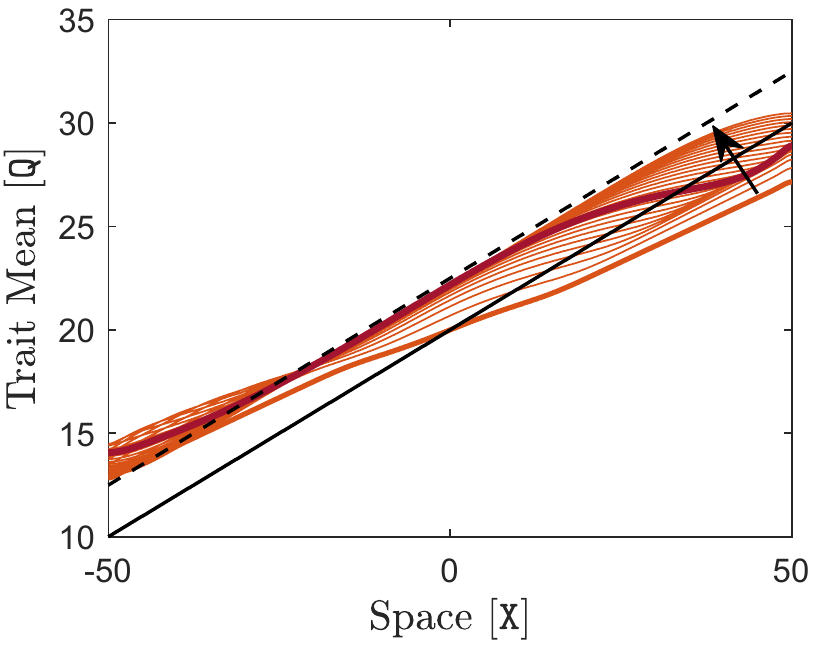}
		\caption{} \label{fig:q_fluctuation_1D_single}
	\end{subfigure}	
	\caption{Impact of periodic environmental fluctuations on a species' range dynamics in a one-dimensional habitat.
	Here, $\rmm = 1$, $\rN = 1$, and the rest of the model parameters take their typical values given in Table \ref{tb:Parameters}.
	The period of abrupt fluctuations in the trait optimum is $2 \, \tT$. 
	At the beginning of each period, $\rQ$ is shifted up by a factor of $5\, \tQ$ and remains at this value for the first half of the period. 
	Then it is shifted down by the same factor to the initial value and remains at this value for the second half of the period.	 
	Curves are shown at every $2 \, \tT$, and the thick orange curves indicate the initial curves at $t=0 \,\tT$. 
	Arrows show the direction of evolution in time.
	In each graph, a sample curve at $t = 20 \, \tT$ is highlighted in red.
	The solid black line in (b) shows the initial value of $\rQ$, and the dashed line in (b) shows this initial value shifted up by a factor of $5/2 \, \tQ$.   
	} \label{fig:Fluctuation_1D_single}
\end{figure}

The impact of the climate change described above is transient and gradually dissolves as the species adapts itself to the new environmental gradient.
However, the impacts of climate change can last longer and become more severe if rapid changes in climate occur more frequently.
For instance, Figure \ref{fig:Fluctuation_1D_single} shows the results obtained when abrupt fluctuations of magnitude $5 \, \tQ$ occur periodically in the trait optimum.
If the period of these fluctuations is sufficiently small, as in Figure \ref{fig:Fluctuation_1D_single}, the abundance and expansion speed of the species is steadily affected by the fluctuations.
The species' range margins, particularly the right margin, advance slower under the climate fluctuations and the species' density remains significantly below its environmental carrying capacity.     
Moreover, note that the pattern of fluctuations will shape the longterm profile of the population's trait mean.
Here, fluctuations are in the form of a square wave, with $\rQ$ shifting up by a factor of $5 \, \tQ$ for the first half of the fluctuations' period, and then shifting back to its initial value for the second half.
Regulated by the evolutionary force of natural selection, the trait mean then converges to a line which is above the initial $\rQ$ by a factor equal to half of the periodic shift, that is, $5/2 \, \tQ$.     
This can be seen in Figure \ref{fig:q_fluctuation_1D_single}.

\begin{figure}[t!]
	\centering 
	\begin{subfigure}[t]{0.4\textwidth} 
		\includegraphics[width= 1\textwidth]{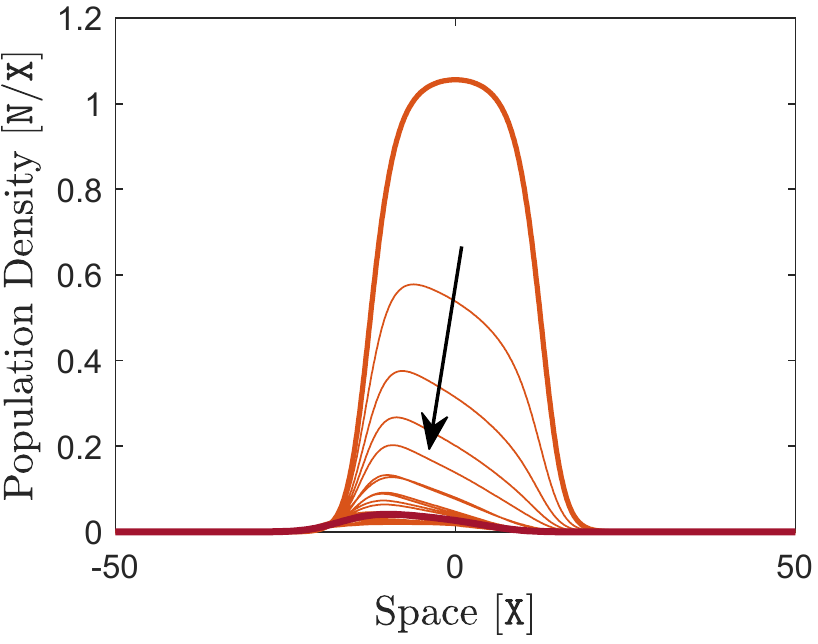}
		\caption{} \label{fig:n_fluctuation_extinction_1D_single}
	\end{subfigure}
	\hfil
	\begin{subfigure}[t]{0.4\textwidth} 
		\includegraphics[width= 1\textwidth]{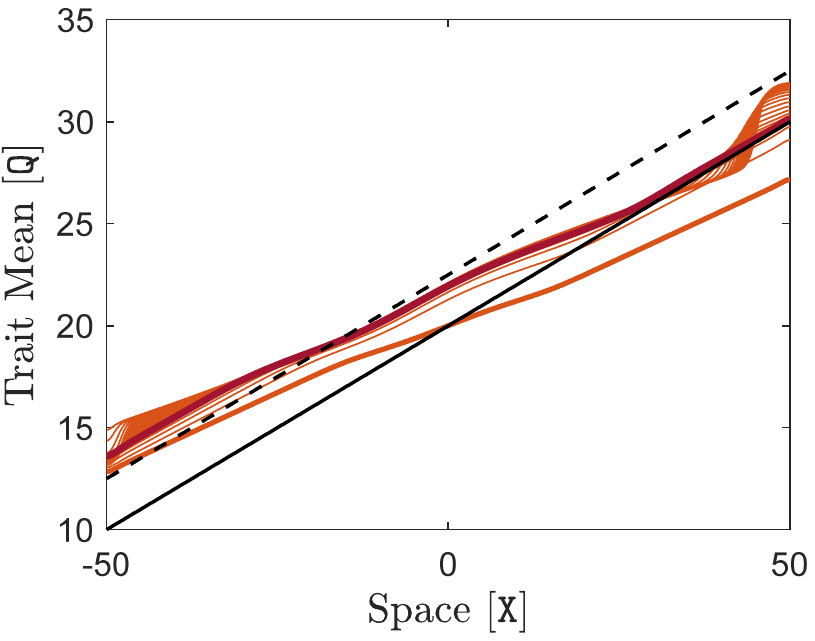}
		\caption{} \label{fig:q_fluctuation_extinction_1D_single}
	\end{subfigure}	
	\caption{Population extinction caused by periodic abrupt fluctuations in the environmental trait optimum for a species under strong phenotypic selection, $\rS = 0.6 \, {\tQ}^{-2}/{\tT}$.
	The rest of the model parameters take their typical values given in Table \ref{tb:Parameters}.
		Environmental fluctuations are modeled as described in Figure \ref{fig:Fluctuation_1D_single}.
		The thick orange curves are initial curves at $t=0 \,\tT$.
		Sample curves at $t = 10 \, \tT$ are highlighted in red.
		In (a), curves are shown at $20$ logarithmically distributed time samples, with the first curve after the initial curve being shown at $t=0.1 \,\tT$. 
		The arrow shows the direction of evolution in time.
		In (b), curves are shown at every $2 \, \tT$.
		The solid black line shows the initial value of $\rQ$, and the dashed line shows this initial value shifted up by a factor of $5/2 \, \tQ$.} \label{fig:FluctuationExtinction_1D_single}
\end{figure}

The impacts of the periodic fluctuations described above can be more severe if the fluctuations have larger amplitudes or the species evolves under stronger phenotypic selections.
For instance, if we repeat the computations of Figure \ref{fig:Fluctuation_1D_single} with a stronger selection, $\rS = 0.6 \, {\tQ}^{-2}/{\tT}$, we obtain the results shown in Figure \ref{fig:FluctuationExtinction_1D_single}. 
It can be seen that, in this case, the population fails to withstand the climate fluctuations and becomes extinct.
Figure \ref{fig:q_fluctuation_extinction_1D_single} shows that the natural selection is still regulating the trait mean at the best possible value, that is  $5/2 \, \tQ$ above the initial $\rQ$.
But, this does note give the population the fitness required for survival under such a strong selection.        

\section{Range dynamics of two competing species} \label{sec:TwoSpecies}

Interspecific competition is a determining factor in forming and limiting species' ranges in a community of ecologically similar species.
To show general predictions of the model on the role of interspecific competition in the coevolutionary range dynamics of a group of species, we investigate the solutions of the model for two competing species over both a one-dimensional and a two-dimensional  habitat.
The typical values given in Table \ref{tb:Parameters} are used as the parameter values for both species, unless otherwise stated.
As in Section \ref{sec:SingleSpecies}, in the one-dimensional studies of Sections \ref{sec:Typical_Two} and \ref{sec:Competitive exclusion}, the habitat is set as $\Omega = (-50\; \tX,50\; \tX)$ with reflecting boundary conditions, the trait optimum $\rQ$ is considered to be linearly increasing over $\Omega$, and the rest of the parameters are assumed to be constant.
In Section \ref{sec:MarginalCoexistence}, a smaller habitat $\Omega = (-20\; \tX,50\; \tX)$ is used to reduce the computational cost of the simulations.
The layout of the two-dimensional problem presented in Section \ref{sec:MultipleCompetitionFactors} is described there.
The details of the numerical scheme and discretization parameters used to compute the solutions are given in Appendix B.
The implementation of the numerical simulations in MATLAB R2021a is provided in Supplementary Material 1.

\subsection{Range limits established by interspecific competition} \label{sec:Typical_Two}

It is shown by \citet{Case:Naturalist:2000} that interspecific competition can effectively limit the range of two competing species that would expand indefinitely in the absence of competition.
Here, we show that this result is not significantly affected by the evolution of the intraspecific phenotypic variance, and similar predictions still hold when the constant variance assumption is removed.
For this, we consider two populations of species that are initially distributed allopatrically and are both perfectly adapted to the environment at their center.
Similar to the initial values considered for the single species of Section \ref{sec:Typical_Single}, we set $\dx q_1(x,0) = \dx q_2(x,0) = 0.6\, \rd_x \rQ$ and $v_1(x,0) = v_2(x,0) = 1 \; \tQ^2$ for all $x\in \Omega$.

\begin{figure}[t!]
	\centering 
	\begin{subfigure}[t]{0.4\textwidth} 
		\includegraphics[width= 1\textwidth]{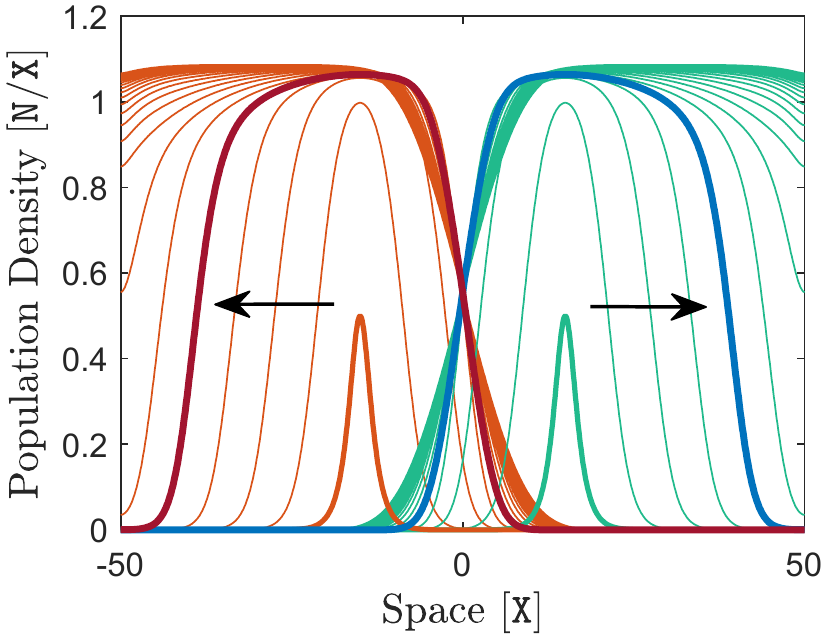}
		\caption{} \label{fig:n_nominal_1D_Two}
	\end{subfigure}
	\hfil
	\begin{subfigure}[t]{0.4\textwidth} 
		\includegraphics[width= 1\textwidth]{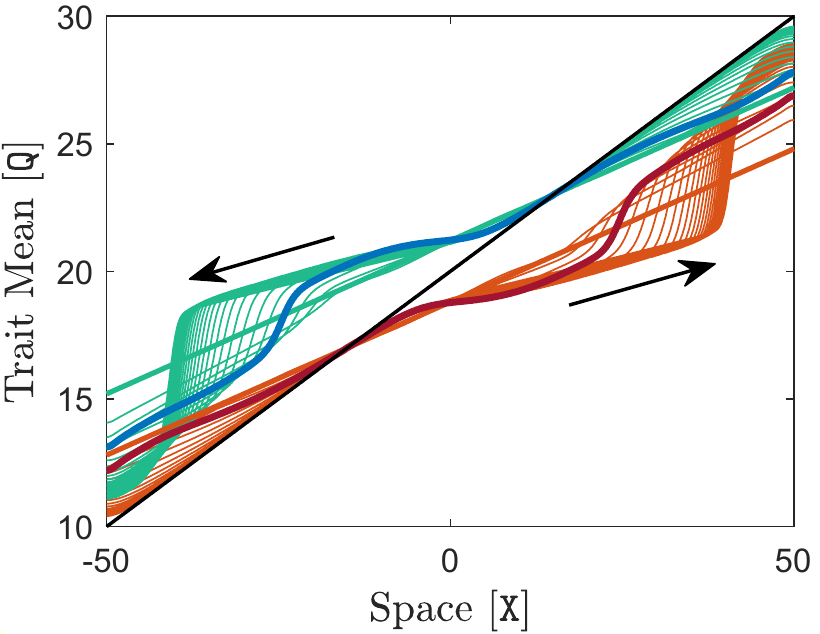}
		\caption{} \label{fig:q_nominal_1D_Two}
	\end{subfigure}
	\vfil
	\begin{subfigure}[t]{0.4\textwidth} 
		\includegraphics[width= 1\textwidth]{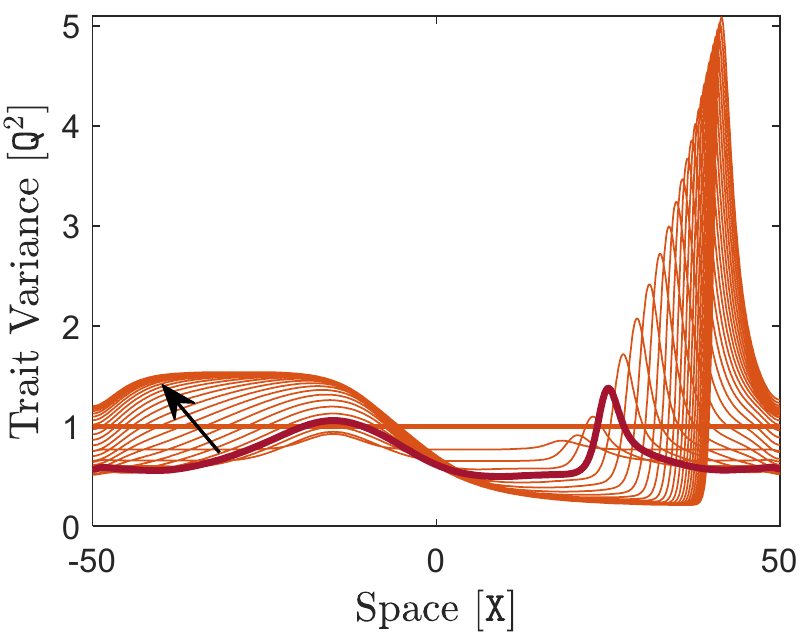}
		\caption{} \label{fig:v_nominal_1D_Two}
	\end{subfigure}
	\hfil
	\begin{subfigure}[t]{0.4\textwidth} 
		\includegraphics[width= 1\textwidth]{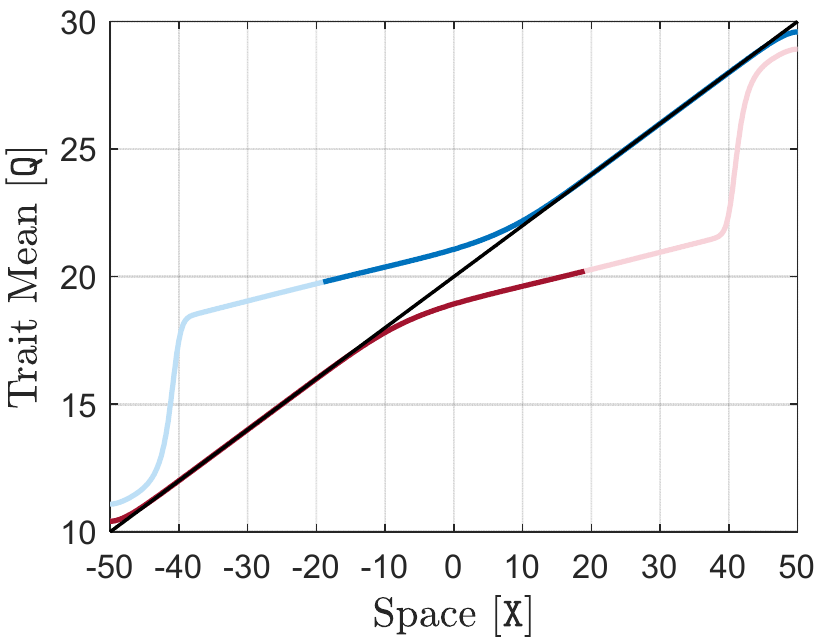}
		\caption{} \label{fig:q_nominal_final_1D_Two}
	\end{subfigure}	
	\caption{Range dynamics of two species with the same parameters and allopatric initial distributions. 
		Here,  $\rmm = 1$, $\rN = 2$,  and the rest of the model parameters for both species are the same and equal to the typical values given in Table \ref{tb:Parameters}.
		In (a)--(c), curves are shown at every $2 \, \tT$ and arrows show the direction of evolution in time.
		For the $1$st species in (a)--(c), curves are shown in orange, thick orange curves indicate the initial curves at $t=0 \,\tT$, and a sample curve at $t = 8 \, \tT$ is highlighted in red.
		For the $2$nd species in (a)--(c), curves are shown in green, thick green curves indicate the initial curves at $t=0 \,\tT$, and a sample curve at $t = 8 \, \tT$ is highlighted in blue.
		In (c), trait variance curves of the $2$nd species are not shown since they are the same as the curves of $1$st species reflected about the origin.
		The solid black line in (b) and (d) shows the environmental trait optimum $\rQ$.
		In (d), final curves of the species' trait mean at $t = 50 \, \tT$ ar shown.
		The curves are made transparent over the regions where population densities are approximately zero, as the values of trait mean over these regions are not biologically meaningful.
	}
	\label{fig:Nominal_1D_Two}
\end{figure}

Figure \ref{fig:Nominal_1D_Two} shows the solutions of the model for $T = 50 \, \tT$.
Each species expands its range to the edge of the habitat on the side where the species initially resides.
However, in the middle of the habitat where the two species meet, they prevent each other's progress.
At the beginning, when the species are apart from each other, their range dynamics is basically the same as the dynamics of a solitary species shown in Section \ref{sec:Typical_Single}, that is, their trait mean gradually converges to the trait optimum as the species adapt and advance to new areas.
Therefore, when the species first meet at the middle of the habitat, the individuals of either species that are near the interface of the two populations have relatively close phenotypes.
As a result, a strong interspecific competition is initiated between these individuals, according to the competition kernel specified by equation \eqref{eq:CompetitionKernel} in Appendix \ref{sec:Assumptions}.
This competition decreases the fitness of the populations at their interface.
Hence, the density of the populations declines over their interface, and this in turn intensifies the effect of asymmetric gene flow within each population at the interface.
By the same process as described for a single species in Section \ref{sec:Typical_Single}, this asymmetric gene flow gradually flattens the trait mean curve of each population over the interface, so that the curves depart from the trait optimum curve in opposite directions.       
This can be seen through the highlighted curves in Figure \ref{fig:q_nominal_1D_Two}.

The interaction described above between gene flow and interspecific competition continues until the level of maladapted phenotypes in each species' peripheral populations inside the interface becomes so extreme that it prevents local adaptation and hence stops species' range expansion through the interface.
Consequently, a region of sympatry is  formed between the species at the middle of the habitat, over which the species' population density monotonically declines to zero.
Figure \ref{fig:n_nominal_1D_Two} shows the formation of the associated range limits.
Moreover, as shown in Figure \ref{fig:q_nominal_final_1D_Two}, the species exhibit significant character displacement in sympatry.   
These observations are consistent in general with those given by \citet{Case:Naturalist:2000}.   
Note that smaller values of the variance of individuals' phenotypic utilization, $\rV_i$, restrict the interspecific competition to individuals with closer phenotypes.
As a result, the overall interspecific competition becomes weaker and the region of sympatry becomes wider.
The width of the region of sympatry increases also with shallower environmental gradients, which result in less extreme asymmetry in the gene flow.
If the environmental gradient is zero, the two species of Figure \ref{fig:Nominal_1D_Two} eventually become sympatric over the entire available habitat.

The highlighted curve in Figure \ref{fig:v_nominal_1D_Two} shows that the phenotypic variance within each species evolves to a bell-shaped curve.
This is a consequence of asymmetric gene flow at both edges of the species' range, as explained for the single species of Section \ref{sec:Typical_Single}.
Note that the large peaks in the curves shown in Figure \ref{fig:v_nominal_1D_Two} occur where the population has an infinitesimal density. 
Such regions are not practically considered within the range of the species, and values of trait mean and variance over these regions are of no biological meaning.

\subsection{Competitive exclusion} \label{sec:Competitive exclusion}

Generically, the region of sympatry originated by interspecific competition, as described above in Section \ref{sec:Typical_Two}, does not remain stationarily centered between the two species. 
The stationary region that is seen in Figure~\ref{fig:n_nominal_1D_Two} is simply due to the identical choice of parameter values for both species. 
Any differences in the parameters that change the competition balance between the two species cause the interface to constantly move towards the competitively weaker species.
If the imbalance is large enough, the weaker species goes extinct after being ultimately pushed to the boundary of the habitat.  
However, the dynamics of this competitive exclusion process is relatively slow.
For instance, if we repeat the computations of Figure \ref{fig:Nominal_1D_Two}, but this time with $\rR_1 = 3\, \tT^{-1}$ and $\rR_2 = 2\, \tT^{-1}$, we see that a region of sympatry is  formed quickly in the middle of the habitat within $10 \, \tT$, but it then moves rather slowly towards the right boundary.
The second species reaches the right boundary and begins to decline in density approximately at $t = 270 \, \tT$.
It practically becomes extinct at about $t = 460 \, \tT$.
This evolutionary dynamics can indeed be much slower if the difference between $\rR_1$ and $\rR_2$ is made smaller.

\begin{figure}[t!]
	\centering 
	\begin{subfigure}[t]{0.4\textwidth} 
		\includegraphics[width= 1\textwidth]{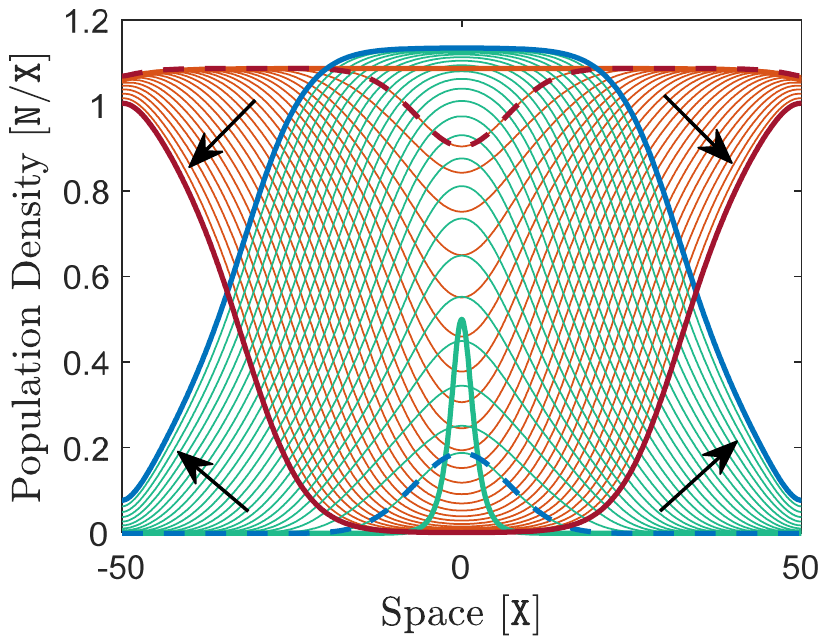}
		\caption{} \label{fig:n_fastGrowing_1D_Two}
	\end{subfigure}
	\hfil
	\begin{subfigure}[t]{0.4\textwidth} 
		\includegraphics[width= 1\textwidth]{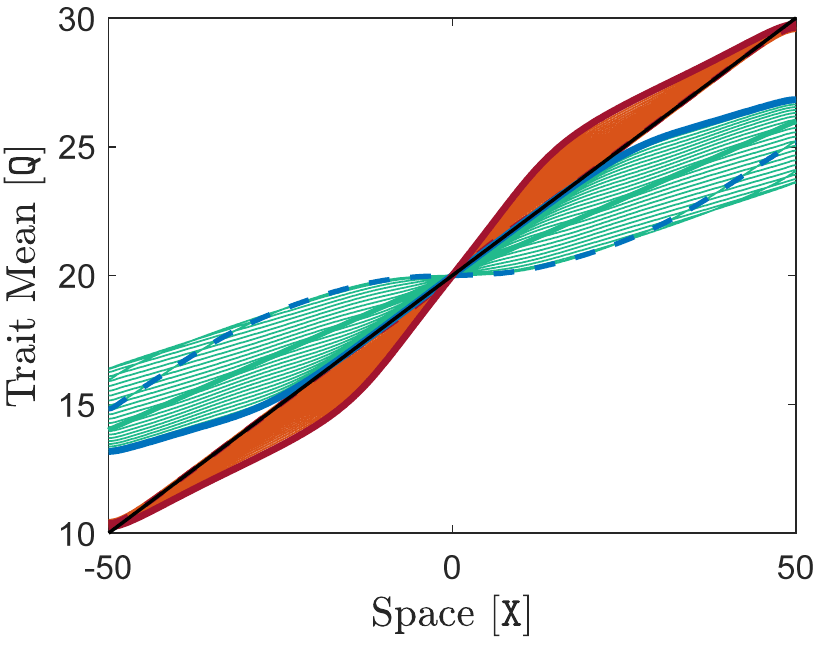}
		\caption{} \label{fig:q_fastGrowing_1D_Two}
	\end{subfigure}	
	\caption{Competitive exclusion of an established species by the invasion of a faster growing species ($\rmm = 1$, $\rN = 2$).
		Here, $\rR_1 = 2 \, \tT^{-1}$, $\rR_2 = 2.5 \, \tT^{-1}$, and the rest of the parameters take their typical values given in Table \ref{tb:Parameters}.
		Curves are shown at every $20 \, \tT$, and arrows show the direction of evolution in time after $t=20 \,\tT$.
		For the $1$st species, curves are shown in orange, thick orange curves are the initial curves at $t=0 \,\tT$, dashed red curves are shown at $t=20 \,\tT$, and solid red curves are shown at the final time $t = 500 \, \tT$.
		For the $2$nd species, curves are shown in green, thick green curves are the initial curves at $t=0 \,\tT$, dashed blue curves are shown at $t=20 \,\tT$, and solid blue curves are shown at the final time $t = 500 \, \tT$.
		The solid black line in (b) shows the environmental trait optimum $\rQ$.
		Note that, in (b), the curves of $1$st species at $t=0 \,\tT$ and $t=20 \,\tT$ are not clearly visible as they are very close to the trait optimum.
	} \label{fig:CompetitiveExclusion}
\end{figure}

A similar, but biologically more interesting competitive exclusion occurs when an invasive species is introduced to a habitat that is already occupied by a well-adapted species. 
To see this, we assume that the final population of the solitary species of Section \ref{sec:Typical_Single} at $t = 50\, \tT$ is occupying the habitat here at $t=0\, \tT$, labeled as the $1$st species.
Moreover, we assume a faster growing $2$nd species with $\rR_2 = 2.5\, \tT^{-1}$ is introduced at $t=0 \, \tT$ at the center of the habitat, with the same initial conditions used for the species of Section \ref{sec:Typical_Single}. 
With these initial populations, we compute the solutions over the time horizon of  $T = 500 \,\tT$.
The results are shown in Figure \ref{fig:CompetitiveExclusion}.
It can be seen that, after a transient initial reduction in its density, at about $t = 20 \, \tT$ the introduced species starts growing constantly and expanding its range.
The established species, which is significantly weaker than the invasive species due to its smaller maximum growth rate,  declines constantly in density over the expanding range of the invasive species.
When this species' central population at the origin becomes extinct, two separate regions of sympatry are formed on opposite sides of the habitat.
From this point on, these regions move in opposite directions towards the boundary of the habitat, and the introduced species eventually establishes itself over the entire habitat by fully excluding the preexisting species.  
This eco-evolutionary process, however, occurs very slowly.

\subsection{Stable marginal coexistence} \label{sec:MarginalCoexistence}

\begin{figure}[t!]
	\centering 
	\begin{subfigure}[t]{0.4\textwidth} 
		\includegraphics[width= 1\textwidth]{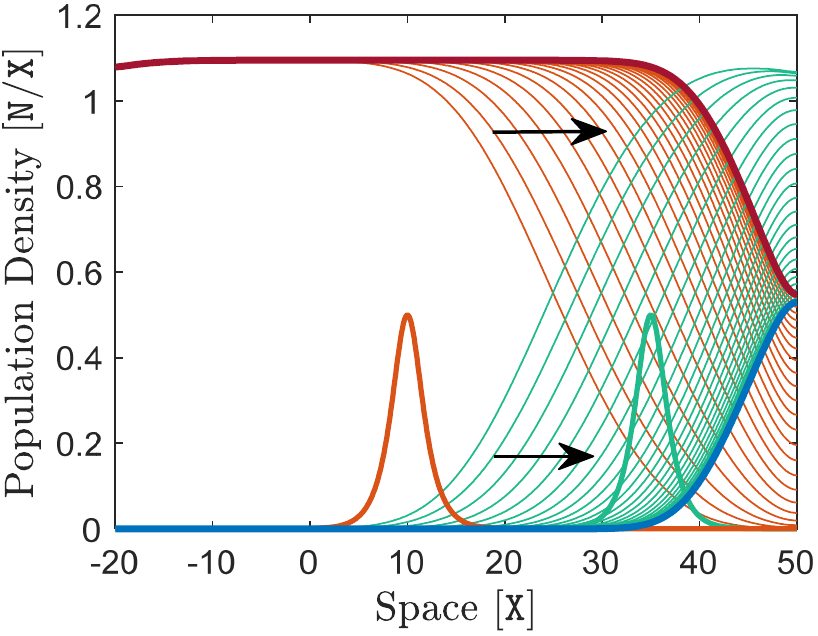}
		\caption{} \label{fig:n_edge_stable_1D_Two}
	\end{subfigure}
	\hfil
	\begin{subfigure}[t]{0.4\textwidth} 
		\includegraphics[width= 1\textwidth]{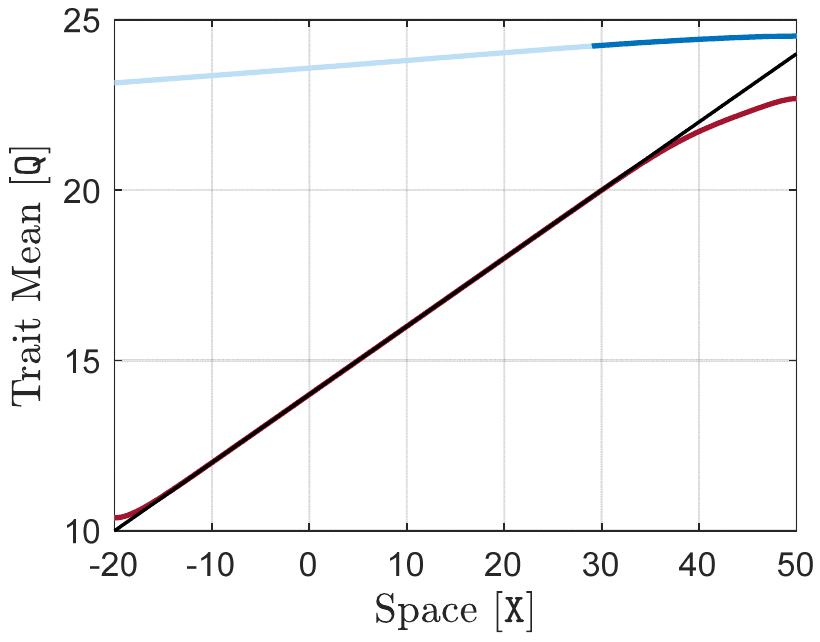}
		\caption{} \label{fig:q_edge_stable_final_1D_Two}
	\end{subfigure}
	\caption{Formation of an evolutionary stable marginal coexistence of two species at the vicinity of habitat's boundary ($\rmm = 1$, $\rN = 2$).
		Here, $\rR_1 = 2.1 \, \tT^{-1}$, $\rR_2 = 2 \, \tT^{-1}$, and the rest of the parameters take their typical values given in Table \ref{tb:Parameters}.
		In (a), curves are shown at every $200 \, \tT$ and arrows show the direction of evolution in time after the region of sympatry formed between the species starts moving towards the right boundary.
		For the $1$st species, curves are shown in orange, the thick orange curve indicates the initial curve at $t=0 \,\tT$, and the final curve at $t = 5000 \, \tT$ is highlighted in red.
		For the $2$nd species, curves are shown in green, the thick green curve indicates the initial curve at $t=0 \,\tT$, and the final curve at $t = 5000 \, \tT$ is highlighted in blue.
		In (b), final curves of the species' trait mean at $t = 5000 \, \tT$ are shown in red and blue, and the solid black line shows the environmental trait optimum $\rQ$.
		The curves in (b) are made transparent over the regions where population densities are approximately zero, as the values of trait mean over these regions are not biologically meaningful.
	}
	\label{fig:Edge_stable_1D_Two}
\end{figure}

If the two species of Section~\ref{sec:Typical_Two} differ from each other, but the difference between them is rather minor, then a stable equilibrium can exist in which the weaker species is not entirely excluded from the habitat, but it appears only over a limited extent adjacent to the boundary.
To see this, we consider two populations of species which differ only in their maximum growth rate, with $\rR_1 = 2.1 \, \tT^{-1}$ and $\rR_2 = 2 \, \tT^{-1}$.
We initialize a simulation with these two populations being allopatrically distributed and perfectly adapted to the environment everywhere, that is, 
$\dx q_1(x,0) = \dx q_2(x,0) = \rd_x \rQ$. 
Moreover, we set $v_1(x,0) = v_2(x,0) = 1 \; \tQ^2$ for all $x\in \Omega$.
Figure \ref{fig:Edge_stable_1D_Two} shows the solutions of the model with these initial populations over a long time horizon of $T = 5000 \, \tT$. 

It can be seen in Figure \ref{fig:Edge_stable_1D_Two} that an evolutionary stable marginal region of sympatry is eventually formed at the vicinity of the right boundary. This is because when the weaker species is ultimately pushed to the boundary, the level of maladaptive gene flow to its inner peripheral population declines, as there is no inward flux of phenotypes through the boundary.
Moreover, the overall reduction in the species' population density at the boundary also reduces the effect of intraspecific competition within the population. 
The peripheral population of the other species, however, does not experience a significant change in the level of gene flow it receives from the species' core areas. 
Since the interaction between the two species is mainly through their peripheral individuals, the relative advantage that the weaker species gains from the reduction in the amount of maladapted phenotypes and intraspecific competition can compensate for the species' minor weakness in interspecific competition.
As a result, the two species reach a steady state in which the weaker species survives marginally.
Convergence to this equilibrium, however, is very slow.  
Note that this stable range equilibrium also exists similarly under the assumption of constant phenotypic variance, as discussed by \citet{Case:Naturalist:2000}. 

The evolutionary dynamics described above at the vicinity of the habitat's boundary can be affected by the boundary conditions of the problem.
The reflective boundary condition we considered in the simulation above does not impose any reduction on the population density of the species when they reach the boundary.
An absorbing boundary condition, in contrast, allows for reduction of the population densities.
As a result, it might be imagined that under an absorbing boundary condition the weaker species of Figure \ref{fig:Edge_stable_1D_Two} will fail to survive, even marginally.
Although this is indeed a valid possibility, it does not imply that the stable marginal coexistence observed in Figure \ref{fig:Edge_stable_1D_Two} is simply an artifact of the reflecting boundary condition. 
In fact, depending on the overall effect of the different eco-evolutionary factors involved in the range dynamics near the boundary, the weaker species may still achieve sufficient fitness that compensates for its population reduction through the boundary as well. 
To see these possibilities, below we consider a, perhaps more realistic, situation in which the boundary of the habitat is set by a physical barrier.

\begin{figure}[t!]
	\centering 
	\begin{subfigure}[t]{0.4\textwidth} 
		\includegraphics[width= 1\textwidth]{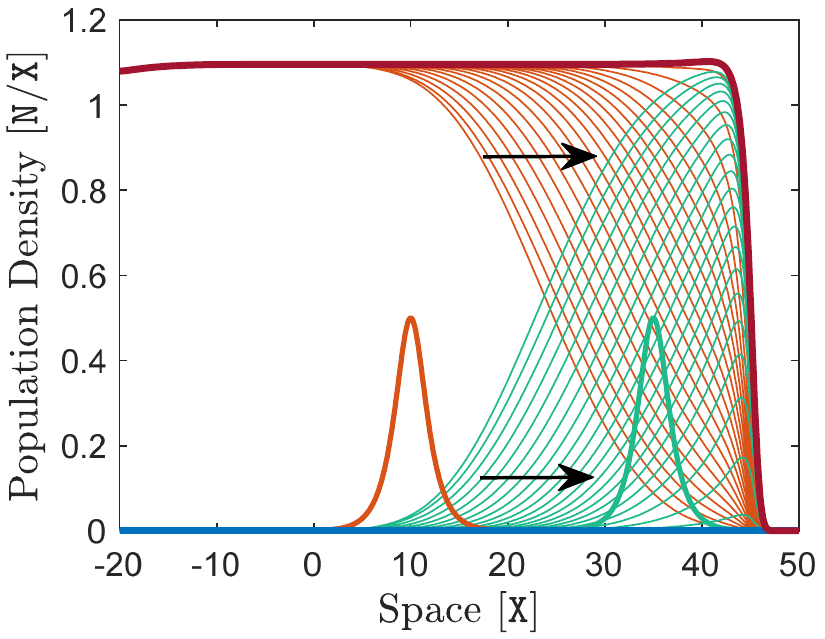}
		\caption{} \label{fig:n_sharpGradient_extinct_1D_Two}
	\end{subfigure}
	\hfil
	\begin{subfigure}[t]{0.4\textwidth} 
		\includegraphics[width=1\textwidth]{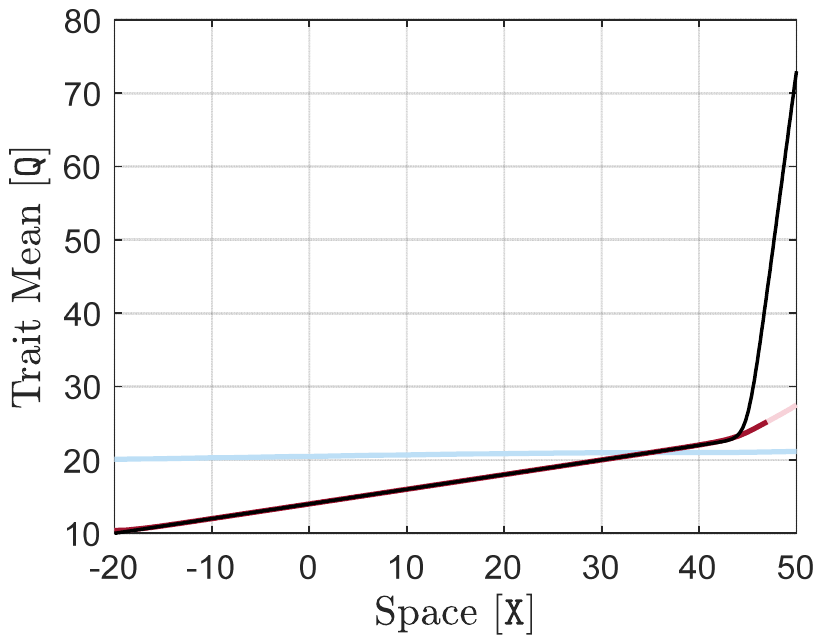}
		\caption{} \label{fig:q_sharpGradient_extinct_final_1D_Two}
	\end{subfigure}
	\vfill 
	\rule{1\textwidth}{0.5pt}
	\begin{subfigure}[t]{0.4\textwidth} 
		\includegraphics[width= 1\textwidth]{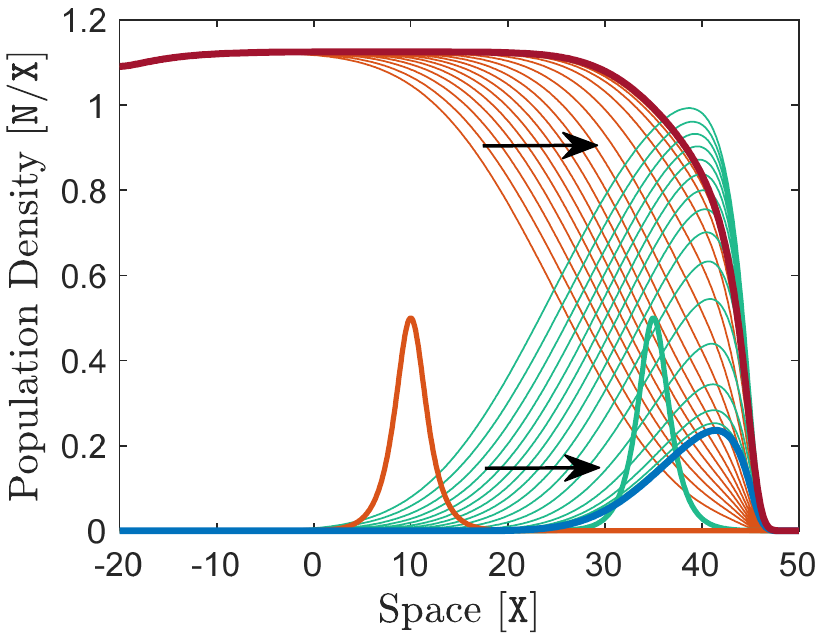}
		\caption{} \label{fig:n_sharpGradient_stable_1D_Two}
	\end{subfigure}
	\hfil
	\begin{subfigure}[t]{0.4\textwidth} 
		\includegraphics[width=1\textwidth]{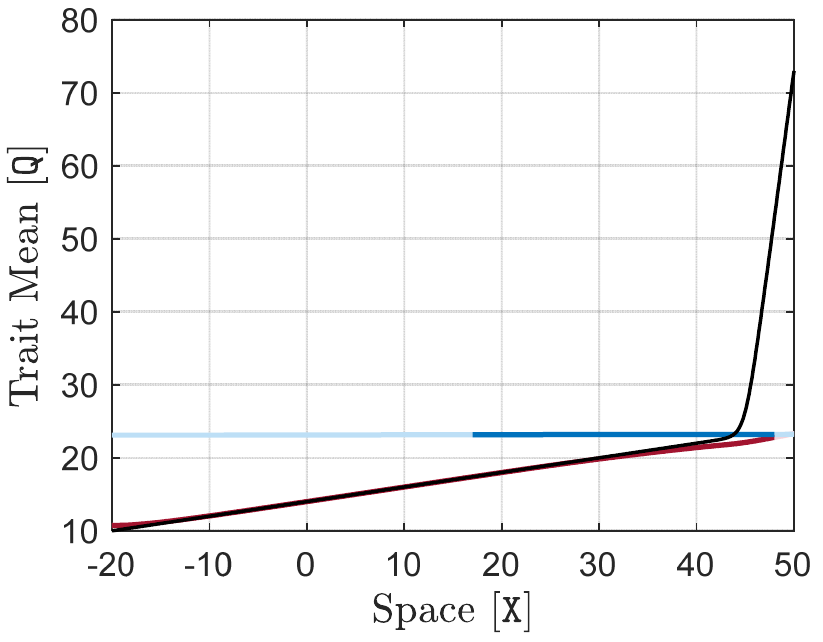}
		\caption{} \label{fig:q_sharpGradient_stable_final_1D_Two}
	\end{subfigure}
	\caption{
	Range dynamics of two species at the vicinity of a geographic region with extreme environmental gradient ($\rmm = 1$, $\rN = 2$).
	For the results shown in both panels, $\rR_1 = 2.1 \, \tT^{-1}$ and $\rR_2 = 2 \, \tT^{-1}$.
	The results shown in the upper panel are obtained with $\rD_1 = \rD_2 = 1\, \tX^2 / \tT$, whereas the results shown in the lower panel are obtained with $\rD_1 = \rD_2 = 5\, \tX^2 / \tT$.  
	The rest of the parameters take their typical values given in Table \ref{tb:Parameters} for the results shown in both panels.
	The same description as given in Figure \ref{fig:Edge_stable_1D_Two} holds here for the curves and arrows, with the only difference being that here the curves in (a) and (c) are shown at every $100 \, \tT$. 
	}
	\label{fig:SharpGradient_1D_Two}
\end{figure}

We assume that the right boundary of the habitat is set by a physical barrier at $x = 45 \, \tX$, which we model by an abrupt change in the gradient of the optimal trait from the typical value of $0.2 \, \tQ/\tX$ to the extreme value of $10 \, \tQ/\tX$. 
The results of Section \ref{sec:Gradients_Single} predict no chance of survival for the species over this region of extreme environmental gradient.
We repeat the computations of Figure \ref{fig:Edge_stable_1D_Two} with the same species, and the same reflecting boundary conditions. 
However, we note that here the reflecting boundary condition has no impact on the range dynamics of species at the vicinity of the barrier, as the population density of the species will be infinitesimal at the boundary.
The simulation results, over a time horizon of $T = 5000\, \tT$, are shown in the upper panel of Figure \ref{fig:SharpGradient_1D_Two}.

Unlike the results shown in Figure~\ref{fig:Edge_stable_1D_Two}, we see that no marginal region of coexistence is formed at the vicinity of the barrier in Figure~\ref{fig:n_sharpGradient_extinct_1D_Two}.
That is, the population loss that the weaker species suffers from in this case---due to the diffusion of its peripheral individuals to the region of extreme environmental gradient, where they cannot survive---eventually brings the species to extinction.
However, to show that a stable marginal region of sympatry can still be formed in the present habitat layout, we repeat the simulations described above, but this time with larger species dispersal of $\rD_1 = \rD_2 = 5 \, \tX^2/\tT$. 
This increases the effect of gene flow in the overall balance of eco-evolutionary factors, so that the advantage that the weaker species gains from the reduction in the level of maladapted phenotypes at the vicinity of the barrier can compensate for both its population loss through the barrier and its minor weakness in interspecific competition.
The resulting evolutionarily stable region of coexistence is shown in the lower panel of Figure \ref{fig:SharpGradient_1D_Two}.

\subsection{Effect of multiple competition factors} \label{sec:MultipleCompetitionFactors}

Competing species in nature may differ in many ecological and evolutionary parameters, which may also vary both over space and time.
The overall effect of these parameter differences dynamically changes the competition balance between the species, and hence the chance of survival, invasion success, and spread of a species within a community.
To demonstrate this---to some extent---we investigate a more complicated, but still intuitively understandable example of the range dynamics of two interacting species in a two-dimensional habitat.
Specifically, we assume that the habitat is already occupied by a well-adapted species with a spatially heterogeneous carrying capacity.
We then introduce a new species into the habitat which has a spatially homogeneous carrying capacity and consists of less specialized individuals, as compared with the established species.     
We investigate the establishment success or failure of the new species and whether or not it can be affected by changes in the species' dispersal.

We consider the habitat $\Omega = (-50\, \tX, 50\, \tX) \times (-50\, \tX, 50\, \tX)$ with an optimal cline given as $\rQ(x) = 20 + 0.2 x_1$.  
That is, the trait optimum is assumed to be linearly growing with a slope of $\rd_x \rQ = 0.2 \, \tQ/\tX$ along the $x_1$-axis, whereas it is assumed to be constant along the $x_2$-axis.
Boundary conditions are set to be reflecting at $x_1=-50$ and $x_1=50$, and periodic across $x_2=-50$ and $x_2=50$, as described in Remark \ref{rmk:BoundaryConditions}.
The parameter values that are not specified below are set to be constant and equal to the typical values given in Table \ref{tb:Parameters}. 

\begin{figure}[t!]
	\centering 
	\includegraphics[width= 1\textwidth]{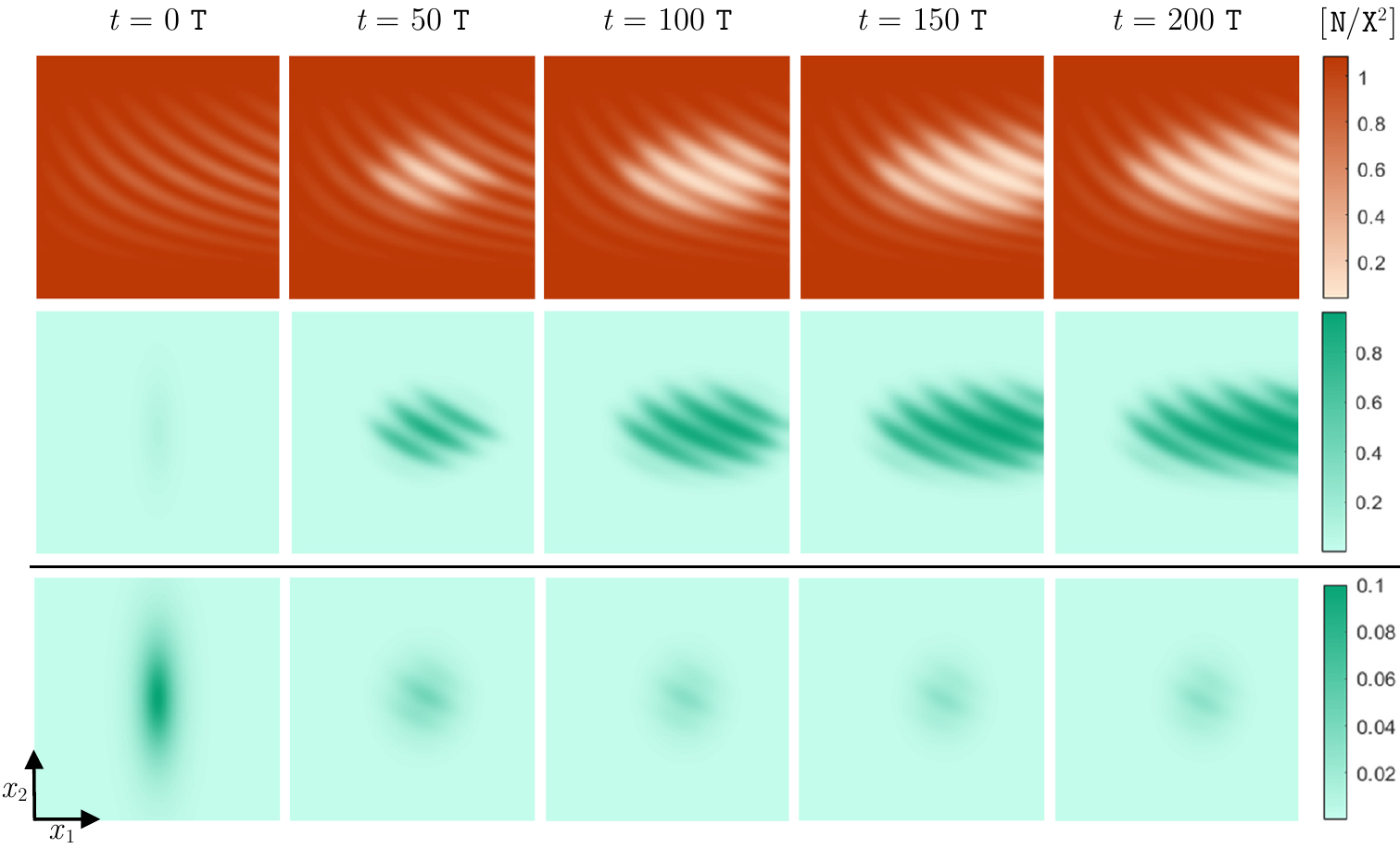}
	\caption{Effect of multiple competition factors on the establishment success of a newly introduced species in a two-dimensional habitat that is preoccupied by another species, ($\rmm = 2$, $\rN = 2$).
		Population density of the preoccupying species is shown in orange, whereas population density of the introduced species is shown in green.
		Carrying capacity $\rK_1$ of the environment for the preoccupying species ($1$st species) is heterogeneous and varies from $0.65 \, \tN/{\tX}^2$ to $1 \, \tN/{\tX}^2$, whereas for the introduced species ($2$nd spices) the carrying capacity is homogeneous and equal to $\rK_2 = 1 \, \tN/{\tX}^2$. 
		For the $1$st species, $\rV_1 = 4 \, \tQ^2$ and $\rD_1 = I_2 \, {\tX}^2/\tT$, where $I_2$ denotes the $2 \times 2$ identity matrix.
		For the $2$nd species, $\rV_2 = 6 \, \tQ^2$ and $\rD_2$ takes different values for the results shown in the upper and lower panels, as specified below.
		The rest of the parameters for both species take their typical values given in Table \ref{tb:Parameters}.
		Upper panel: With $\rD_2 = I_2 \, {\tX}^2/\tT$, the newly introduced species succeeds in establishing itself over the regions where it is favored by the overall effect of the competition factors.
		Lower panel: With $\rD_2 = 6 I_2 \, {\tX}^2/\tT$, the newly introduced species fails to establish itself in the habitat in $200 \, \tT$.
		Note the difference in the scales of the color bars in the upper and lower panels.
		Population density of the $1$st species is not shown in the lower panel as its initial profile is not significantly affected during the evolution of the $2$nd species over the simulation time horizon. 
	} \label{fig:HeterogeneousCapacity}
\end{figure}

The preexisting species, which is labeled as $1$st species hereafter, is assumed to have a spatially heterogeneous carrying capacity $\rK_1$ that varies over $\Omega$ within a range of values from $0.65\, \tN/\tX^2$ to $1\, \tN/\tX^2$. 
The spatial pattern of $\rK_1$ is approximately the same as the spatial pattern of the $1$st species's initial population density, shown at $t=0 \, \tT$ in Figure \ref{fig:HeterogeneousCapacity}. 
This is because, as described below, the initial population of this species is almost at a fully adapted steady state and occupies the entire habitat nearly to its full capacity.    
The introduced species, which is labeled as $2$nd species hereafter, is assumed to have a spatially homogeneous carrying capacity of $\rK_2 = 1 \, \tN/\tX^2$ over the entire geographic space. 
Moreover, this species is composed of individuals which are more generalist than the individuals of the $1$st species, with $\rV_2 = 6\, \tQ^2$.

To initialize the population of the $1$st species that preoccupies the habitat, we first consider only a solitary species with the same parameters as specified above for the $1$st species, and perform the following preliminary computations.   
We begin with a local population of this solitary species located at the center of the habitat, and let it evolve for a sufficiently long period of $T = 100 \, \tT$, so that the species has enough time to fully adapt to the environment and spread throughout the habitat.
Since the typical environmental gradient considered here is not steep, the results of Section \ref{sec:Gradients_Single} imply that the species eventually occupies the entire geographic space almost to its full capacity.
Therefore, at the end of the computations, the spatial pattern of the species' population density closely follows the spatial pattern of the carrying capacity $\rK_1$, and the species' trait mean converges approximately to $\rQ$.
We pick the final solutions obtained at $t = 100\, \tT$, and use them as the initial conditions for the preoccupying species of current study at $t = 0\, \tT$.  
    
We assume that the $2$nd species is introduced at $t = 0\, \tT$ over a fairly wide region at the center of the habitat, but with a relatively low density, as shown in Figure \ref{fig:HeterogeneousCapacity}. 
Note that this initial distribution is the same in both upper and lower panels of Figure \ref{fig:HeterogeneousCapacity}, but it is more visible in the lower panel due to the scale of the color bar.    
Finally, we set $q_2((0,0),0) = \rQ((0,0))$, $\dx q_2(x,0) = 0.6\, \rd_x \rQ$, and $v_2(x,0) = 1$.

Since $\rK_2 \geq \rK_1$ everywhere in $\Omega$, the introduced species would be expected to successfully establish itself and exclude the preoccupying species from most areas of the habitat, if the two species were composed of equally specialized individuals.
However---since we assume no limit on the amount of resources that can be utilized by individuals with any phenotypic value---the first species which is composed of more specialized individuals has a competition advantage over the second species.  
That is, the introduced species would have no chance of survival and would quickly become extinct if it didn't have a greater carrying capacity.
Therefore, whether or not the introduced species will eventually succeed in establishing itself over a region depends on the balance between these two opposite competition factors on that region, as well as on the interacting dynamic effects of gene flow.
To see this, we compute the solutions for a time horizon of $T = 200\, \tT$.
The computed population densities are shown in the upper panel of Figure \ref{fig:HeterogeneousCapacity} at every $50\, \tT$.
It can be seen that the introduced species successfully establishes itself by gradually excluding the preestablished species from the areas where the difference between the carrying capacities is sufficiently large
to compensate for the utilization disadvantage of the introduced species. 

The establishment success of the introduced species can also be affected by other factors, besides the competition factors.
In particular, it can be affected by the level of maladaptation created by gene flow.
To see this, first note that the introduced species cannot quickly adapt itself to the regions over which $\rK_1$ is not sufficiently small and hence the $1$st species dominates. 
These regions are relatively close to each other, due to the special pattern of $\rK_1$ considered here.
Now, assume that the individuals of the introduced species migrate over significantly longer distances,
so that phenotypes can be effectively shared between the regions populated by poorly adapted individuals and their adjacent regions populated by better adapted individuals.
As a result, the mean trait of the introduced species becomes relatively uniform over the range of the species and deviates largely from the optimum.
Depending on the value of $\rD_2$, this process can significantly decrease the establishment speed of the $2$nd species, or can even totally prevent it from happening.   
For instance, as shown in the lower panel of Figure \ref{fig:HeterogeneousCapacity}, the introduced species fails to establish itself in $200 \, \tT$ if we set 
$\rD_2 = 6 I_2 \, {\tX}^2/\tT$, where $I_2$ denotes the $2 \times 2$ identity matrix.

\section{Discussion} \label{sec:Discussion}

The present study is part of a larger effort to explain why species can form stable range boundaries in the absence of environmental discontinuities. 
We focused on two important factors, competition and maladaptation to an environmental gradient, that are commonly thought as possible causes of species's range limits.   
Our specific goal was to reconcile the conclusions of \citet{Barton:BookChapter:2001}, who modeled a single species in an environmental gradient with variable genetic variance, and those of \citet{Case:Naturalist:2000}, who modeled two competing species in an environmental gradient with constant genetic variance. 

\subsection{Range dynamics in the presence of competition and an environmental gradient}

Toward the goal of our study, we developed a model that resembles the competition model of \citet{Case:Naturalist:2000}, but with variable (genetic) trait variance for each species. 
We provided a rigorous mathematical derivation of the model, as well as a detailed discussion on its parameters and their biologically reasonable ranges of values.
We computationally studied the solutions of the model to investigate its predictions in a number of evolutionary regimes, indicating the effects of gene flow, environmental gradients, interspecific competitions, climate change, and species dispersal on species' eco-evolutionary range dynamics. 
Our simulations show behavior that contrasts strongly with that of the seminal model of \citet{Kirkpatrick:Naturalist:1997}, of which the models developed by \citet{Case:Naturalist:2000} and \citet{Barton:BookChapter:2001} are extensions---most strikingly, we do not find range pinning of a single species.
Instead, we find behavior broadly consistent with that of \citet{Case:Naturalist:2000}, indicating that the conclusions in that study are robust to the removal of constraints on genetic variance. 

\subsection{Comparison with single-species models}

To some extent, the single-species reduction of our model showed consistent results with the work of \citet{Barton:BookChapter:2001},
that the species' phenotypic variance increases proportionally as the environmental gradient increases, so that the species' adaptation and range expansion is still possible even at steep environmental clines.
However, we showed that the species' expansion speed reduces as the environmental gradient increases.
Moreover, the species' ability to expand its range does not necessarily imply that the species will eventually spread all over the habitat in full capacity.
Stabilizing selection tends to decrease genetic variance, both directly through the term $-\rS v^2$ in \eqref{eq:TraitVariance_Single}, and indirectly by imposing a genetic load on the fitness of the species.
This genetic load appears as the term $-\frac{\rS}{2} v$ in the fitness function given in \eqref{eq:PopulationDensity_Single}. 
As a result, the species suffers from substantial loss of fitness when $v$ takes large values along steep optimal clines, and hence its equilibrium population density lies significantly below its ecological carrying capacity.
At extreme gradients, as we showed, the species fails to survive and becomes extinct.
For linear clines, this extinction was estimated to occur at any gradient beyond $|\rd_x \rQ|_{\max} = \sqrt{(2\rR^2/\rS \rD) - \rU/2\rD}$.

In comparison with the equations of \citeauthor{Barton:BookChapter:2001}'s model \citep[equ. 16]{Barton:BookChapter:2001}, our equations \eqref{eq:PopulationDensity_Single}--\eqref{eq:TraitVariance_Single} include additional nonlinear terms that model the effect of intraspecific competition. 
These nonlinear terms affect the shape of the wave amplitude and speed curves discussed in Section \ref{sec:Gradients_Single}.
However, the main reason why our results---showing species extinction at steep but finite environmental gradients, even with mutational forcing---differ from those of \citet{Barton:BookChapter:2001} is most likely due to the approximations made in the analysis performed by \citeauthor{Barton:BookChapter:2001}.
To see this, let $\rA$ and $\rB$ be, respectively, the scaled selection strength and the scaled optimum gradient defined by \citet{Barton:BookChapter:2001}.
The approximate analysis of \citeauthor{Barton:BookChapter:2001} based on a simplified version of the equations \citep[equ. 10]{Barton:BookChapter:2001} led  to the conclusion that the genetic load generated by inflation of the genetic variance reduces the equilibrium population density at a rate approximately proportional to $\exp(-\rB/\sqrt{2})$, while the species is still allowed to occupy an indefinitely wide range. 
However, if we perform our equilibrium analysis of Section \ref{sec:Gradients_Single} on the original equations given by \citet[equ. 16]{Barton:BookChapter:2001} with logistic density dependent fitness, we obtain a critical value, 
$\rB_{\max} = 2 - \rA^2/2$, of the scaled environmental gradient above which the species fails to survive.
This would then be consistent with our results.

\subsection{Comparison with two-species models}
 
Comparison between the results presented in Section \ref{sec:Typical_Two} and those presented by \citet{Case:Naturalist:2000} shows that the evolution of trait variance does not substantially change the dynamics of the range limit formed at the interface of two competing species.
The interaction between interspecific competition and gene flow can still effectively prevent species' range expansion when they meet each other.
The species exhibit character displacement and coexist in sympatry over a region formed between them.
Although the results show that in practice this region of sympatry does not remain stationary in time and moves towards the competitively weaker species---and can eventually result in exclusion of this species---the dynamics of this movement is expected to be very slow, so that the competitively formed range limits may appear to be stationary in experimental measurements. 

\subsection{Spatial profile of trait variance}

The spatial profile of trait variance, as it evolves in time according to \eqref{eq:PopulationDensity}--\eqref{eq:TraitVariance}, is consistent with experimental measurements performed by \citet{Takahashi:MolecularEcology:2016} on two similar species of damselflies along a latitudinal temperature gradient.
These measurements show that genetic variation is relatively constant and high within well-adapted central populations, whereas it drastically declines at species' range margins where significant phenotype-environment mismatches are observed. 
However, the results presented here do not necessarily support the conclusion made by \citet{Takahashi:MolecularEcology:2016}, which suggests that the lack of genetic variation at species' range margin is responsible for preventing adaptation and range expansion.
The bell-shaped profile of trait variance shown in Section \ref{sec:SingleSpecies} is quickly formed as the initial population grows and adapts to the environment, so that its trait mean converges to the trait optimum at the population center.
The sharp decline in the trait variance at the periphery of the species' range is mainly due to the flattened curve of trait mean over these regions, which in turn is caused by asymmetric gene flow from the core.
This general profile of trait variance is maintained as the species expands it range and eventually occupies the entire habitat, even with steep environmental gradients.  
Therefore, based on the predictions of our model, the significant decline in genetic variation at the range margin, compared with the core, does not necessarily identify it as a main factor preventing range expansion.
Note that the results presented in Section \ref{sec:Typical_Two} show that intraspecific trait variance also declines significantly at the interface between two competing species, where species' range expansion is indeed prevented but mainly as a result of interaction between gene flow and interspecific competition.

\subsection{Numerical and mathematical remarks}

It is worth commenting on features of our model that make it delicate to simulate numerically and challenging to investigate mathematically. 
First, we note that the mathematical equations of the model and their derivation assume that $n_i(x,t) \neq 0$ for all $(x,t) \in \Omega \times [0,T]$; otherwise the terms $\dx \log n_i(x,t)$ in \eqref{eq:TraitMean} and \eqref{eq:TraitVariance} will present singularities.
Therefore, at least an infinitesimal population density must be considered for all initial populations on the entire domain $\Omega$.
This consideration for the initial population, however, does not necessarily prevent potential numerical singularities that may arise during the evolution of the species when population density of a species becomes extremely small at some points.
Using finer spatial discretization meshes and smaller time steps for the numerical scheme, as well as choosing better adapted initial populations and smaller geographic spaces, can resolve such numerical singularity problems in many simulations.
However, specifically designed numerical treatment will be required for certain problems, for instance when a species undergoes an extinction regime over a long simulation time horizon.
None of the simulations results presented in this paper, however, required such a specific numerical treatment.

Next, we note that the basic eco-evolutionary behaviors demonstrated in this paper do not necessarily provide a comprehensive picture of the dynamics of the model.
It would be fruitful to carry out a rigorous mathematical analysis of the model to establish existence or nonexistence of other evolutionary regimes.
Investigating whether or not there exist sets of biologically plausible parameter values that will still result in the formation of an evolutionarily stable limited range for a solitary species is of particular interest.
As illustrated in the example of Section \ref{sec:MultipleCompetitionFactors}, for more realistic problems which involve multiple species of different biological and genetic characteristics in a two-dimensional geographic space, the community range dynamics predicted by the model can be quite complicated.
Although rigorous mathematical analysis of the model  may not be tractable for such problems, numerically computed solutions of the model under different conditions can result in valuable insights.
Exploring the impact of environmental heterogeneity and patchiness on the geographic structure of a community of species can be an example of a potentially interesting computational study.    

Finally, we note that, for a single species, the numerical singularity problem described above can be technically resolved by restructuring the model and representing it as a type of PDE system with moving boundary.
In this new model structure, the equations \eqref{eq:PopulationDensity}--\eqref{eq:TraitVariance} will be defined on the evolving range of species given as $\Omega_t(n):=\{ x \in \Gamma: n(x,t) \geq 0 \}, t \in [0,T]$, where $\Gamma$ denotes the available geographic space. 
Note that this implies $n=0$ in $\Gamma \setminus \Omega_t$. 
Appropriate boundary conditions will be required on the moving boundary $\partial \Omega_t$ for variables $q$ and $v$.
The equation of the evolution of the moving boundary can also be derived using commonly used velocity conditions.
The resulting system of singular parabolic differential equations would, like the original system, be a rewarding problem to be analyzed mathematically.

\subsection{Future research directions}

There is a large body of empirical work involving transplants beyond range boundaries that investigate species' (mal)adaptation at range margins \citep{Angert:AnnualReview:2020}.
	The key question, though, is what causes the range boundaries to exist where they do. 
	A test of the ``genetic swamping'' hypothesis---that gene flow from the range's center inhibits adaptation at the margins---requires estimates of gene flow (and hence
	dispersal) as well as of fitness and of the optimal phenotype as a function of space. 
	Thus, although many transplant experiments have added to our understanding of local adaptation or failure to adapt at range margins, few experiments have accounted for
	all the factors necessary to test the genetic swamping hypothesis.

Theoretical models of evolutionary range dynamics ultimately must help explain empirical phenomena. 
To identify the causes of observed range dynamics for a given species, practitioners must choose the most appropriate model from a potentially large family of models. 
A case of particular interest is range pinning: the Kirkpatrick-Barton family of models, including \citet{Case:Naturalist:2000} and \citet{Barton:BookChapter:2001}, can be used to decide whether genetic swamping has set range limits for a population in nature. Given the paucity of real-life cases where this has been shown to occur
\citep{Angert:AnnualReview:2020, Colautti:MolecularEcology:2015, Bridle:ProcRoyalSociety:2009, Benning:Naturalist:2019, Willi:Naturalist:2019, Micheletti:MolecularEcology:2020, Paul:Naturalist:2011}, it is reasonable to be conservative when concluding that genetic swamping has set the range limits for a species.

Since stochasticity often promotes range pinning 
\citep{Bridle:EcologyLetters:2010, Polechova:PNAS:2015}, it is therefore reasonable to use deterministic models when testing for genetic swamping as the cause of a range limit. 
However, it would be misguided to only consider models that added a single extra feature to the KB model, since multiple additional factors surely coexist in many, if not all, natural systems. 
This justifies the construction and careful study of models incorporating multiple extra features. 
We have done so in the present work by incorporating both competition and evolving trait variance in our model.
We believe that further studies of this type, incorporating other demographic, genetic, environmental or ecological factors that may influence range dynamics, would be well justified. 
Such studies should not only facilitate the application of these models to testing the genetic swamping hypothesis, but also inform the development of practical models that help predict the outcomes of specific invasions and the effectiveness of possible control measures.

\section{Conclusion}

It is intuitively plausible that, as our results suggest, arbitrarily high levels of 
genetic 
variance will not always promote range expansion. However, the absence of range pinning in our single-species model is noteworthy, since it harmonizes with the conclusions of both \citet{Barton:BookChapter:2001} and \citet{Case:Naturalist:2000}.
It suggests that, when additive genetic variance is not held constant by fiat, the phenomenon of range pinning via ``genetic swamping'' identified by \citet{Kirkpatrick:Naturalist:1997} will not occur. 
However, viewing this finding as conclusive would ignore the growing body of theoretical work that has built on the results of \citet{Kirkpatrick:Naturalist:1997}, identifying conditions that promote or inhibit range pinning. 
Most notably, genetic drift and other stochastic effects are absent from our model, as they are from the works of \citet{Kirkpatrick:Naturalist:1997}, \citet{Case:Naturalist:2000}, and \citet{Barton:BookChapter:2001}. 
Studies with individual-based models, which implicitly feature not only stochasticity but also variable trait variance, suggest that a system with these factors in combination may exhibit range pinning through genetic swamping.
However, such models are difficult to analyze, and their simulation may involve hidden factors such as a discrete spatial grid that would promote range pinning in the numerical results but not in the underlying model. 
Mathematical analysis of stochastic differential equations may provide a middle ground for understanding the effects of stochasticity in an environmental gradient. 
As with the present study and other theoretical work, such analyses should help guide empirical studies, so far quite rare \citep{Micheletti:MolecularEcology:2020, Angert:AnnualReview:2020, Colautti:MolecularEcology:2015, Benning:Naturalist:2019}, that will be the ultimate test of the hypothesis that genetic swamping can induce range pinning in the absence of competition.

\section*{Acknowledgements} 
The authors would like to thank J. Goodman and the Courant Institute of Mathematical Sciences, New York University, for their hospitality during part of the preparation of this research.

\appendix \normalsize
	
\section*{Appendix A: Model Derivation} 
\addtocounter{section}{1} 

To derive the equations of the model given by \eqref{eq:PopulationDensity}--\eqref{eq:NonlinearTerms}, 
we first formulate the intrinsic growth rate of the individuals within each species, 
which determines the local dynamics of the evolution of the species.
For this, at position $x \in \Omega$ and time $t \in [0,T]$, let $\phi_i(x,t,p)$ denote the relative frequency of a quantitative phenotypic trait with phenotype value $p \in \bbR$ within the $i$th species.
Moreover, let $\alpha_{ij}(p,p')$ denote the competition kernel that captures the per capita effect of phenotype $p'$ in the $j$th species on the frequency of phenotype $p$ in the $i$th species.
The exact definition of this competition kernel is given in Section \ref{sec:CompetionKernels} below. 
Finally, let $g_i(x,t,p)$ denote the intrinsic growth rate of the population of individuals with phenotype $p$ within the $i$th species.

\subsection{Intrinsic growth rates}  \label{sec:GrowthRate}
For a community of $\rN$ competing species, we define the intrinsic growth rate of each species as \citep[equ. (2)]{Case:Naturalist:2000},
\begin{align} \label{eq:GrowthRate}
	g_i(x,t,p) &:= \rR_i(x) \left(1 - \frac{1}{\rK_i(x)} \sum_{j=1}^{\rN} n_j(x,t) \int_{\bbR} \alpha_{ij}(p, p') \phi_j(x,t, p') \rd p' \right) 
	-\frac{\rS }{2} (p - \rQ(x))^2, \quad i=1, \dots, \rN. 
\end{align}
The first term in \eqref{eq:GrowthRate} is a Lotka-Volterra model of competing species in which the convolution term with $j=i$
expresses the effect of intraspecific phenotypic competition on the frequency of phenotype $p$, whereas the 
convolution terms with $j \neq i$ account for the effect of interspecific competition from the phenotypes in the other species.
The second term in \eqref{eq:GrowthRate}
incorporates the effect of directional and stabilizing selection on individuals with phenotype $p$.
A local population of a species at position $x$ that has a phenotypic trait value different from the environmental optimal value $\rQ(x)$ can only reach an equilibrium density that is lower than its carrying capacity.
This penalizing effect of the phenotypic selection is made stronger by choosing larger values for $\rS$. 

\subsection{Competition kernels} \label{sec:CompetionKernels}

As proposed by \citet[equ. (3)]{Case:Naturalist:2000}, we obtain the competition kernels $\alpha_{ij}$ in \eqref{eq:GrowthRate} using the MacArthur-Levins overlap formula between resource utilization curves of each species \citep{Macarthur:Naturalist:1967}, along with the total resource consumption law given by \citet[equ. (24.50)]{Roughgarden:PopulationGenetics:1979}.
Suppose the environmental resources vary continuously along a resource axis denoted by variable $r$.
Moreover, suppose that individuals with phenotype $p$ within each species possesses a resource utilization curve of the form
\begin{equation*}
 \xi_{i,p}(r) := \Psi e^{\kappa p} \psi_{i,p}(r), \quad i=1, \dots, \rN,
\end{equation*}
where $\psi_{i,p}(r)$ is a probability density function, which gives the probability density that the individuals obtain a unit of resource from point $r$. 
The term $\Psi e^{\kappa p}$ gives the total amount of resource consumed by an individual with phenotype $p$.
This power law is proposed by \citet[equ. (24.50)]{Roughgarden:PopulationGenetics:1979} based on the assumption 
that energy consumption by an individual is proportional to its weight.
This interpretation, however, does not necessarily hold for the general trait-based model presented in this paper,
and this specific form of resource consumption form is mainly adopted for the simplicity of derivations and for providing the model with the flexibility of incorporating asymmetric intraspecific competitions with $\kappa \neq 0$.

The resource utilization functions $\xi_{i,p}(r)$ are used to obtain the competition kernels $\alpha_{ij}$ by the following overlap formula \citep[equ. (24.5)]{Roughgarden:PopulationGenetics:1979}
\begin{equation*}
\alpha_{ij}(p,p') := \frac{ \int_{\bbR} \xi_{i,p}(r) \xi_{j,p'}(r) \rd r }{\int_{\bbR} \xi_{i,p}^2(r) \rd r}.
\end{equation*}
Calculation of $\alpha_{ij}(p,p')$ based on this formula involves having precise information about resource values.
However, it is convenient to assume that the resource axis can be identified by phenotype axis, as proposed by \citet[equ. (24.51)]{Roughgarden:PopulationGenetics:1979}. 
For this, let $r_p$ denote the point on the $r$-axis from which individuals of phenotype $p$ obtain their average amount  of resources.
We assume that $r_p$ does not depend on which species the individuals belong to, that is, 
$r_p = \int_{\bbR} r \psi_{i,p}(r) \rd r$ for all $i= 1,\dots,\rN$.
We further assume that there is a smooth one-to-one map $I:p \mapsto r_p$, which can be used to identify the $r$-axis with the $p$-axis, that is, for every $r \in \bbR$ there exist a unique phenotype $\tilde{p}\in\bbR$ such that 
$r=I(\tilde{p})\equiv \tilde{p}$. 
Therefore, the resource utilization functions and competition kernels can be written only based on trait values, as
\begin{alignat}{3}
	 \xi_{i,p}(\tilde{p}) &:= \Psi e^{\kappa p} \psi_{i,p}(\tilde{p}),  &\quad i=1, \dots, \rN, & \label{eq:UtilizationFunction}\\
	 \alpha_{ij}(p,p') &:= \frac{ \int_{\bbR} \xi_{i,p}(\tilde{p}) \xi_{j,p'}(\tilde{p}) \rd \tilde{p} }{\int_{\bbR} \xi_{i,p}^2(\tilde{p}) \rd \tilde{p}},   &\quad i=1, \dots, \rN,  &\quad j=1, \dots, \rN.\label{eq:OverlapFormula}
\end{alignat}
Note that, by the definition of $I$ we have $\int_{\bbR} \tilde{p} \psi_{i,p}(\tilde{p}) \rd \tilde{p} = p$.
We refer to $\xi_{i,p}$ in \eqref{eq:UtilizationFunction} as \emph{phenotype utilization function} of individuals with phenotype $p$ within the $i$th species.

The identification stated above can represent the empirical relationship between  functional response traits and environments.
In addition to simplifying the mathematical derivations, this identification allows estimation of the parameters of the utilization functions using trait-based approaches to niche quantification \citep{Violle:PlantEcology:2009, Ackerly:EcologyLetters:2007}. 
This was further discussed in Section~\ref{sec:Parameters}.

\subsection{Changes due to mutation}
Let $\nu(\delta p)$ denote the probability density that by mutation a phenotype $p$ changes to a phenotype $p + \delta p$.
We use the equation provided by \citet{Kimura:PNAS:1965}, but at the level of phenotypic effects, to approximately model the rate of mutational changes in the frequency of phenotypes as 

\begin{equation}\label{eq:Mutationeffect}
	\dt^{(\rM)} \phi_i(x,t,p):= -\eta \phi_i(x,t,p) 
	+ \eta \int_{\bbR} \nu(p-p') \phi_i(x,t,p') \rd p', \quad i=1,\dots,\rN,
\end{equation}
where $\eta \geq 0$ is the mutation rate per capita per generation.
The first term in \eqref{eq:Mutationeffect} gives the reduction rate in the frequency of phenotype $p$ within the $i$th species due to mutation to other phenotypic values.
The second term gives the growth rate in the frequency of phenotype $p$ due to mutations to $p$ from other phenotypic values within the $i$th species.

\subsection{Model assumptions}  \label{sec:Assumptions}

As stated in Section \ref{sec:Discription}, the following major assumptions are made on the populations' dispersal and reproduction, and on the elements of the intrinsic growth rates and competition kernels described above.
These assumptions are used in Section \ref{sec:EquationsDerivation} to derive the equations of the model based on  \eqref{eq:GrowthRate}.

	\begin{enumerate}[i)]
		\item Each species disperses in the habitat by diffusion.
		\item \label{assmpt:Selection} Nonlinear environmental selection for the optimal phenotype $\rQ(x)$ is stabilizing for all $x\in \Omega$. 			
		\item Frequency of trait values follows a normal distribution for all $x \in \Omega$ and $t \in [0,T]$, that is,
		\begin{equation} \label{eq:TraitDistribution}
			\phi_i(x,t,p) := \frac{1}{\sqrt{2 \pi v_i(x,t)}} \exp\left( -\frac{(p - q_i(x,t))^2}{2 v_i(x,t)} \right), \quad i =1,\dots, \rN.
		\end{equation}
	\item{Within each species, the reproduction rate of individuals with phenotype $p$ depends on the population density of individuals with the same phenotype $p$.}
		 \item Phenotype utilization distribution $\psi_{i,p}$ in \eqref{eq:UtilizationFunction} is normal, that is,  
		\begin{equation} \label{eq:UtilizationDistribution}
		\psi_{i,p}(\tilde{p}) = \frac{1}{\sqrt{2 \pi \rV_i}} \exp\left( -\frac{(\tilde{p} - p)^2}{2 \rV_i} \right), \quad i =1,\dots, \rN.
		\end{equation}
		Therefore, competition kernels given by \eqref{eq:OverlapFormula} can be calculated as 
		\begin{equation} \label{eq:CompetitionKernel}
			\alpha_{ij}(p,p') = \Lambda_{ij} \exp(\kappa^2 \bar{\rV}_{ij}) \exp \left( -\frac{(p - p' + 2 \kappa \bar{\rV}_{ij})^2}
			{4 \bar{\rV}_{ij}} \right), \quad i=1, \dots, \rN,  \quad j=1, \dots, \rN,
		\end{equation}
		where $\Lambda_{ij}:=\sqrt{\smash[b] { \rV_i / \bar{\rV}_{ij}}}$ with $\bar{\rV}_{ij} := \frac{1}{2}(\rV_i + \rV_j)$, as in \eqref{eq:NonlinearTerms}.
		\item \label{assmpt:MutationKernel}The mutation kernel $\nu$ in \eqref{eq:Mutationeffect} is the probability density function of a probability distribution with constant zero mean and constant variance $\rV_{\scM}$. 
		That is, in particular,  $\nu$ is independent of population density, trait mean, or baseline trait variance.
	\end{enumerate}

\begin{remark}[Symmetric competition kernel] 
	The MacArthur-Levins overlap formula \eqref{eq:OverlapFormula} gives asymmetric competition kernels \eqref{eq:CompetitionKernel}, wherein 
	$\alpha_{ij}(p,p') \neq \alpha_{ji}(p,p')$ when $\rV_i \neq \rV_j$ or $\kappa \neq 0$.  
	A symmetric alternative to \eqref{eq:OverlapFormula} is proposed in the literature \citep{Pianka:AnnualReviewEcolology:1973}, which can be written as
	\begin{equation*}
		\alpha_{ij}(p,p') := \frac{ \int_{\bbR} \xi_{i,p}(\tilde{p}) \xi_{j,p'}(\tilde{p}) \rd \tilde{p} }{\left( \int_{\bbR} \xi_{i,p}^2(\tilde{p}) \rd \tilde{p} \int_{\bbR} \xi_{i,p'}^2(\tilde{p}) \rd \tilde{p} \right)^{\frac{1}{2}} },   \quad i=1, \dots, \rN,  \quad j=1, \dots, \rN.
	\end{equation*}
	With the normal density function $\psi_{i,p}$ given in \eqref{eq:UtilizationDistribution}, this symmetric overlap formula yields	
	\begin{equation} \label{eq:CompetitionKernel_Symmetric}
		\alpha_{ij}(p,p') = \Lambda_{ij} \exp \left( -\frac{(p - p')^2}
		{4 \bar{\rV}_{ij}} \right), \quad i=1, \dots, \rN,  \quad j=1, \dots, \rN,
	\end{equation}
	where $\Lambda_{ij}:=\sqrt{\mathring{\rV}_{ij} /\bar{\rV}_{ij}} $ with $\mathring{\rV}_{ij} := \sqrt{\rV_i \rV_j}$ and $\bar{\rV}_{ij} := \frac{1}{2}(\rV_i + \rV_j)$.
	In Section \ref{sec:EquationsDerivation}, however, the asymmetric kernels \eqref{eq:CompetitionKernel} are used  to derive the equations of the model given in \eqref{eq:NonlinearTerms}.
	Note that, \eqref{eq:CompetitionKernel} can be easily transformed to \eqref{eq:CompetitionKernel_Symmetric} by setting $\kappa=0$ and replacing $\rV_i$ with $\mathring{\rV}_{ij}$.
	\qed
\end{remark}

\subsection{Derivation of equations} \label{sec:EquationsDerivation}

The derivation of equations \eqref{eq:PopulationDensity}--\eqref{eq:NonlinearTerms} begins with the following equation
\begin{align} \label{eq:PhenPopuVariation}
	n_i(x,t+\tau) \phi_i(x,t&+\tau,p) - n_i(x,t)\phi_i(x,t,p) \nonumber\\
	&= \tau \big[ \divergence (\rD_i(x) \dx (n_i(x,t)\phi_i(x,t,p)))  
	+ g_i(x,t,p) n_i(x,t)\phi_i(x,t,p) \nonumber\\
	& + n_i(x,t) \dt^{(\rM)} \phi_i(x,t,p)
	\big], \quad i=1, \dots, \rN, 
\end{align}
wherein, within each species the variation in the population density of individuals with phenotype $p$ over a small time interval $\tau \to 0$ is assumed to result from the contributions of three factors, namely, the diffusive migration of individuals to and from neighboring locations, the intrinsic growth of the population, and the mutational changes in the relative frequency of $p$.

Integrating both sides of \eqref{eq:PhenPopuVariation} with respect to $p$ over $\bbR$, we obtain
\begin{equation} \label{eq:PopulationVariation}
	n_i(x,t+\tau) - n_i(x,t) = \tau \big[ \divergence (\rD_i(x) \dx (n_i(x,t))) +  G_i(x,u(x,t)) n_i(x,t) \big],
\end{equation} 
where 
\begin{equation} \label{eq:GG}
	G_i(x,u(x,t)) := \int_{\bbR} g_i(x,t,p) \phi_i(x,t,p) \rd p
\end{equation}
denotes the mean value of the intrinsic growth rate of the population of individuals with phenotype $p$ within the $i$th species. 
Note that in writing \eqref{eq:GG} we have used \eqref{eq:Mutationeffect} to obtain $\int_{\bbR} \dt^{(\rM)} \phi_i(x,t,p) \rd p =0$.
Moreover, we have implicitly presumed that the mean value of $g_i(x,t,p)$ can be written in terms of $x$ and the variables of the model, $u$.
This is indeed true by the calculations that follow below.
In addition, note that \eqref{eq:PopulationDensity} is obtained immediately by dividing both sides of \eqref{eq:PopulationVariation} by $\tau$ and taking the limit as $\tau \to 0$, provided $G_i(x,u)$ is shown to be given by \eqref{eq:G}. 

Next, to derive \eqref{eq:TraitMean}, we multiply both sides of \eqref{eq:PhenPopuVariation} by $p$ and integrate the result with respect to $p$ over $\bbR$. 
Note that the zero-mean assumption (\ref{assmpt:MutationKernel}) on the mutation distribution, along with \eqref{eq:Mutationeffect}, gives $\int_{\bbR} p \dt^{(\rM)} \phi_i(x,t,p) \rd p =0$.
Therefore, we obtain
\begin{align}  \label{eq:TraitMeanVariation}
	n_i(x,t&+\tau) q_i(x,t+\tau) - n_i(x,t)q_i(x,t) \nonumber \\
	&= \tau \left[ \divergence (\rD_i(x) \dx (n_i(x,t)q_i(x,t)))  
	+ n_i(x,t) \int_{\bbR} p g_i(x,t,p) \phi_i(x,t,p) \rd p \right], 
\end{align}
which further implies, after dividing by $\tau$ and taking the limit as $\tau \to 0$, that
\begin{equation} 
	\dt (n_i(x,t) q_i(x,t))= \divergence (\rD_i(x) \dx (n_i(x,t)q_i(x,t))) + n_i(x,t) \int_{\bbR} p g_i(x,t,p) \phi_i(x,t,p) \rd p.
\end{equation}
Now, using the chain rule on the left hand side of the above equation and substituting \eqref{eq:PopulationDensity} into the result, we obtain
\begin{align*}
	\dt q_i(x,t)&= \frac{1}{n_i(x,t)} \Big[ \divergence (\rD_i(x) \dx (n_i(x,t)q_i(x,t))) 
	- q_i(x,t) \big( \divergence(\rD_i(x) \dx n_i(x,t)) + G_i(x,u(x,t)) n_i(x,t) \big) \Big]\\
	& \quad +  \int_{\bbR} p g_i(x,t,p) \phi_i(x,t,p) \rd p.
\end{align*} 
For the first term within the brackets we can write 
\begin{align*}
	\divergence( \rD_i(x) \dx (n_i(x,t)q_i(x,t)) ) &= 
	\divergence( q_i(x,t) \rD_i(x) \dx n_i(x,t) ) + \divergence( n_i(x,t) \rD_i(x) \dx q_i(x,t) )  \\
	&=q_i(x,t) \divergence(\rD_i(x) \dx n_i(x,t) ) + 2 \inner{\bbR^{\rmm}}{\dx n_i(x,t)}{\rD_i(x) \dx q_i(x,t)} \\
	&\quad + n_i(x,t) \divergence(\rD_i(x) \dx q_i(x,t) ).
\end{align*}
Therefore, it follows that
\begin{equation*}
	\dt q_i(x,t)= \divergence(\rD_i(x) \dx q_i(x,t)) + 2 \inner{\bbR^{\rmm}}{\frac{\dx n_i(x,t)}{n_i(x,t)}}{\rD_i(x) \dx q_i(x,t)}
	\!\!+ H_i(x,u(x,t)),
\end{equation*} 
where
\begin{equation} \label{eq:HH}
	H_i(x,u(x,t)):= \int_{\bbR} p g_i(x,t,p) \phi_i(x,t,p) \rd p - G_i(x,u(x,t)) q_i(x,t).
\end{equation}
This gives \eqref{eq:TraitMean}, provided we show $H_i(x,u)$ can be given by \eqref{eq:H}.

Finally, to derive \eqref{eq:TraitVariance}, we multiply both sides of \eqref{eq:PhenPopuVariation} by $(p-q_i(x,t+\tau))^2$ and integrate the result with respect to $p$ over $\bbR$. 
For the mutation term in \eqref{eq:PhenPopuVariation}, it follows from \eqref{eq:Mutationeffect} and assumption (\ref{assmpt:MutationKernel}) that $\int_{\bbR} (p-q_i(x,t+\tau))^2 \dt^{(\rM)} \phi_i(x,t,p) \rd p = \rU$, where $\rU := \eta \rV_{\scM}$.
Therefore, we obtain
\begin{align*}
	n_i(x,t+\tau) v_i(x,t+\tau) &= \int_{\bbR} (p-q_i(x,t+\tau))^2 \big[ \tau \divergence (\rD_i(x) \dx (n_i(x,t)\phi_i(x,t,p)))  \\
	&\hspace{10.5em} + (1 + \tau g_i(x,t,p)) n_i(x,t)\phi_i(x,t,p) \big] \rd p 
	\mathbf{ + \tau n_i(x,t) \rU} \\
	&= \tau \int_{\bbR} p^2 \big[ \divergence (\rD_i(x) \dx (n_i(x,t)\phi_i(x,t,p))) 
	+ g_i(x,t,p) n_i(x,t) \phi_i(x,t,p) \big] \rd p \\
	&\quad	+ n_i(x,t) \int_{\bbR} p^2 \phi_i(x,t,p) \rd p - n_i(x,t+\tau) q_i^2(x,t+\tau) \mathbf{ + \tau n_i(x,t) \rU} \\
	&= \tau \left[ \divergence (\rD_i(x) \dx (n_i(x,t) v_i(x,t))) + \divergence (\rD_i(x) \dx (n_i(x,t) q_i^2(x,t))) \right]  \\
	&\quad + \tau n_i(x,t) \int_{\bbR} p^2 g_i(x,t,p) \phi_i(x,t,p) \rd p + n_i(x,t) v_i(x,t) 	\\
	&\quad - \left( n_i(x,t+\tau) q_i^2(x,t+\tau) - n_i(x,t) q_i^2(x,t) \right) \mathbf{ + \tau n_i(x,t) \rU}. 
\end{align*}
Dividing both sides of the above equation by $\tau$ and taking the limit as $\tau \to 0$, we obtain  
\begin{align} \label{eq:1252019355}
	\dt(n_i(x,t)v_i(x,t)) &= \divergence (\rD_i(x) \dx (n_i(x,t) v_i(x,t))) + \divergence (\rD_i(x) \dx (n_i(x,t) q_i^2(x,t))) \nonumber\\
	&\quad +  n_i(x,t) \int_{\bbR} p^2 g_i(x,t,p) \phi_i(x,t,p) \rd p - \dt(n_i(x,t) q_i^2(x,t)) \mathbf{ +n_i(x,t) \rU}. 
\end{align}
Note that, 
\begin{align*}
	\divergence (\rD_i(x) \dx (n_i(x,t) q_i^2(x,t))) &= q_i^2(x,t)\divergence (\rD_i(x) \dx n_i(x,t) )
	+ 2 n_i(x,t) q_i(x,t) \divergence (\rD_i(x) \dx q_i(x,t)) \\
	& \quad + 2 n_i(x,t) \inner{\bbR^{\rmm}}{\dx q_i(x,t)}{\rD_i(x) \dx q_i(x,t)} \\
	&\quad + 4 q_i(x,t) \inner{\bbR^{\rmm}}{\dx n_i(x,t)}{\rD_i(x) \dx q_i(x,t)}.
\end{align*}
Moreover, $\dt(n_i(x,t) q_i^2(x,t))$ can be calculated using the chain rule and equations \eqref{eq:PopulationDensity} and \eqref{eq:TraitMean}, wherein $G_i(x,u)$ and $H_i(x,u)$ are given by \eqref{eq:GG} and \eqref{eq:HH}, respectively.
Therefore, \eqref{eq:1252019355} gives
\begin{align} \label{eq:1252019756}
	\dt(n_i(x,t)v_i(x,t)) &= \divergence (\rD_i(x) \dx (n_i(x,t) v_i(x,t))) 
	+ n_i(x,t) \bigg[2 \inner{\bbR^{\rmm}}{\dx q_i(x,t)}{\rD_i(x) \dx q_i(x,t)} \nonumber\\
	&\quad + \int_{\bbR} p^2 g_i(x,t,p) \phi_i(x,t,p) \rd p 
	-2 q_i(x,t) \int_{\bbR} p g_i(x,t,p) \phi_i(x,t,p) \rd p  \nonumber\\
	&\quad + G_i(x,u(x,t)) q_i^2(x,t) \mathbf{ +\rU}\bigg]. 
\end{align}
Now, note that 
\begin{align*}
	\divergence (\rD_i(x) \dx (n_i(x,t) v_i(x,t))) &= v_i(x,t) \divergence(\rD_i(x) \dx n_i(x,t) ) + 2 \inner{\bbR^{\rmm}}{\dx n_i(x,t)}{\rD_i(x) \dx v_i(x,t)} \\
	&\quad + n_i(x,t) \divergence(\rD_i(x) \dx v_i(x,t) ).
\end{align*}
Therefore, using the chain rule on the left hand side of \eqref{eq:1252019756} and substituting \eqref{eq:PopulationDensity} into the result, we obtain
\begin{align*} 
	\dt v_i(x,t) &= \divergence(\rD_i(x) \dx v_i(x,t) ) 
	+ 2 \inner{\bbR^{\rmm}}{\frac{\dx n_i(x,t)}{n_i(x,t)}}{\rD_i(x) \dx v_i(x,t)}\\
	&\quad + 2 \inner{\bbR^{\rmm}}{\dx q_i(x,t)}{\rD_i(x) \dx q_i(x,t)} + W_i(x,u(x,t)),
\end{align*}
where
\begin{align} \label{eq:WW}
	W_i(x,u(x,t)) &:= \int_{\bbR} p^2 g_i(x,t,p) \phi_i(x,t,p) \rd p 
	-2 q_i(x,t) \int_{\bbR} p g_i(x,t,p) \phi_i(x,t,p) \rd p \nonumber \\
	&\quad - G_i(x,u(x,t)) \left( v_i(x,t) - q_i^2(x,t) \right) \mathbf{ +\rU}. 
\end{align}
This gives \eqref{eq:TraitVariance} provided $W_i(x,u)$ is shown to be given by \eqref{eq:W}.

Now, to complete the derivation of \eqref{eq:PopulationDensity}--\eqref{eq:TraitVariance},
it remains to show that $G_i(x,u)$, $H_i(x,u)$, and $W_i(x,u)$ can be given by \eqref{eq:G}, \eqref{eq:H}, and \eqref{eq:W}, respectively. 
For simplicity of exposition, the dependence of functions $n_i$, $q_i$, $v_i$, $g_i$, and $\phi_i$ on variables $x$ and $t$, as well as the dependence of $\rR_i$, $\rK_i$, and $\rQ$ on $x$, are
not explicitly shown in the rest of this section.

We begin with calculating $g_j(p)$.
Using \eqref{eq:TraitDistribution} and \eqref{eq:CompetitionKernel}, the integral in \eqref{eq:GrowthRate} can be written as
\begin{align} \label{eq:Convloution}
	\int_{\bbR} \alpha_{ij}(p,p') \phi_j(p') \rd p' 
	&= \frac{\Lambda_{ij}\exp(\kappa^2 \bar{\rV}_{ij})}{\sqrt{2\pi v_j}} 
	\int_{\bbR} \exp \left( -\frac{(p -p' + 2 \kappa \bar{\rV}_{ij})^2}{4\bar{\rV}_{ij}} \right) 
	\exp\left( -\frac{(p' - q_j)^2}{2 v_j} \right) \rd p' \nonumber\\
	&= \frac{\Lambda_{ij}\exp(\kappa^2 \bar{\rV}_{ij})}{\sqrt{2\pi v_j}}  \hat{M}_j(u) \int_{\bbR} \hat{A}_j(p',u) \rd p', 
\end{align}
where
\begin{align} \label{eq:Btilde}
	\hat{M}_j(u) &:=\exp \left( -\frac{ -\dfrac{[2 v_j (p + 2 \kappa \bar{\rV}_{ij}) + 4 \bar{\rV}_{ij} q_j ]^2 }{2 v_j + 4 \bar{\rV}_{ij}} + 2 v_j (p + 2 \kappa \bar{\rV}_{ij})^2 + 4 \bar{\rV}_{ij} q_j^2 }{(4\bar{\rV}_{ij}) (2 v_j)} \right) \nonumber\\
	&= \exp \left( -\frac{[q_j - (p + 2 \kappa \bar{\rV}_{ij})]^2}{2 v_j + 4 \bar{\rV}_{ij}} \right),
\end{align}
and
\begin{equation} \label{eq:Atilde}
	\hat{A}_j(p',u) :=\exp \left( -\frac{(p' - \hat{\mu}_j(u))^2}{2 \hat{\sigma}_j^2(u)} \right), 
\end{equation}
with \begin{equation} \label{eq:muTildesigmaTilde}
	\hat{\mu}_j(u):= \frac{2 v_j (p + 2 \kappa \bar{\rV}_{ij}) + 4 \bar{\rV}_{ij} q_j}{2 v_j + 4 \bar{\rV}_{ij}}, \quad
	\hat{\sigma}_j^2(u):= \frac{1}{2} \frac{(4 \bar{\rV}_{ij})(2v_j)}{2 v_j + 4 \bar{\rV}_{ij}}.
\end{equation}
Note that $\int_{\bbR} \hat{A}_j(p',u) \rd p' = \sqrt{2 \pi}\hat{\sigma}_j(u)$.
Therefore, substituting the results into \eqref{eq:GrowthRate}, we obtain
\begin{align} \label{eq:GrRateCalculated}
	g_i(p) = \rR_i - \frac{\rR_i}{\rK_i} \sum_{j=1}^{\rN} \frac{\sqrt{2 \bar{\rV}_{ij}} \Lambda_{ij} \exp(\kappa^2 \bar{\rV}_{ij})}{\sqrt{\smash[b]{v_j + 2 \bar{\rV}_{ij}}}} 
	\exp \left( -\frac{[q_j - (p + 2 \kappa \bar{\rV}_{ij})]^2}{2 v_j + 4 \bar{\rV}_{ij}} \right) n_j 
	-\frac{\rS}{2}(p-\rQ)^2.
\end{align}
Now, we substitute \eqref{eq:GrRateCalculated} into \eqref{eq:GG} to calculate $G_i(x,u)$.
Note that, 
\begin{align}
	\int_{\bbR} \rR_i \phi_i(p) \rd p &= \rR_i,   \label{eq:G_First}\\
	\int_{\bbR} \frac{\rS}{2} (p-\rQ)^2 \phi_i(p) \rd p 
	&= \frac{\rS}{2 } \int_{\bbR} (p^2 - 2 p \rQ  + \rQ^2) \phi_i(p) \rd p
	=  \frac{\rS}{2} \left[ (v_i + q_i^2) - 2 q_i \rQ  + \rQ^2) \right] \nonumber\\
	&= \frac{\rS}{2} \left[ (q_i - \rQ)^2 + v_i \right]. \label{eq:G_Second}
\end{align}
Moreover, the integral associated with the term inside the summation in \eqref{eq:GrRateCalculated} can be calculated using similar calculation as given for \eqref{eq:Convloution}.  
Specifically, as compared with \eqref{eq:Convloution}--\eqref{eq:muTildesigmaTilde}, 
\begin{equation} \label{eq:G_Third}
	\frac{1}{\sqrt{2 \pi v_i}} 
	\int_{\bbR} \exp \left( -\frac{[q_j - (p + 2 \kappa \bar{\rV}_{ij})]^2}{2 v_j + 4 \bar{\rV}_{ij}} \right)
	\exp\left( -\frac{(p - q_i)^2}{2 v_i} \right) \rd p 
	= \frac{M_{ij}(u)}{\sqrt{2 \pi v_i}} \int_{\bbR} A_{ij}(p,u) \rd p,
\end{equation}
where $M_{ij}(u)$ is given in \eqref{eq:NonlinearTerms} and
\begin{equation} \label{eq:AhatBhat}
	A_{ij}(p,u) :=\exp \left( -\frac{(p - \mu_{ij}(u))^2}{2 \sigma_{ij}^2(u)} \right), 
\end{equation}
with
\begin{align} \label{eq:muSigma}
	\mu_{ij}(u):= \frac{2 v_i (q_j - 2 \kappa \bar{\rV}_{ij}) + (2 v_j + 4 \bar{\rV}_{ij}) q_i}{2 v_i + (2 v_j + 4 \bar{\rV}_{ij})}, \quad
	\sigma_{ij}^2(u):= \frac{1}{2} \frac{(2 v_j + 4 \bar{\rV}_{ij})(2v_i)}{2 v_i + (2 v_j + 4 \bar{\rV}_{ij})}.
\end{align}
Now, note that $\int_{\bbR} A_{ij}(p,u) \rd p = \sqrt{2 \pi} \sigma_{ij}(u)$.
Therefore, substituting \eqref{eq:GrRateCalculated} into \eqref{eq:GG}, using \eqref{eq:G_First}--\eqref{eq:muSigma}, and letting $C_{ij}(u)$ be defined as in \eqref{eq:NonlinearTerms}, 
the mean growth rate $G_i(x,u)$ is obtained as given by \eqref{eq:G}.

Next, we substitute \eqref{eq:GrRateCalculated} into \eqref{eq:HH} to calculate $H_i(x,u)$.
We can write
\begin{align}
	\int_{\bbR} p \rR_i \phi_i(p) \rd p &= \rR_i q_i,   \label{eq:H_First}\\
	-\int_{\bbR} p \frac{\rS}{2}(p-\rQ)^2 \phi_i(p) \rd p 
	&= -\frac{\rS}{2} \int_{\bbR} (p^3 - 2 p^2 \rQ  + p \rQ^2 ) \phi_i(p) \rd p  \nonumber\\
	&=  - \frac{\rS}{2} \left[ (3 v_i q_i + q_i^3) - 2 (v_i + q_i^2) \rQ  + q_i \rQ^2 ) \right].  \label{eq:H_Second}
\end{align}
Note that \eqref{eq:H_Second} is equal to $E_i(x,u)$ as defined  in \eqref{eq:NonlinearTerms}. 
Moreover, as compared with \eqref{eq:Convloution}--\eqref{eq:muTildesigmaTilde},
\begin{equation} \label{eq:H_Third}
	\frac{1}{\sqrt{2 \pi v_i}} 
	\int_{\bbR} p \exp \left( -\frac{[q_j - (p + 2 \kappa \bar{\rV}_{ij})]^2}{2 v_j + 4 \bar{\rV}_{ij}} \right)
	\exp\left( -\frac{(p - q_i)^2}{2 v_i} \right) \rd p 
	= \frac{M_{ij}(u)}{\sqrt{2 \pi v_i}} \int_{\bbR} p A_{ij}(p,u) \rd p,
\end{equation}
where $M_{ij}(u)$ and $A_{ij}(p,u)$ are given by \eqref{eq:NonlinearTerms} and \eqref{eq:AhatBhat}, respectively.
Therefore, we have $\int_{\bbR} p A_{ij}(p,u) \rd p = \sqrt{2 \pi} \sigma_{ij}(u) \mu_{ij}(u)$, 
where $\mu_{ij}(u)$ and $\sigma_{ij}(u)$ are given by \eqref{eq:muSigma}.
Now, letting $L_{ij}(u)$ be defined as in \eqref{eq:NonlinearTerms}, the equation \eqref{eq:HH} along with \eqref{eq:GrRateCalculated} and \eqref{eq:muSigma}--\eqref{eq:H_Third} gives $H_i(x,u)$ as in \eqref{eq:H}. 

Finally, we use \eqref{eq:WW} with  \eqref{eq:GrRateCalculated} and \eqref{eq:HH} to calculate $W_i(x,u)$.
Note that,
\begin{align}
	\int_{\bbR} p^2 \rR_i \phi_i(p) \rd p &= (v_i + q_i^2) \rR_i ,   \label{eq:W_First}\\
	-\int_{\bbR} p^2 \frac{ \rS}{2} (p-\rQ)^2 \phi_i(p) \rd p 
	&= - \frac{\rS}{2} \int_{\bbR} (p^4 - 2 p^3 \rQ  + p^2 \rQ^2 ) \phi_i(p) \rd p  \nonumber\\
	&=  -\frac{\rS}{2} \left[ (3 v_i^2 + 6 v_i q_i^2 + q_i^4) - 2 (3 v_i q_i + q_i^3) \rQ  + (v_i + q_i^2) \rQ^2 ) \right]  =: \hat{Y}_i(x,u), \label{eq:W_Second}
\end{align}
and, as compared with \eqref{eq:Convloution}--\eqref{eq:muTildesigmaTilde},
\begin{equation} \label{eq:W_Third}
	\frac{1}{\sqrt{2 \pi v_i}} 
	\int_{\bbR} p^2 \exp \left( -\frac{[q_j - (p + 2 \kappa \bar{\rV}_{ij})]^2}{2 v_j + 4 \bar{\rV}_{ij}} \right)
	\exp\left( -\frac{(p - q_i)^2}{2 v_i} \right) \rd p 
	= \frac{M_{ij}(u)}{\sqrt{2 \pi v_i}} \int_{\bbR} p^2 A_{ij}(p,u) \rd p,
\end{equation}
where $M_{ij}(u)$ and $A_{ij}(p,u)$ are given by \eqref{eq:NonlinearTerms} and \eqref{eq:AhatBhat}, respectively.
It follows that $\int_{\bbR} p^2 A_{ij}(p,u) \rd p = \sqrt{2 \pi} \sigma_{ij}(u) [\sigma_{ij}^2(u) + \mu_{ij}^2(u)]$, 
where $\mu_{ij}(u)$ and $\sigma_{ij}(u)$ are given by \eqref{eq:muSigma}.
Therefore, the first integral in \eqref{eq:WW} can be written as
\begin{equation} \label{eq:12920191204}
	\int_{\bbR} p^2 g_i(p) \phi_i(p) \rd p = (v_i + q_i^2) \rR_i 
	-\frac{\rR_i}{\rK_i} \sum_{j=1}^{\rN} \hat{P}_{ij}(u) M_{ij}(u) C_{ij}(u) + \hat{Y}_i(x,u)
\end{equation}
where $\hat{Y}_i(x,u)$ is given by \eqref{eq:W_Second} and
\begin{equation}\label{eq:Shat}
	\hat{P}_{ij}(u) := \frac{v_i(v_j + 2 \bar{\rV}_{ij})}{v_i + v_j + 2 \bar{\rV}_{ij}} + L_{ij}^2(u).
\end{equation}
Moreover, the second integral in \eqref{eq:WW} can be calculated immediately using \eqref{eq:HH} and \eqref{eq:H}.
The result along with \eqref{eq:12920191204} gives $W_i(x,u)$ as in \eqref{eq:W}, wherein
$P_{ij}(u) = \hat{P}_{ij}(u) - 2 q_i L_{ij}(u)$ and 
$Y_i(x,u) = \hat{Y}_i(x,u) - 2 q_i E_i(x,u) \mathbf{ +\rU}$.
Note that, using \eqref{eq:W_Second} and \eqref{eq:Shat}, we obtain $P_{ij}(u)$ and $Y_i(x,u)$ as given in \eqref{eq:NonlinearTerms}. 
This completes the derivation of the model.

Finally, the homogeneous Neumann boundary conditions \eqref{eq:ReflectingBC} are obtained by assuming no phenotypic flux through the boundaries, that is,
\begin{equation} \label{eq:71720201144}
\rD_i(x) \dx( n_i(x,t) \phi_i(x,t,p) ) = 0, \quad i=1,\dots, \rN, \quad \text{for all } (x,t)\in \{a,b\} \times [0,T] \text{ and all } p \in \bbR.
\end{equation} 
Integrating this condition with respect to $p$ over $\bbR$ and noting that in general $\rD_i$ is nonzero on the boundary, we obtain the boundary condition $\dx n_i = 0$ as in \eqref{eq:ReflectingBC}.
Moreover, it follows from multiplying \eqref{eq:71720201144} by $p$, integrating the result over $\bbR$, and using the condition $\dx n_i = 0$, that $\dx(n_i q_i) = n_i \dx q_i = 0$ on the boundary.
This gives the boundary condition $\dx q_i = 0$ given in \eqref{eq:ReflectingBC}, since $n_i$ is not required to be zero on the boundary under a no-flux condition.
The boundary condition $\dx v_i = 0$ given in \eqref{eq:ReflectingBC} is obtained similarly by multiplying \eqref{eq:71720201144} by $(p - q_i(x,t))^2$ and integrating the result over $\bbR$.

\section*{Appendix B: Numerical methods and discretization parameters} 

The numerical solutions presented in Sections \ref{sec:SingleSpecies} and \ref{sec:TwoSpecies} have been computed using an Alternating Direction Implicit (ADI) scheme with two stabilizing correction stages, as presented by \citet[equ. (20)]{Hundsdorfer:AppNumMath:2002}.
The parameter $\theta$ in the formulation of this scheme is set to $\theta = 1/2$.
The function $F$ in the formulation has two components when we consider a one-dimensional geographic space.
One of these components is associated with the terms involving spatial derivatives, and the other component is associated with the reaction terms. 
For the two-dimensional space considered in Section \ref{sec:MultipleCompetitionFactors}, the function $F$ has three components, first component associated with derivatives with respect to $x_1$, second component associated with derivatives with respect to $s_2$, and third component associated with the reaction terms.
We treated all components in both spatial dimensions implicitly.  
For further details of this numerical method see the results developed by \citet{Hundsdorfer:AppNumMath:2002} and \citet{Hout:AppNumMath:2007}.

In each iteration of the scheme, instead of solving the nonlinear algebraic equations of the scheme using Newton's method, we have solved the linearized version of these equations. 
This is in fact equivalent to performing only one Newton iteration.
The required changes in the formulations to incorporate this linearization step is also provided by \citet{Hundsdorfer:AppNumMath:2002}.
The computational time that is saved by solving linearized equations can then be used to allow for smaller time steps, which can in turn compensate for the loss of accuracy due to the linearization. 
Linearizing the scheme has the advantage of providing a better control over the total computation time of the simulations, as the computation time will then become almost linearly proportional to the time steps.

Finally, we have used fourth-order centered difference approximations for both first and second derivatives in each spatial direction.
In one-dimensional space, we have considered a uniform discretization mesh of size $\Delta x = 0.1 \, \tX$, as well as uniform time steps of length $\Delta t = 0.002 \, \tT$.
For the two-dimensional problem of Section~\ref{sec:MultipleCompetitionFactors}, we have used a rectangular mesh of size $\Delta x_1 \times \Delta x_2 = 0.5 \times 0.5 \, \tX^2$, and a time step of $\Delta t = 0.01 \, \tT$.

\bibliographystyle{abbrvnat}      
\bibliography{References}

\end{document}